\documentclass[a4paper,fleqn,usenatbib]{mnras}

\usepackage{newtxtext,newtxmath}

\usepackage[T1]{fontenc}
\usepackage{ae,aecompl}

\usepackage{graphicx}	
\usepackage{amsmath}	
\usepackage{amssymb}	

\usepackage[usenames]{xcolor} 
\usepackage{color}
\usepackage{ulem}

\title[Thermal torque effects on the migration of growing low-mass planets]{Thermal torque effects on the migration of growing low-mass planets}

\author[Guilera et al.]{O. M. Guilera,$^{1,2,3,4}$\thanks{E-mail: oguilera@fcaglp.unlp.edu.ar}
N. Cuello$^{3,4}$, M. Montesinos$^{4,5,6}$, M. M. Miller Bertolami$^{1,2}$
\newauthor M. P. Ronco$^{3,4}$, J. Cuadra$^{3,4}$ \& F. S. Masset$^{7}$\\ \\
$^{1}$ Instituto de Astrof\'{\i}sica de La Plata, CONICET-UNLP, La Plata, Argentina. \\
$^{2}$ Facultad de Ciencias Astron\'omicas y Geof\'{\i}sicas, UNLP, La Plata, Argentina. \\
$^{3}$ Instituto de Astrof\'{\i}sica, Pontificia Universidad Catolica de Chile, Santiago, Chile. \\
$^{4}$ N\'ucleo Milenio de Formaci\'on Planetaria (NPF), Chile. \\
$^{5}$ Instituto de F\'{\i}sica y Astronom\'{\i}a, Universidad de Valpara\'{\i}so. Valpara\'{\i}so, Chile. \\
$^{6}$ Chinese Academy of Sciences South America Center for Astronomy, National Astronomical Observatories, CAS, Beijing, China \\
$^{7}$ Instituto de Ciencias F\'{\i}sicas, Universidad Nacional Aut\'onoma de M\'exico, Av. Universidad s/n, 62210 Cuernavaca, Mor., Mexico\\
} 
\date{Accepted XXX. Received YYY; in original form ZZZ}

\pubyear{2019}

\begin{document}
\label{firstpage}
\pagerange{\pageref{firstpage}--\pageref{lastpage}}
\maketitle

\begin{abstract}
As planets grow the exchange of angular momentum with the gaseous component of the protoplanetary disc produces a net torque resulting in a variation of the semi-major axis of the planet. For low-mass planets not able to open a gap in the gaseous disc this regime is known as type I migration. Pioneer works studied this mechanism in isothermal discs finding fast inward type I migration rates that were unable to reproduce the observed properties of extrasolar planets. In the last years, several improvements have been made in order to extend the study of type I migration rates to non-isothermal discs. Moreover, it was recently shown that if the planet's luminosity due to solid accretion is taken into account, inward migration could be slowed down and even reversed. In this work, we study the planet formation process incorporating, and comparing, updated type I migration rates for non-isothermal discs and the role of planet's luminosity over such rates. We find that the latter can have important effects on planetary evolution, producing a significant outward migration for the growing planets.
\end{abstract}



\section{Introduction}
\label{Intro}

In the classical core accretion mechanism \citep{Safronov1969,Mizuno1980,Stevenson1982, bodenheimer_pollack1986, GL1992}, the formation of both terrestrial and giant planets occurs through subsequent stages. Immediately after the collapse of the primordial gas cloud a circumstellar disc is formed. The dust of the protoplanetary disc quickly settles into the disc mid-plane, where it coagulates to form larger solid structures known as planetesimals. Then planetesimals coaggregate to form planetary embryos, which can then grow to form planets.

At the moment, there are two main scenarios for planetesimal formation: collisional growth models, wherein a gradual growth by continuous collisions is invoked; or gravoturbulent models, wherein the gravitational collapse of dust accumulation is invoked \citep[see][and references therein for an update review]{Blum2018}. The first scenario predicts that most of the dust particles coagulate to form planetesimals, which continue growing by mutual accretion until planetary embryos of about the mass of the Moon form. At this stage, these are the only bodies that continue growing by the accretion of planetesimals in a process often called oligarchic growth regime \citep{IdaMakino1993,KokuboIda1998,Ormel.et.al2010}. In the second scenario, dust particles are concentrated in high pressure regions in sufficiently high amount to be able to trigger the streaming instability. In this case, the planetesimals are directly formed by the gravitational collapse of the dust aggregates \citep{YoudinGoodman2005,Johansen.et.al.2007}.

Once planetesimals are formed they can efficiently accrete small particles and quickly form massive solid embryos \citep{Ormel&Klahr2010,J&L2010,Lambrechts&Johansen2012,Lambrechts.et.al.2014, Chambers2016}. There are two necessary conditions for this process: massive planetesimals combined to a high concentration of small particles.

Dust particles and small planetesimals experience a fast radial-drift. This is because the gaseous component of the disc rotates at a sub-Keplerian velocity, while dust and planetesimals orbit at Keplerian velocities. This difference causes a drag force on the solid bodies, which results in a loss of angular momentum. As bodies grow and become planetary embryos, the drag force becomes negligible. However, planetary embryos are able to locally modify the gas density structure. This effect leads to an exchange of angular momentum, which results in a net torque that makes the planets migrate within the disc. Different regimes of migration have been studied during the last decades, starting with the pioneer works of \citet{LinPapaloizou1986} and \citet{Ward1997}. Planet--disc interactions are key in the orbital evolution of forming planets, leading to potentially large variations in their semi-major axis. For planets not massive enough to open a gap in the gaseous disc, the migration regime is usually known as type I migration. Early models of type I migration were developed for isothermal discs \citep{Tanaka.et.al.2002}. Such models found very high inward type I migration rates, so planets migrate too fast towards the central star. Therefore, in order to reproduce observations, many authors had to reduce by up to a factor of 0.001 these previous migration rates \citep{Ida-Lin2004, Alibert2005, Miguel2011}.

Numerical studies including further physical processes --- magnetic fields \citep{Guilet.et.al.2013} and non-isothermal protoplanetary discs for instance ---  have shown that type I migration can be slowed down, and even reversed. In hydrodynamical simulations of non-isothermal discs --- in 2D and 3D --- it was found that the outward type I migration strongly depends on the planetary mass and semi-major axis, but also on the disc structure and thermodynamics \citep{PM2006, BM2008, PP2008, Kley.et.al.2009}. Later, \citet{MC2010} and \citet{paardekooper.etal2010, paardekooper.etal2011} derived semi-analytical prescriptions for type I migration rates in non-isothermal discs, which can be incorporated in planet formation models. In fact, the migration rates derived by \citet{paardekooper.etal2010, paardekooper.etal2011} are the most often employed in planet formation models that consider non-isothermal discs \citep{Fortier2013,Alibert.et.al.2013,CL2014,Dittkrist.et.al.2015,Bitsch.et.al.2015b,Bitsch.et.al.2015,CL2016,Guilera.et.al.2017,Izidoro2017,KB2018,Ndugu2018}. It is worth noting that the type I migration rate prescriptions given by \citet{MC2010} and \citet{paardekooper.etal2010, paardekooper.etal2011} were derived from the results of 2D hydrodynamical simulations. Although they account for most of the mechanisms that contribute to the total torque over the planet, these prescriptions do not match the more physical 3D radiative hydrodynamic simulations with enough accuracy \citep{Kley.et.al.2009, BK2011}.

More recently, however, \citet{jm2017} updated the torque formula of \citet{MC2010} by means of 3D numerical simulations. This resulted in an improvement in the derived type I migration prescriptions for both the Lindblad and the corotation torques with respect to the aforementioned works. These new type I migration recipes were compared with the results from the 3D radiative hydrodynamical simulations from \citet{BK2011} and \citet{Lega2015}, showing good agreement.

Using 3D radiative hydrodynamical simulations, \citet{Lega.et.al.2014} found that in the proximity to a low-mass planet the gas is cooler and denser than it would be if it behaved adiabatically. Such effect is asymmetric, generating an additional negative {\it cold torque}, which generates an inward migration. Additionally, \citet{Benitez-llambay.et.al.2015} found that the heat released by a planet due to accretion of solids is diffused in the nearby disc, generating two asymmetric hot and low-density lobes. This effect produces a {\it heating torque} that can become positive, depending on the amount of heat released by the planet. Then, \citet{Chrenko2017} also performed radiative hydrodynamical simulations considering growing planets immersed in the disc. They also found planet outward migration if the heat released into the disc by the planets due to the accretion of solids is taken into account. However, neither of these works derived analytical prescriptions that reproduced such phenomena.

It is important to note that the effects found by \citet{Lega.et.al.2014} and \citet{Benitez-llambay.et.al.2015} were not considered by \citet{jm2017} for the derivation of their migration recipes. However, these two phenomena were studied analytically in a novel work by \citet{masset2017}, who proposed a semi-analytical prescription for the {\it thermal torque}, defined by the author as the sum of the two aforementioned effects. This torque can be incorporated in planet formation models.

In this work, we update our planet formation model {\scriptsize PLANETALP} \citep{Ronco.et.al.2017, Guilera.et.al.2017} by implementing the type I migration rates derived by \citet{jm2017}, along with the thermal torque derived by \citet{masset2017}. We aim to study the impact of these new prescriptions on planet formation. In doing so, we pay particular attention to the thermal torque. This work is organised in the following way: in Sect.~\ref{Sec2}, we describe our planet formation model and the improvements to the type I migration rates. Readers not interested in the technical details of the simulations can directly jump to  Sect.~\ref{sec3}, where we present the migration maps obtained with these new prescriptions and the resulting impact on planet formation. Finally, we discuss our findings and draw our conclusions in Sect.~\ref{sec4}.

\section{Description of {\small PLANETALP}}
\label{Sec2}

In a series of previous works \citep{Guilera2010, Guilera2014, Guilera.et.al.2017, Guilera.Sandor.2017, Ronco.et.al.2017}, we developed a planet formation code called {\scriptsize PLANETALP}. It models the formation of a planetary system immersed in a protoplanetary disc that evolves in time. The protoplanetary disc is modelled by two components: a gaseous and a solid one. Planets immersed in the disc start to grow by the accretion of solid material. Once they are massive enough, they are also able to accrete the surrounding gas. The growing planets are allowed to migrate under type I and type II migration along the disc. Even though {\scriptsize PLANETALP} allow us to calculate the simultaneous formation of several planets, in this work we will only calculate the formation of only one planet per disc.  

Below, we describe the most relevant features of {\scriptsize PLANETALP}, especially the new type I migrations rates incorporated in this work. 

\subsection{The gaseous disc}
\label{Sec2-1}

In the latest version of {\scriptsize PLANETALP} \citep{Guilera.et.al.2017}, the disc gaseous component is modelled by a standard 1D+1D viscous accretion disc. We consider an axisymmetric, irradiated disc in hydrostatic equilibrium. To calculate the vertical structure of the disc, we solve for each radial bin the structure equations \citep{Papaloizou.Terquem.1999,Alibert2005, Migaszewski2015}
\begin{align}
  \frac{\partial P}{\partial z} & =  -\rho \Omega^2 z, \nonumber\\  
  \frac{\partial F}{\partial z} & =  \frac{9}{4} \rho \nu \Omega^2, \label{eq1-sec2-1} \\  
  \frac{\partial T}{\partial z} & =  \nabla \frac{T}{P}\frac{\partial P}{\partial z}. \nonumber 
\end{align}
Where $P$, $\rho$, $F$, $T$ and $z$ represent the pressure, density, radiative heat flux, temperature, and vertical coordinate of the disc, respectively. We note $\Omega$ the Keplerian frequency at a given radial distance and $\nu= \alpha c_s^2/\Omega$ the viscosity \citep{Shakura1973}, with $\alpha$ a free parameter (we varied $\alpha$ between $10^{-4}$ and $10^{-2}$) and $c_s^2= P/\rho$ the square of the local isothermal sound speed. In our model, convection sets in when the standard Schwarzschild criterion is fulfilled. In convective regions, the vertical temperature gradient $\nabla= d \log T / d \log P$ is calculated considering the mixing length theory, following the prescription of \cite{2012sse..book.....K}; while in purely radiative regions, the temperature gradient $\nabla_{\rm rad}$ is given by $\nabla_{\text{rad}}= 3\kappa \rho F/16 \sigma \Omega^2 z T^4$, where $\sigma$ is the Stefan-Boltzmann constant and $\kappa$ is the local Rosseland mean opacity \citep{Bell.Lin.1994}. We adopt the equation of state of an ideal diatomic gas $P= \rho k T / \mu m_{\text{H}}$, where $k$ is the Boltzmann constant, $\mu$ the mean molecular weight, $m_{\text{H}}$ the mass of the hydrogen atom, and $\nabla_{\text{ad}}$ the adiabatic gradient. Since the main component of the gas disk is molecular hydrogen, it is reasonable to assume as in previous works that $\mu= 2$ and $\nabla_{\text{ad}}= 2/7$. 

To solve the system of Eqs.~\ref{eq1-sec2-1}, we consider that the boundary conditions at the surface of the disc $H$ as in \cite{Papaloizou.Terquem.1999}, \cite{Alibert2005} and \cite{Migaszewski2015}. Namely, $P_s= P(z=H)$, $F_s= F(z=H)$, and $T_s= T(z=H)$. These quantities are given by:
\begin{align}
  P_s & =  \frac{\Omega^2 H \tau_{\text{ab}}}{\kappa_s},\nonumber \\ 
  F_s & =  \frac{3 \dot{M}_{\text{st}} \Omega^2}{8\pi}, \label{eq2-sec2-1}\\  
  0   & =  2 \sigma \left( T_s^4 + T_{\text{irr}}^4 - T_{\text{b}}^4 \right) - \frac{9 \alpha \Omega k (T_s^4 + T_{\text{irr}}^4)^{1/4}}{8 \kappa_s \mu m_{\text{H}}} - F_s, \nonumber
\end{align}
where $\tau_{\text{ab}}= 10^{-2}$ is the optical depth, $T_{\text{b}}= 10$~K is the background temperature, and $\dot{M}_{\text{st}}$ is the equilibrium accretion rate. The temperature associated with the stellar irradiation is given by:
\begin{equation}
  T_{\text{irr}}= T_{\star} \left[ \frac{2}{3\pi} \left( \frac{R_{\star}}{R} \right)^3 + \frac{1}{2} \left( \frac{R_{\star}}{R} \right)^2 \left( \frac{H}{R} \right) \left( \frac{d\log H}{d\log R} - 1 \right) \right]^{0.5},
  \label{eq3-sec2-1} 
\end{equation}
where $R$ is the radial coordinate, and $d\log H/d\log R= 9/7$ \citep{CG1997}. We note that here we are not considering the self-shadowing of the disc \citep{Baillie2015,Baillie2016,Bitsch.et.al.2015b,Migaszewski2015}. The central object is assumed to be a $1\,M_\odot$ protostar with $R_{\star}= 2~R_{\odot}$ and $T_{\star}= 4000$~K \citep{Baraffe2015}. We note that the stellar luminosity does not evolve in time in our model. In addition to the surface boundary conditions, in the mid-plane of the disc the heat flux should vanish: $F(z=0)= F_0 =0$. The system of Eqs.~\ref{eq2-sec2-1} is solved by a multidimensional Newton-Raphson algorithm. The vertical structure of the disc is solved following \citet{Alibert2005} and \citet{Migaszewski2015}, using a shooting method with a Runge-Kutta-Fehlberg integrator. 

Then, the time evolution of the gas surface density $\Sigma_{\text{g}}$ is calculated solving the classical diffusion equation of \cite{Pringle1981}:
\begin{equation}
  \frac{\partial \Sigma_{\text{g}}}{\partial t}= \frac{3}{R}\frac{\partial}{\partial R} \left[ R^{1/2} \frac{\partial}{\partial R} \left( \bar{\nu} \Sigma_{\text{g}} R^{1/2}  \right) \right],  
\label{eq4-sec2-1}
\end{equation}
where $\bar{\nu}$ is the mean viscosity. Therefore, to solve Eq.~\ref{eq4-sec2-1}, we need to know the radial profile of $\bar{\nu}$ at each time-step as a function of the corresponding $\Sigma_{\text{g}}$, i.e. $\bar{\nu}= \bar{\nu}(\Sigma_{\text{g}}, \,R)$. To do this, we solve the vertical structure of the disc only at the beginning of the simulation, for a fixed value of the $\alpha$-parameter and varying $\dot{M}_\text{st}$ from $10^{-16}~\text{M}_{\odot}/\text{yr}$ to $10^{-4}~\text{M}_{\odot}/\text{yr}$ using a dense grid between these values. For each $\dot{M}_\text{st}$ --solving the vertical structure of the disc-- we calculate the radial profiles of $\Sigma_{\text{g}}$ and $\bar{\nu}$ assuming that at each radial bin the disc is in a steady state (i.e. $\dot{M}_\text{st}= 3\pi\nu\Sigma_\text{g}$). Then, from the initial gas surface density radial profile we interpolate at each radial bin the corresponding value of $\dot{M}_\text{st}$. With this information we can then interpolate the value of the corresponding mean viscosity at each radial bin. Given the initial gas surface density and mean viscosity radial profiles we solve for the first time-step the diffusion equation, obtaining a new radial profile for the gas surface density, and repeating the aforementioned mechanism. Finally, we note that Eq.~\ref{eq4-sec2-1} is solved using an implicit Crank-Nicholson method considering zero torques as boundary conditions, which is equivalent to consider zero density as boundary condition. We also remark that the radial profiles time evolution of the other relevant thermodynamic quantities at the disc mid-plane (temperature, pressure, opacities, etc.), needed to calculate the planet migration rates, are calculated in the same way as the mean viscosity radial profiles \citep[see the extended version on the arXiv of][for details]{Migaszewski2015}.

Throughout this work, the initial surface density of the disc is assumed to be equal to
\citep{Andrews2010}:
\begin{equation}
\Sigma_{\text{g}}(R) = \Sigma^{0}_{\text{g}}\left(\dfrac{R}{R_{\text{c}}}\right)^{-b}e^{-(R/R_{\text{c}})^{2-b}},
\label{eq1-sec3-1}
\end{equation}  
where $R_{\text{c}}= 39$~au is the characteristic disc radius,
$b= 1$, and $\Sigma_{\text{g}}^{0}$ is a normalisation constant. The latter is a function of the disc mass $\text{M}_{\text{d}}$. In all our simulations, the characteristic disc radius and the exponent $b$ are kept fixed. For our fiducial model, we assume a disc with a mass of $0.05~\text{M}_{\odot}$, and we adopt a value of $\alpha=10^{-3}$ for the Shakura-Sunyaev $\alpha$-viscosity parameter \citep{Shakura1973}. As we are interested in the early stages of planet formation, we only analyse the first million years of the disc evolution.

\subsection {The solid component of the disc }
\label{Sec2-2}

In {\scriptsize PLANETALP}, the solid component of the disc is characterised by a particle population (planetesimals or smaller particles as pebbles), represented by a solid surface density. The initial solid surface density is given by
\begin{equation}
\Sigma_\text{p}(R)= z_0 \eta_{\text{ice}} \Sigma_\text{g}(R).
\label{eq0-sec2-2}
\end{equation}
Here $z_0$ is the initial dust to gas ratio \citep[in this work we assume $z_0= 0.0153$, the initial solar metallicity][]{Lodders.et.al.2009}, and $\eta_{\text{ice}}$ represents the sublimation of volatiles inside the ice-line. $\eta_{\text{ice}}$ is equal to $1$ and $1/3$ outside and inside the ice-line, respectively. In our model, the ice-line is located at the distance to the central star where the initial temperature at the disc mid-plane drops to 170~K (for the fiducial disc it is located at $\sim 3$~au). We note that the ice-line does not evolve in time, and it is only used to define the initial solid surface density (which decreases as we move far away from the central star).

This population evolves according to the drift due to the nebular gas considering the Epstein, Stokes and quadratic regimes \citep{Rafikov2004,Chambers2008}, and according to the accretion by the embryos (see next section). In general, small particles as pebbles are always in the Epstein regime along the disc; small planetesimals are either in the Stokes or quadratic regimes; and big planetesimals ($\gtrsim 10$~km) are in the quadratic regime along the disc. For each radial bin, the drift velocities of the solid particles are calculated corresponding to the drag regime. In the case of planetesimals, we also take into account ejection/dispersion processes \citep[see][for the approach employed]{Ronco.et.al.2017}. 

As in previous works, we take the particle eccentricities and inclinations out of equilibrium considering the planet gravitational excitations \citep{Ohtsuki.et.al.2002}, and the damping due to the nebular gas drag \citep{Rafikov2004,Chambers2008}. The numerical treatment of these phenomena is described in detail in \citet{Guilera2014} and \citet{Guilera.Sandor.2017}. Although {\scriptsize PLANETALP} allows us to model the collisional evolution of the particle population \citep{Guilera2014, SanSebastian.et.al.2018}, due to its high computational cost, this effect is not taken into account in the present work.

Finally, the time evolution of the solid surface density is calculated by using the following continuity equation:
\begin{eqnarray}
 \frac{\partial}{\partial t} \left[\Sigma_{\text{p}}(R,r_{p_j})\right] &+& \frac{1}{R} \frac{\partial}{\partial R} \left[ R \,v_{\text{mig}}(R,r_{p_j})\,\Sigma_{\text{p}}(R,r_{p_j}) \right] = \nonumber \\ 
&=& \dot{\Sigma}^{\text{tot}}_{\text{p}}(R,r_{p_j}), 
\label{eq1-sec2-2}
\end{eqnarray}  
where $\dot{\Sigma}^{\text{tot}}_{\text{p}}(R,r_{p_j})$ represents the sink terms due to the accretion (and ejection/dispersion in the case of planetesimals) by the planets \citep[see][for numerical details]{Ronco.et.al.2017} and $v_{\text{mig}}(R,r_{p_j})$ is the particle drift velocity. The $r_{p_j}$ dependence emphasises that Eq.~\ref{eq1-sec2-2} is solved independently for each particle size for a given planetesimal distribution. In this case, the total planetesimal surface density is given by: $\Sigma_{\text{p}}(R)= \sum_j \Sigma_{\text{p}}(R,r_{p_j})$. For simplicity, we adopt a single size distribution in this work and we solve Eq.~\ref{eq1-sec2-2} using an implicit Donor cell algorithm considering zero density as boundary conditions.

\subsection{Growth of the planets}
\label{Sec2-3}

In {\scriptsize PLANETALP}, planets grow by the continuous and simultaneous accretion of solid material and the surrounding gas \citep[see][for details]{Ronco.et.al.2017}.

On the one hand, in the case of planetesimals (solid particles with Stokes number\footnote{The Stokes number of a particle is defined as $S_t= \Omega~t_{\text{stop}}$ \citep{1851TCaPS...9....8S,1972fpp..conf..211W,Weidenschilling1977}, being $t_{\text{stop}}$ the stopping time that depends of the drag regime \citep[see][for the adopted stopping times]{Guilera2014}.} greater than unity), the solid cores of the planets grow according to the accretion rate for oligarchic growth derived by \citet{Inaba2001},
\begin{equation}
  \dfrac{{\rm d}\text{M}_{\text{C}}}{{\rm d}t}\bigg|_{\text{psimal}} = \Sigma_{\text{p}} \,(R_{\text{P}}) \, \text{R}_{\text{H}}^{2} \, P_{\text{coll}} \, \Omega_{\text{P}}, 
  \label{eq1-sec2-3}
\end{equation}
where $\text{M}_{\text{C}}$ is the mass of the core, $\Sigma_{\text{p}}(R_{\text{P}})$ the surface density of solids at the planet location $R_\text{P}$, $\text{R}_{\text{H}}$ the planet Hill radius, $P_{\text{coll}}$ the collision probability, and $\Omega_{\text{P}}$ the Keplerian frequency. $P_{\text{coll}}$ is a function of the planet capture radius, considering the drag force experienced by the planetesimals when entering the planetary envelope following \citet{Inaba&Ikoma-2003}, the planet Hill radius, and the relative velocity between the planetesimals and the planet (assuming a coplanar and circular orbit for the latter). There are three regimes for $P_{\text{coll}}$ depending on whether the relative velocities are low, intermediate or high. This in turn depends on the planetesimal eccentricities and inclinations \citep[see][for details]{Guilera2010, Guilera2014}. In general, small planetesimals ($r_p \lesssim 1$~km) are in the low relative velocity regime, while big planetesimals ($r_p \gtrsim 10$~km) are in high relative velocity regime. Thus, small planetesimals have greater $P_{\text{coll}}$ and so, greater accretion rates (typically of the order of $10^{-5}~\text{M}_{\oplus}/\text{yr}$--$10^{-4}~\text{M}_{\oplus}/\text{yr}$) with respect to big planetesimals (which typically have accretion rates of the order of $10^{-6}~\text{M}_{\oplus}/\text{yr}$)\footnote{See \citet{Guilera2010,Guilera2011,Guilera2014} and \citet{Fortier.et.al.2007,Fortier2013} for classical planetesimal accretion rates in the framework of the oligarchic growth depending on planetesimal sizes and disc masses.}. The Eq.~\ref{eq1-sec2-3} is radially integrated along the planetesimal feeding zone, which extends $4.5~\text{R}_{\text{H}}$ at each side of the planet location, using a normalisation function \citep{Guilera2014}. We do not consider some potentially relevant phenomena that could change the planetesimals accretion rates, such as planetesimal resonant trapping or gap opening in the planetesimal disc \citep{Tanaka-Ida1999, Shiraishi-Ida2008}. These phenomena are difficult to incorporate without considering N-body interactions between the planets and the planetesimals.

On the other hand, if the solid particles have Stokes numbers $S_t$ lower or equal than unity we consider them as pebbles, and the accretion rates are given by \citet{Lambrechts.et.al.2014}:
\begin{eqnarray}
  \frac{{\rm d}\text{M}_{\text{C}}}{{\rm d}t}\bigg|_{\text{pebbles}} =
  \begin{cases}
     2 \beta \, \Sigma_{\text{p}}(R_{\text{P}}) \, \text{R}_{\text{H}}^2 \, \Omega_{\text{P}}, ~ \text{if } ~ 0.1 \le \text{$S_t$} < 1, \\
    \\
     2 \beta \, \left( \frac{\text{$S_t$}}{0.1} \right)^{2/3} \, \Sigma_{\text{p}}(R_{\text{P}}) \, \text{R}_{\text{H}}^2 \, \Omega_{\text{P}}, ~ \text{if }~\text{$S_t$} < 0.1, 
  \end{cases}  
  \label{eq2-sec2-3}
\end{eqnarray}
where $\beta= \text{min}(1, \text{R}_{\text{H}}/H_\text{p})$ takes into account a reduction in the pebble accretion rates if the scale-height of the pebbles ($H_\text{p}$) becomes greater than the planet Hill radius. This scale-height is defined as $H_\text{p} = H_{\text{g}}\sqrt{\alpha/(\alpha + S_t)}$ \citep{Youdin2007}, where $H_{\text{g}}$ is the scale-height of the gaseous disc. Eq.~\ref{eq2-sec2-3} is radially integrated along the pebble feeding zone, which extends $1~\text{R}_{\text{H}}$ at each side of the planet location \citep{Ormel&Klahr2010}. While planetesimals can be accreted by a fraction of the Hill sphere of the planet, pebbles can be accreted by the full Hill sphere producing hence higher accretion rates. We do not take into account the pebble isolation mass \citep{Lambrechts.et.al.2014, Bitsch-et-al-2018} or the possible pebble ablation in the planet envelope \citep{Brouwers-et-al2018}. The first phenomenon stops the accretion of pebbles at high planetary masses, while the second could strongly inhibit the formation of massive cores by pebble accretion.  

Regarding the gas accretion, we follow the approach used in \citet{Ronco.et.al.2017} and \citet{Guilera.et.al.2017}. We consider that if the planet core mass reaches a critical mass equal to:
\begin{equation}
  \text{M}_{\text{crit}} = 10\left(\dfrac{\dot{\text{M}}_{\text{C}}}{10^{-6}\,\text{M}_\oplus \text{yr}^{-1}}\right)^{0.25}\text{M}_\oplus,
  \label{eq3-sec2-3}
\end{equation}
where $\dot{\text{M}}_{\text{C}}$ is the solid accretion rate, then the gas accretion rate onto the planet is given by:
\begin{equation}
 \dfrac{{\rm d}\text{M}_{\text{g}}}{{\rm d}t} = \text{min}[\dot{\text{M}}_{\text{KH}},\dot{\text{M}}_{\text{disc}},\dot{\text{M}}_{\text{GP}}],
 \label{eq4-sec2-3}
\end{equation}
where $\dot{\text{M}}_{\text{KH}} = \text{M}_{\text{P}}/\tau_{\text{g}}$, being $\text{M}_{\text{P}}$ the total mass of the planet and $\tau_{\text{g}}= 8.35\times10^{10}\,(\text{M}_{\text{P}}/\text{M}_\oplus)^{-3.65}\,\text{yr}$ the characteristic Kelvin-Helmholtz growth timescale of the envelope \citep{Ronco.et.al.2017}. Additionally, the maximum rate at which the gas can be delivered by the disc onto the planet is equal to:
\begin{equation}
 \dot{\text{M}}_{\text{disc}} = 3\pi\nu \Sigma_{\text{g,P}},
 \label{eq6-sec2-3}
\end{equation}
where $\Sigma_{\text{g,P}}$ is the local gas surface density (hereafter by local we mean at the planet location). Finally, if the planet is able to open a gap in the gaseous disc \citep{TanigawaIkoma2007}, the gas accretion onto the planet $\dot{\text{M}}_{\text{GP}}$ reads as follows:
\begin{equation}
  \dot{\text{M}}_{\text{GP}} = \dot{A}\Sigma_{\text{acc}}, 
  \label{eq7-sec2-3}
\end{equation}
where
\begin{equation}
  \dot{A} = 0.29\left(\dfrac{H_{\text{g,P}}}{R_{\text{P}}}\right)^{-2}\left(\dfrac{\text{M}_{\text{P}}}{\text{M}_{\star}}\right)^{4/3}R^{2}_{\text{P}}\Omega_{\text{P}} 
  \label{eq8-sec2-3}
\end{equation}
with $\Sigma_{\text{acc}} = \Sigma(x=2\text{R}_{\text{H}})$ and,
\begin{equation}
  \Sigma(x) =
  \begin{cases}
    \Sigma_{\text{g,P}}\text{exp}\left[-\left( \dfrac{x}{l} \right)^{-3}\right] & \text{ if $x > x_{\text{m}}$}, \\
    \\
    \!\begin{aligned}
      & \Sigma_{\text{g,P}}\text{exp}\Bigg[-\dfrac{1}{2}\left(\dfrac{x}{H_{\text{g,P}}}-\dfrac{5}{4}\dfrac{x_{\text{m}}}{H_{\text{g,P}}}\right)^{2} \\
    \\
    & + \dfrac{1}{32}\left(\dfrac{x_{\text{m}}}{H_{\text{g,P}}}\right)^{2} - \left(\dfrac{x_{\text{m}}}{l}\right)^{-3}\Bigg] \\
    \end{aligned} & \text{if $x \le x_{\text{m}}$}, 
  \end{cases}
\label{eq9-sec2-3}
\end{equation}
wherein $l$ and $x_{\text{m}}$ are defined as
\begin{equation}
  l = 0.146\left(\dfrac{\nu}{10^{-5}R^{2}_{\text{P}}\Omega_{\text{P}}}\right)^{-1/3}\left(\dfrac{\text{M}_{\text{P}}}{10^{-3}\text{M}_{\star}}\right)^{2/3}R_{\text{P}},
  \label{eq10-sec2-3}
\end{equation}
and
\begin{equation}
  x_{\text{m}} = 0.207\left(\dfrac{H_{\text{g,P}}}{0.1R_{\text{P}}}\dfrac{\text{M}_{\text{P}}}{10^{-3}\text{M}_{\star}}\right)^{2/5}\left(\dfrac{\nu}{10^{-5}R^{2}_{\text{P}}\Omega_{\text{P}}}\right)^{-1/5}R_{\text{P}},
  \label{eq11-sec2-3}
\end{equation}
being $\text{M}_{\star}$ the stellar mass, and $H_{\text{g,P}}$ the local gaseous disc scale height.

\subsection{Planetary orbital evolution}
\label{Sec2-4}

As we mentioned before, as a planet grows, the exchange of angular momentum with the gaseous disc produces a torque that results in the planet's migration along the disc. If the planet is not able to open a gap in the disc, then we consider that it migrates under type I migration. Its stellocentric distance varies as follows:
\begin{equation}
 \dfrac{{\rm d}R_{\text{P}}}{{\rm d}t}\bigg|_{\text{migI}} = -2 R_{\text{P}}\dfrac{\Gamma}{\mathcal{L}_{\text{P}}},
 \label{eq1-sec2-4}
\end{equation}
where $\Gamma$ is the total torque on the planet, $\mathcal{L}_{\text{P}}$ the planet angular momentum, and $R_{\text{P}}$ the planet semi-major axis (assuming that it always remains in a nearly circular orbit). 

When the planet becomes massive enough to open a gap in the gaseous disc, it enters in the type II migration regime. We follow the criterion derived by \citet{Crida2006} for the transition between type I and type II migration. The treatment of the latter regime in our model is described in detail in \citet{Ronco.et.al.2017}.

\subsubsection{Previous type I migration rates}
\label{sec2-4-1}

In \citet{Guilera.et.al.2017}, we implemented the type I migration rates derived by \citet{paardekooper.etal2010, paardekooper.etal2011} to study the formation of giant planets in wide orbits. The total torque is given by:
\begin{equation}
 \Gamma= \Gamma_{\text{L}} + \Gamma_{\text{C}}.
 \label{eq2-sec2-4}
\end{equation}
where $\Gamma_{\text{L}}$ and $\Gamma_{\text{C}}$ are the Lindblad and corotation torques, respectively. The former is equal to: 
\begin{equation}
 \Gamma_{\text{L}}=  (-2.5 -1.7 \beta' + 0.1 \alpha' ) \, \dfrac{\Gamma_0}{\gamma_{\text{eff}}}
 \label{eq3-sec2-4}
\end{equation}
where $\alpha'= - {\rm d}\ln \Sigma_{\text{g,P}}/{\rm d}\ln R$ and $\beta'= - {\rm d}\ln T_{\text{P}}/{\rm d}\ln R$, with $T_{\text{P}}$ the local disc temperature at the mid-plane. The normalised torque $\Gamma_0$ is given by: 
\begin{equation}
 \Gamma_0= \left( \dfrac{q}{h} \right)^2 \Sigma_{\text{g,P}} \, R_{\text{P}}^4 \, \Omega_{\text{P}}^2, 
\label{eq4-sec2-4}
\end{equation}
where $q= \text{M}_{\text{P}}/\text{M}_{\star}$ and $h= H_{g,\text{P}}/R_{\text{P}}$. Finally, $\gamma_{\text{eff}}$ reads:
\begin{align}
& \gamma_{\text{eff}}= \nonumber \\
&\dfrac{2\gamma Q}{\gamma Q + \frac{1}{2} \sqrt{2\sqrt{(\gamma^2Q^2+1)^2-16Q^2(\gamma-1)} + 2\gamma^2Q^2 - 2}}, 
\label{eq5-sec2-4}
\end{align}
where $\gamma= 7/5$ is the specific heat ratio for the diatomic gas, and $Q= 2 \chi_{\text{P}} \Omega_{\text{P}} / 3 h c_{s,\text{P}}^2$. $c_{s,\text{P}}$ and $\chi_{\text{P}}$ are the local sound speed and the local thermal diffusion coefficient at the disc mid-plane, respectively. The latter is computed as follows\footnote{As it was noted by \citet{BK2011}, there is a typo in the Eq.~34 in \citet{paardekooper.etal2011}, the factor in the numerator has to be 16 instead of 4.}:
\begin{equation}
\chi_{\text{P}}= \dfrac{16 (\gamma-1) \sigma T_{\text{P}}^4}{3\kappa_{\text{P}} \rho_{g,{\text{P}}} P_{\text{P}}},
\label{eq6-sec2-4}
\end{equation}
where $\kappa_{\text{P}}$, $\rho_{g,{\text{P}}}$, and $P_{\text{P}}$ are the local opacity, local volumetric gas density, and local pressure of the disc also at the mid-plane (respectively).

The corotation torque has two main contributions: a barotropic component and an entropy one. Therefore, 
\begin{equation}
\Gamma_{\text{C}}= \Gamma_{\text{C,bar}} + \Gamma_{\text{C,ent}}. 
\label{eq6bis-sec2-4}
\end{equation}
The detailed expressions of these two terms are described in Appendix~\ref{apex-1}. It is important to note that the type I migration rates from \citet{paardekooper.etal2011} are the most used rates in planet formation models that consider the evolution of non-isothermal discs. 

\subsubsection{Updated type I migration rates}
\label{sec2-4-2}

In this work we incorporate for the first time the new type I migration rates recently proposed by \citet{jm2017} and \citet{masset2017}. We consider that the total torque on the planet is given by:
\begin{equation}
  \Gamma= \Gamma_{\text{Type~I}} + \Gamma_{\text{thermal}},
  \label{eq1-sec2-4-1}
\end{equation}
where $\Gamma_{\text{Type~I}}$ is the sum of the Lindblad and corotation torques (possibly saturated) on the planet, while $\Gamma_{\text{thermal}}$ is the additional torque on the planet that arises from thermal effects (the perturbation of the flow in the planet's vicinity due to a finite thermal diffusivity and that arising from the heat released in its surroundings by the accreting planet). \citet{jm2017} recently studied the first term of the right side of Eq.~\ref{eq1-sec2-4-1} through 3D hydrodynamical simulations, improving the previous study of \citet{MC2010}. They provide the following expression: $\Gamma_{\text{Type~I}}= \Gamma_{\text{L}} + \Gamma_{\text{C}}$, where the Lindblad torque is now given by:
\begin{equation}
  \Gamma_{\text{L}}= (-2.34 -1.5 \beta' + 0.1 \alpha' ) \Gamma_0 f(\chi_{\text{P}}/\chi_{\text{C}}),
  \label{eq2-sec2-4-1}
\end{equation}
where $\alpha'$, $\beta'$, $\Gamma_0$ and $\chi_{\text{P}}$ are the same quantities aforeintroduced. We see that while the dependence of the Lindblad torque on the radial surface density gradient is the same as in \citet{paardekooper.etal2011}, the dependence on the temperature radial gradient is slightly smaller. The function $f(\chi_{\text{P}}/\chi_{\text{C}})$ plays the role of the inverse of an effective adiabatic index $\gamma_{\text{eff}}$, as in \citet{paardekooper.etal2011} and it reads as follows:
\begin{equation}
  f(x)= \dfrac{\sqrt{x/2} + 1/\gamma}{\sqrt{x/2} + 1},
  \label{eq3-sec2-4-1}
\end{equation}
wherein $x= \chi_{\text{P}}/\chi_{\text{C}}$, and where $\chi_{\text{C}}= R_{\text{P}}^2h^2\Omega_{\text{P}}$ is the critical diffusivity. The corotation torque is now given as the sum of four contributions: three of them associated with disc radial gradients (vortensity, entropy, and temperature) and a fourth contribution which arises from a viscous production of vortensity. The sum of the three contributions arising from radial gradients may seem surprising at first glance, as the disc radial profile is determined by two quantities (e.g. the density and temperature) and two indexes (e.g. $\alpha'$ and $\beta'$). While the first three components of the corotation torque could be compacted into two components, each has different saturation properties, and the result would obfuscate the physical origin of each of the components. This has been discussed by \citet{jm2017} in section~4.3. Thus,
\begin{equation}
\Gamma_{\text{C}}= \Gamma_{\text{C,vor}}+\Gamma_{\text{C,ent}}+\Gamma_{\text{C,temp}}+\Gamma_{\text{C,vv}}.
\label{eq3bis-sec2-4-1}
\end{equation}
These components are also described in detail in  Appendix~\ref{apex-1}. We remark that these new type I migration prescriptions are derived for low-mass planets ($q < 0.2h^3$) and intermediate-mass planets ($0.2h^3 \lesssim q \lesssim 2h^3$). For more massive planets able to deplete significantly the coorbital region this migration recipe could not be accurate enough. But such massive planets should be near the transition to the type~II regime.  

The second term of the right side of Eq.~\ref{eq1-sec2-4-1} was recently studied by \citet{masset2017} from the analytical perspective. This work considers luminous and non-luminous planets and derives a semi-analytical prescription for $\Gamma_{\text{thermal}}$, valid for both regimes. If the planet does not release energy into the disc, two cold and dense lobes on either side of the orbit appear due to the thermal diffusion of the disc. This effect exerts a torque on the planet, comparable in magnitude to the Lindblad torque. \citet{masset2017} defines this as a {\it cold torque} and argues that it corresponds to the {\it cold fingers} reported by \citet{Lega.et.al.2014}. Alternatively, if the planet releases energy into the disc, hot and low-density lobes are generated in the surroundings. In this case, a {\it heating torque} --- of opposite sign with respect to the cold torque --- is exerted on the planet \citep{Benitez-llambay.et.al.2015}. Here, we implement the thermal torque in the same way as proposed by \citet{masset2017}, i.e. as the sum of the cold and the heating torque. The former is given by:
\begin{equation}
\Gamma_{\text{cold}}= -1.61\dfrac{\gamma-1}{\gamma} \dfrac{x_p}{\lambda_c} \dfrac{\Gamma_0}{h},
\label{eq18-sec2-4-1}
\end{equation}
where $x_p= \eta h^2 R_{\text{P}}$, with $\eta= \alpha'/3 + \beta'/6 + 1/2$, and $\lambda_c= \sqrt{\chi_{\text{P}}/q\Omega_{\text{P}}\gamma}$. Alternatively:
\begin{equation}
\Gamma_{\text{heating}}= 1.61\dfrac{\gamma-1}{\gamma} \dfrac{x_p}{\lambda_c} \dfrac{\text{L}_{\text{P}}}{\text{L}_{\text{C}}} \dfrac{\Gamma_0}{h},
\label{eq19-sec2-4-1}
\end{equation}
where the critical luminosity $\text{L}_{\text{C}}$ is given by:
\begin{equation}
\text{L}_{\text{C}}= \dfrac{4\pi G \text{M}_{\text{P}} \chi_{\text{P}} \rho_{g,{\text{P}}}}{\gamma},
\label{eq20-sec2-4-1}
\end{equation}
and the luminosity released by the planet, $\text{L}_{\text{P}}$, due to the accretion of planetesimals reads:
\begin{equation}
\text{L}_{\text{P}}= \dfrac{G \text{M}_{\text{P}}}{\text{R}_{\text{C}}} \dfrac{{\rm d}\text{M}_{\text{C}}}{{\rm d}t},
\label{eq21-sec2-4-1}
\end{equation}
being $\text{R}_{\text{C}}$ the radius of the planet's core. We remark that the total luminosity of the planet is given by the accretion of solids (which is a reasonable approximation until the planet accrete a significant envelope) and that the radius of the planet's core increases as planet grows by the accretion of solid material considering a constant density for the core of $3.2~\text{g}/\text{cm}^3$. Thus, the thermal torque can be written as follows\footnote{The normalised torque ($\Gamma_0$) defined in \citet{masset2017} is not equal that the one defined in this work, so our Eq.~\ref{eq22-sec2-4-1} does not have a factor $h$ multiplying $\Gamma_0$ as in Eq.~146 from \citet{masset2017}.}:
\begin{align}
\Gamma_{\text{thermal}} = & 1.61\dfrac{\gamma-1}{\gamma} \dfrac{x_p}{\lambda_c} \left(\dfrac{\text{L}_{\text{P}}}{\text{L}_{\text{C}}} - 1 \right) \dfrac{\Gamma_0}{h} \nonumber \\
=& 1.61\dfrac{\gamma-1}{\gamma} \eta \left(\dfrac{H_{g,\text{P}}}{\lambda_c}\right) \left(\dfrac{\text{L}_{\text{P}}}{\text{L}_{\text{C}}} - 1 \right) \Gamma_0. 
\label{eq22-sec2-4-1}
\end{align}
Finally, we note that the thermal torque is significant while the heat released by the planet generates an excess of internal energy outside of the Bondi sphere. \citet{masset2017} argues that this condition is satisfied when the time-scale for heat diffusion across the Bondi radius is shorter than the acoustic time. This condition translates into $\text{M}_{\text{P}} < \text{M}_{\text{crit}}^{\text{thermal}}$, where the critical thermal mass is given by  
\begin{equation}
\text{M}_{\text{crit}}^{\text{thermal}}= \chi_{\text{P}} c_{s,\text{P}} /G.
\label{eq23-sec2-4-1}
\end{equation}
If the planet mass is greater than this critical mass, it is not guaranteed that the internal energy injected in the gas near the planet emerges as an excess of internal energy outside of the Bondi sphere. In such case, a cut-off of the thermal torque is expected. From a numerical point of view, when $\text{M}_{\text{P}} > \text{M}_{\text{crit}}^{\text{thermal}}$ we consider that $\Gamma_{\text{thermal}}= 0$. 

\begin{figure*}
    \centering
    \includegraphics[width= 0.31\textwidth, angle= -90]{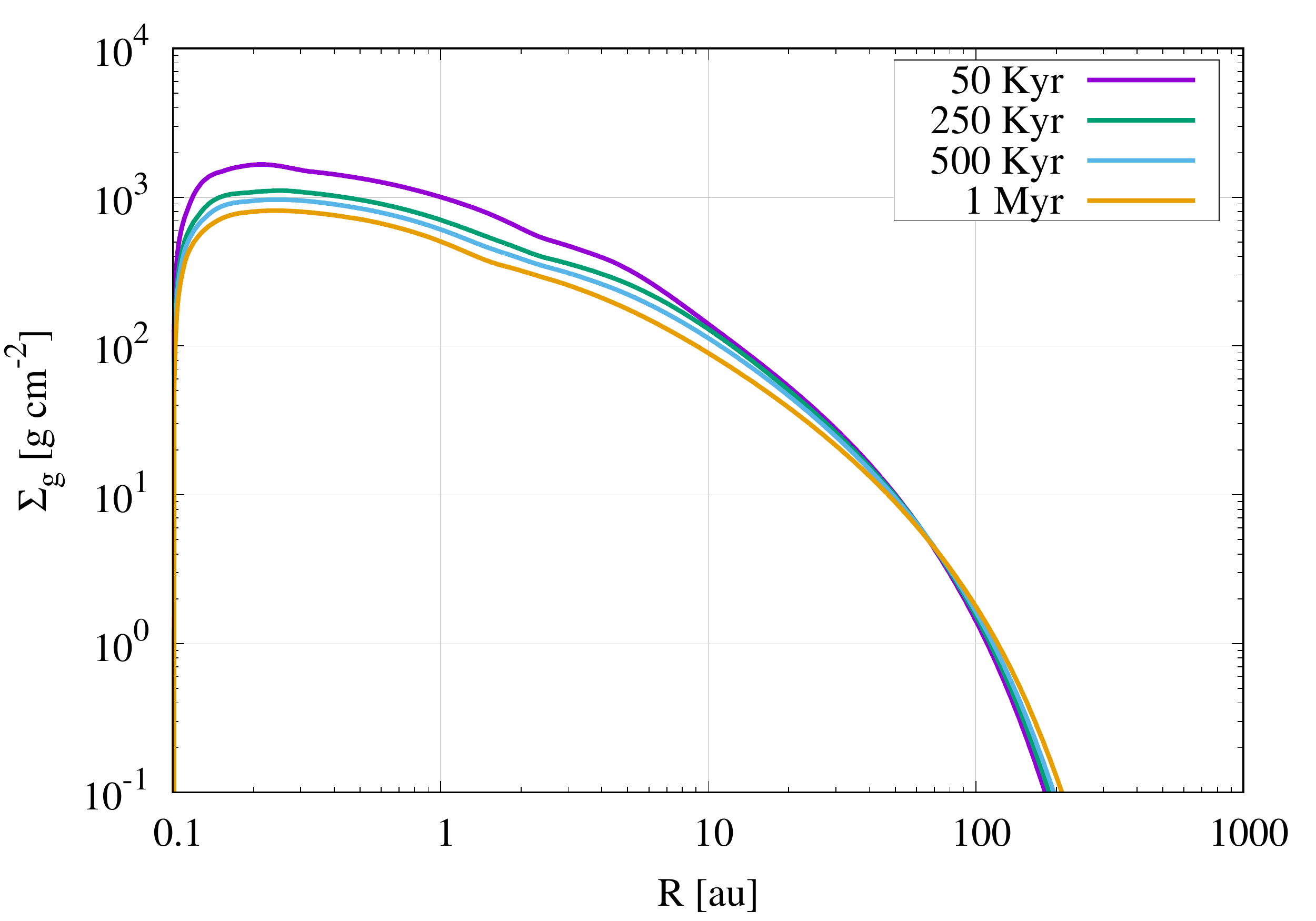} 
    \centering
    \includegraphics[width= 0.31\textwidth, angle=-90]{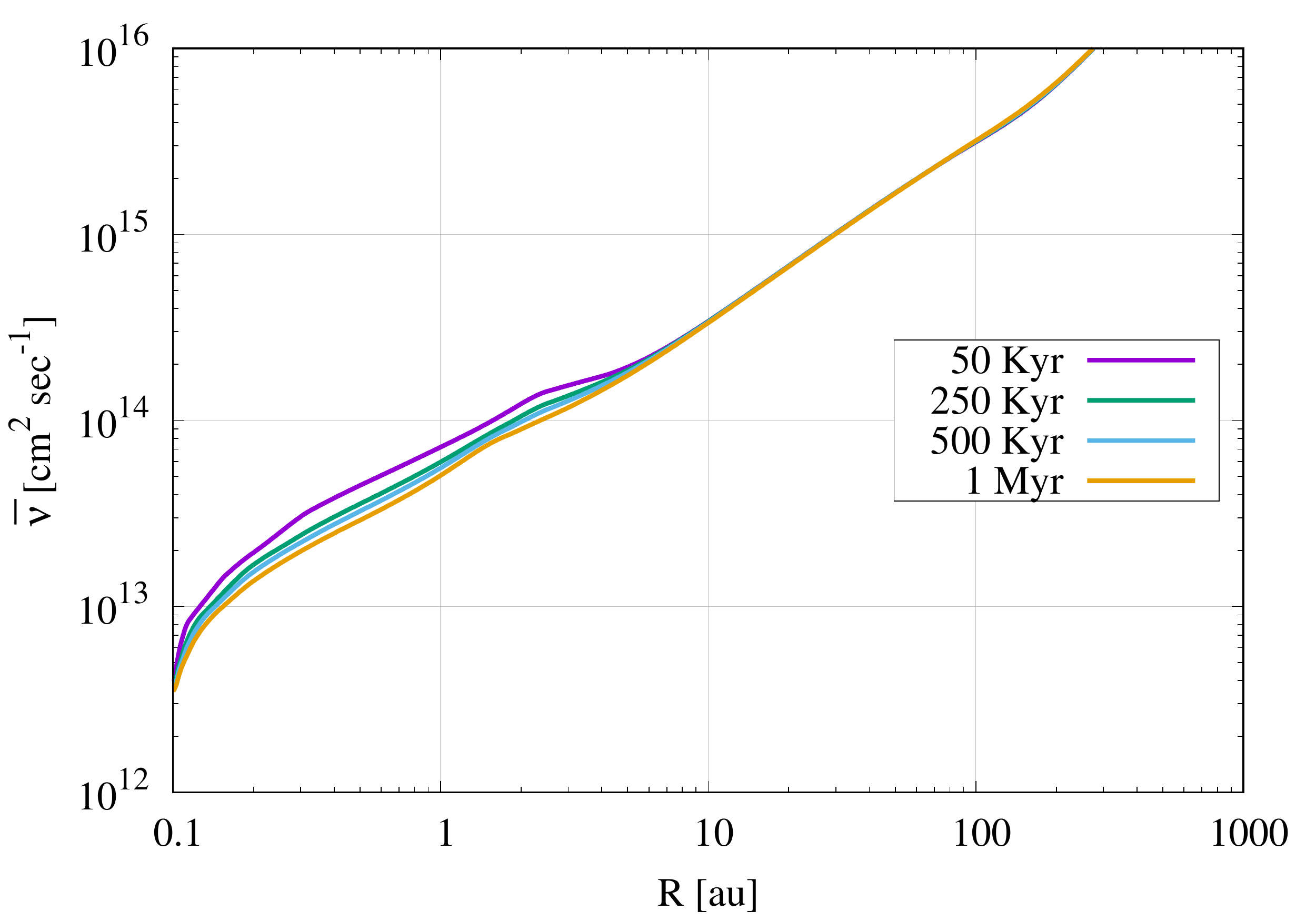} \\
    \centering
    \includegraphics[width= 0.31\textwidth, angle=-90]{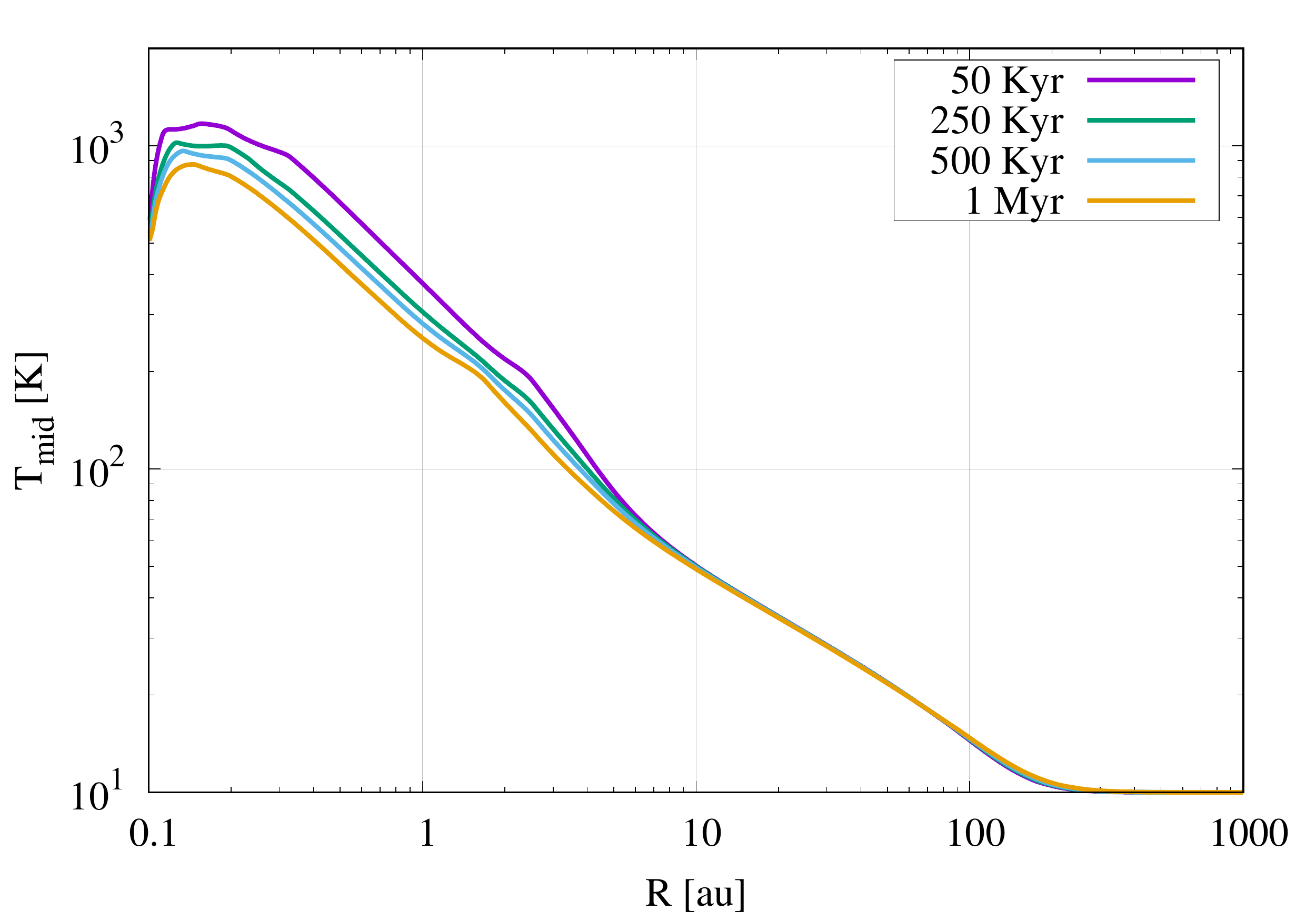} 
    \centering
    \includegraphics[width= 0.31\textwidth, angle=-90]{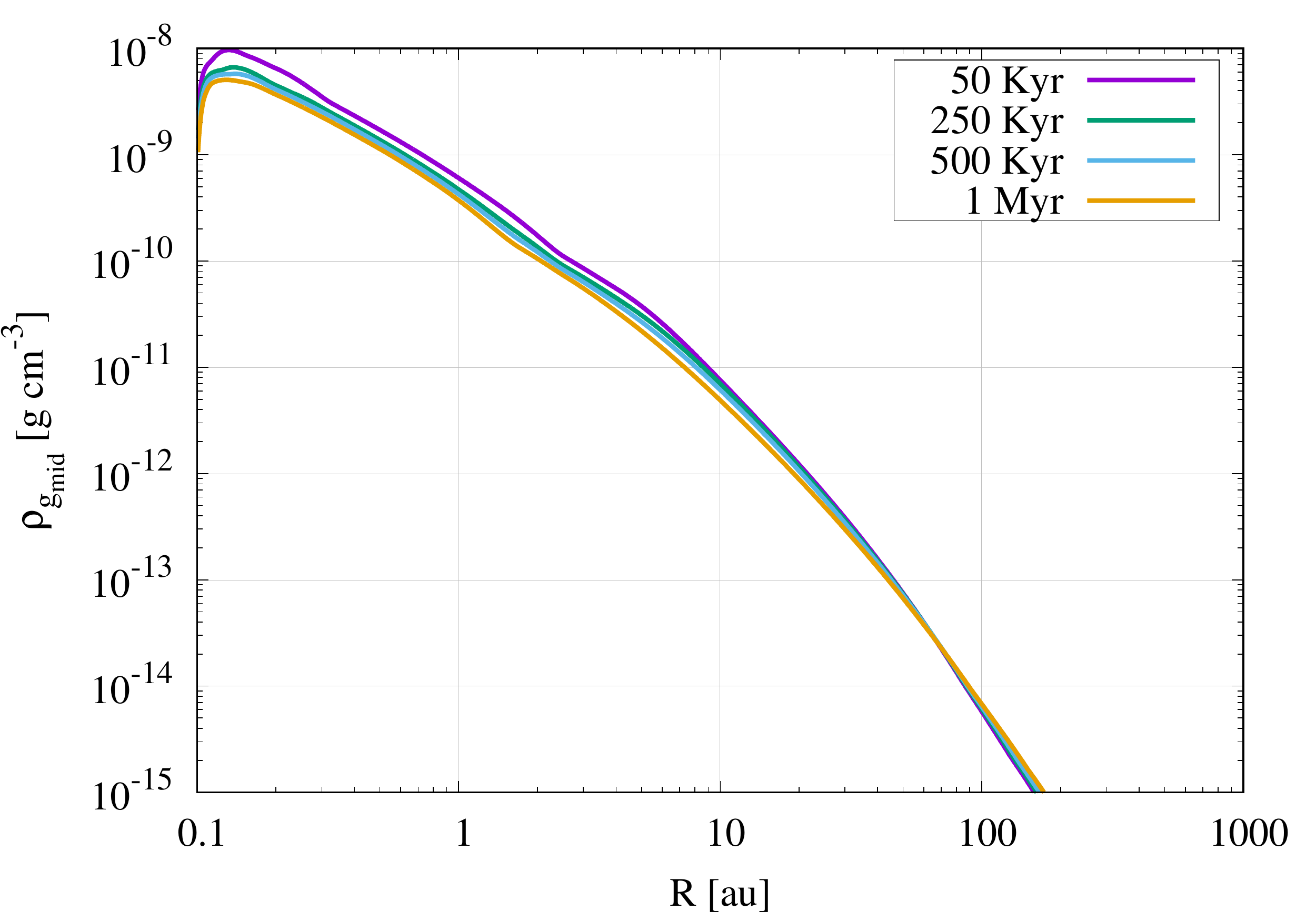} \\
    \centering
    \includegraphics[width= 0.31\textwidth, angle=-90]{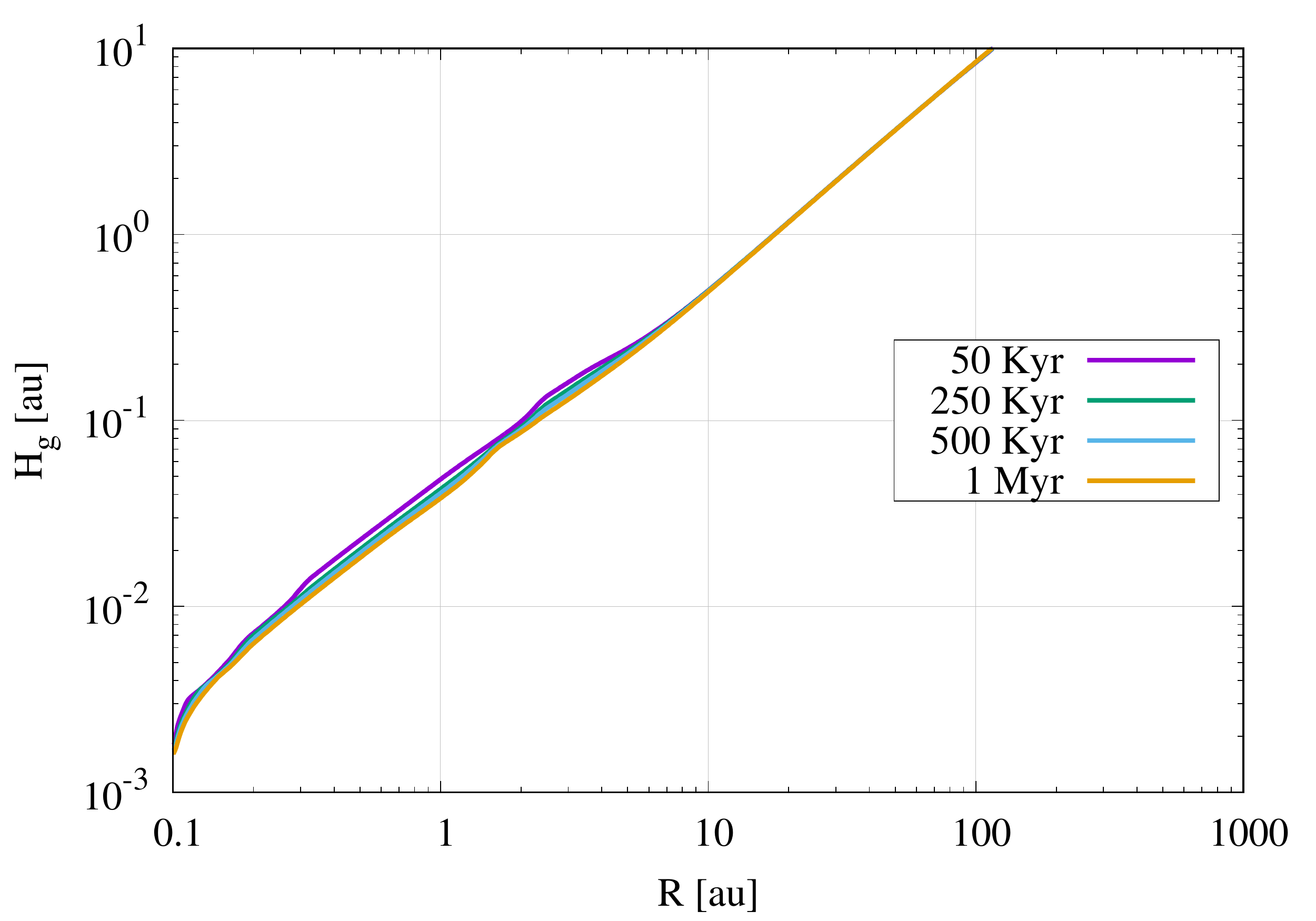} 
    \centering
    \includegraphics[width= 0.31\textwidth, angle=-90]{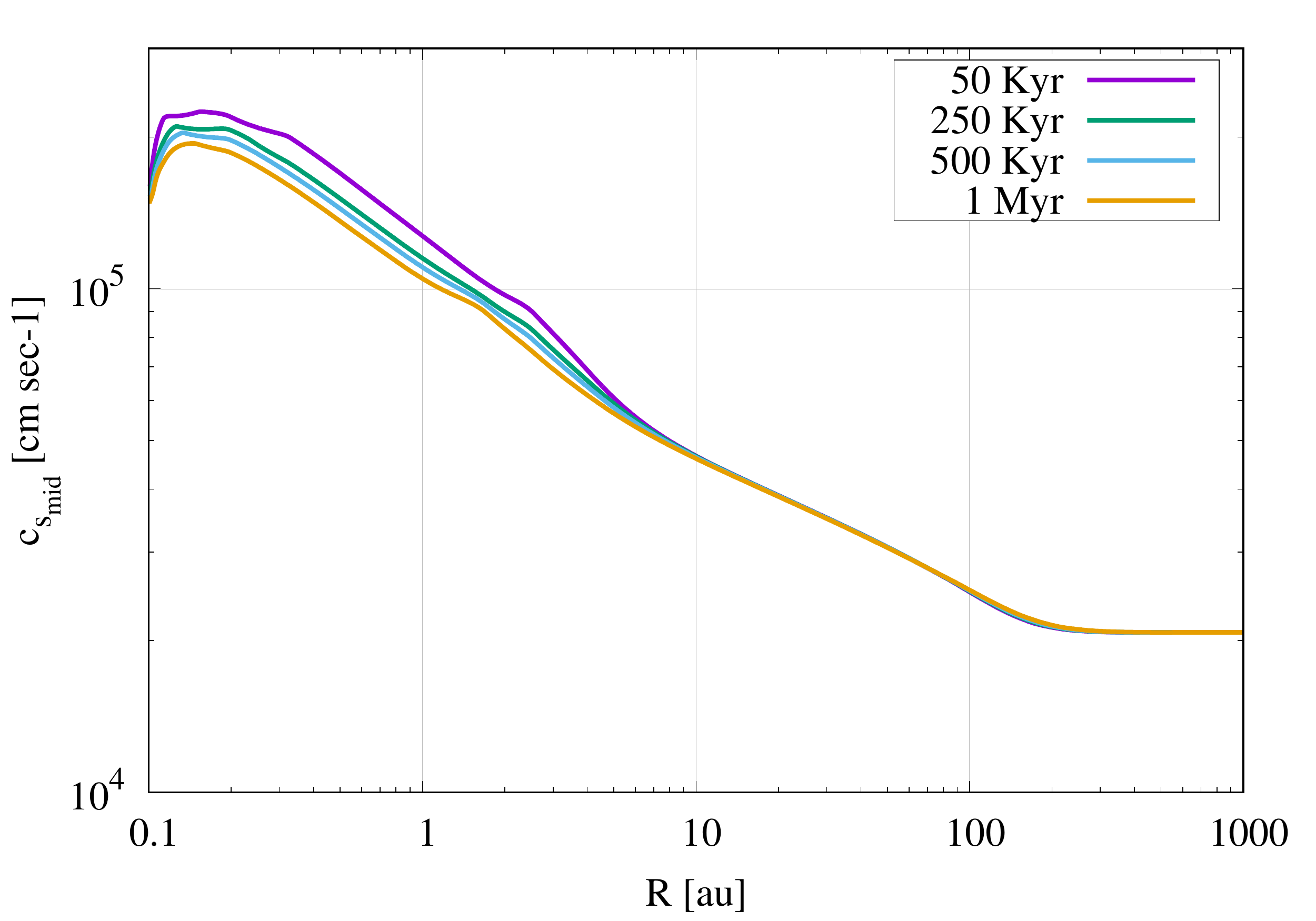} \\  
    \centering
    \includegraphics[width= 0.31\textwidth, angle=-90]{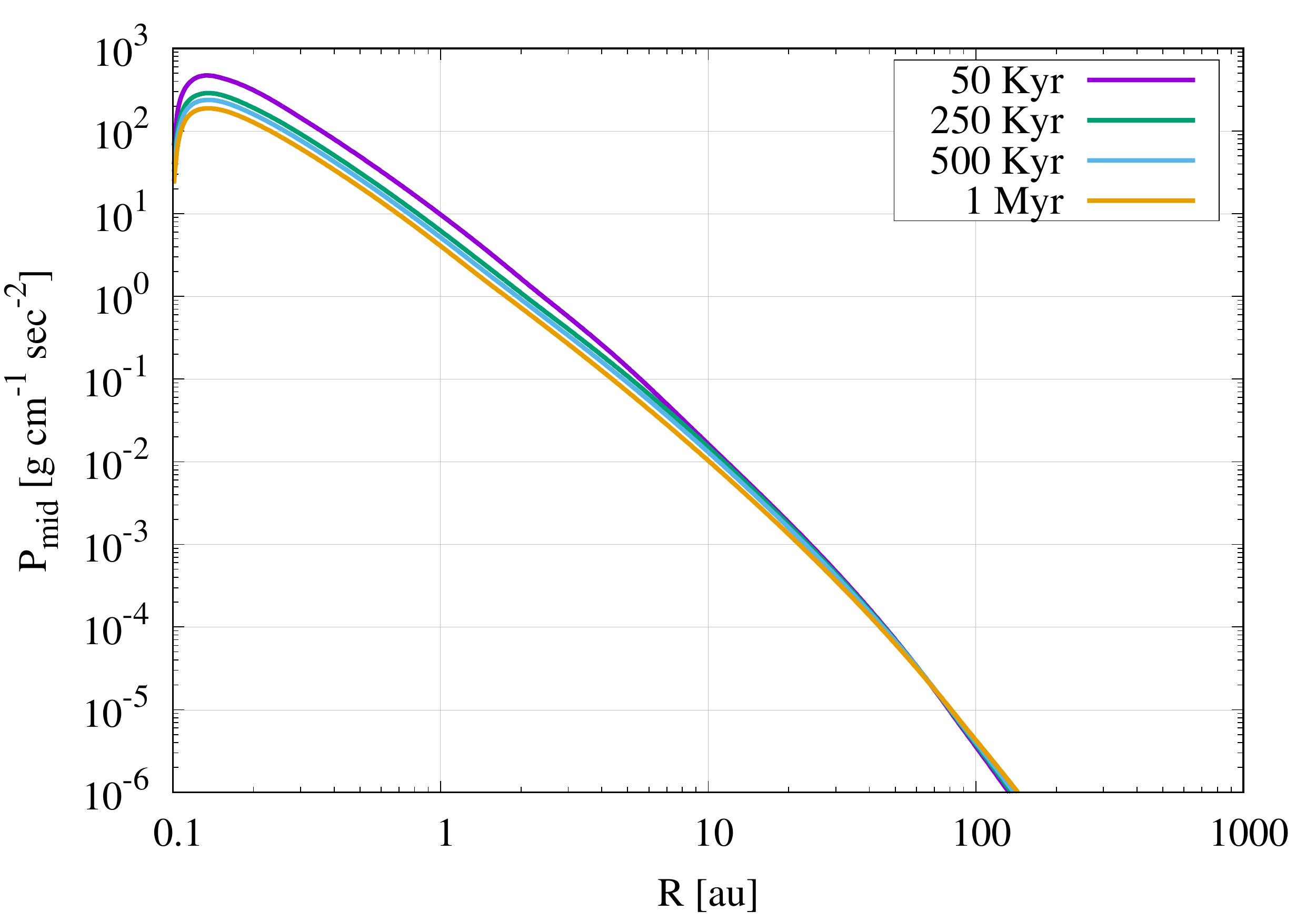} 
    \centering
    \includegraphics[width= 0.315\textwidth, angle=-90]{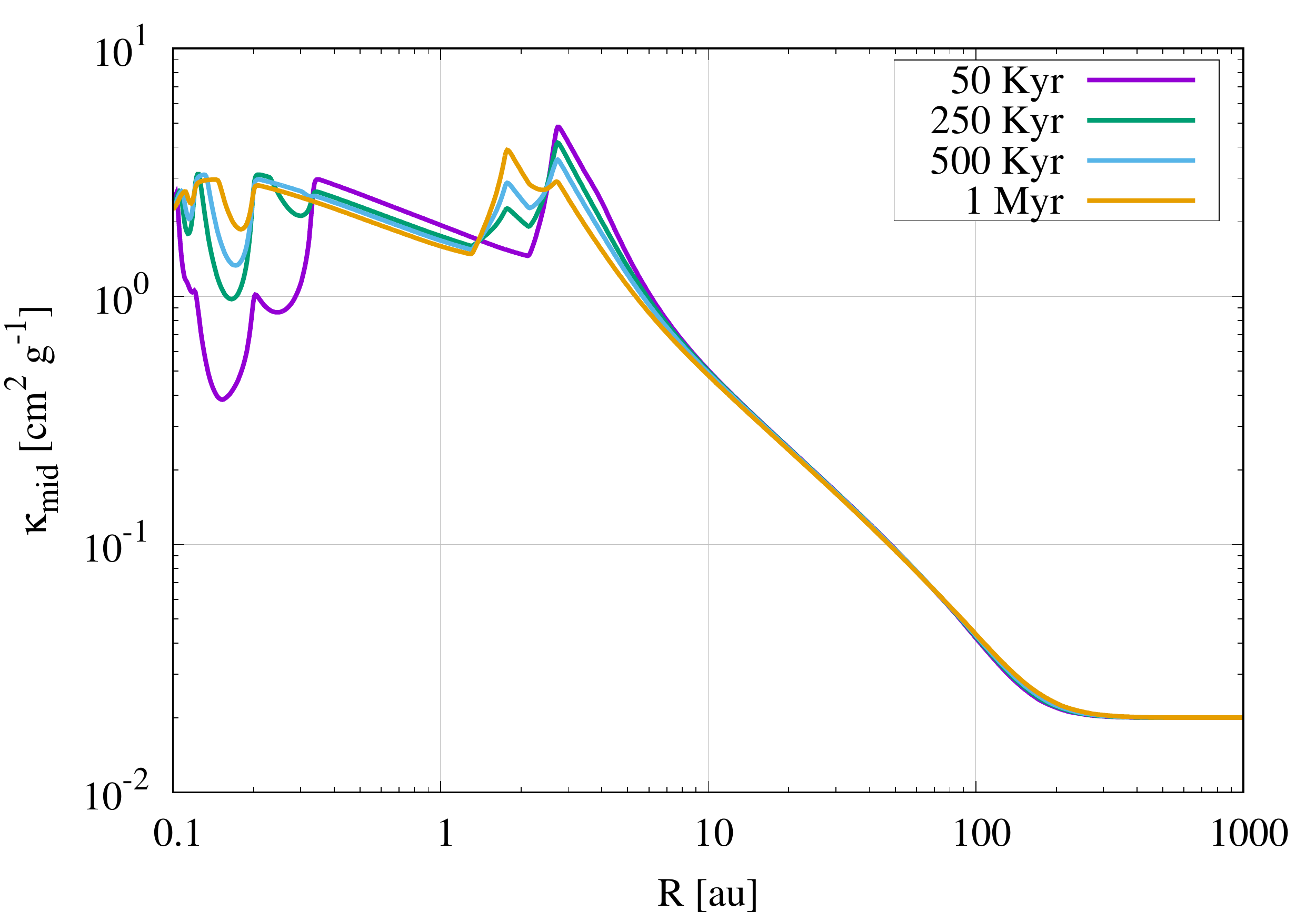} 
  \caption{Disc radial profiles at four different evolutionary times for our fiducial case. We show the main physical variables of the disc that are involved in the type I migration rate calculations. $\Sigma_{\text{g}}$, $\bar{\nu}$, and $\text{H}_{\text{g}}$ represent the disc gas surface density, the mean viscosity, and the disc scale height, respectively. $\text{T}_{\text{mid}}$, $\rho_{\text{g}_{\text{mid}}}$, $\text{c}_{\text{s}_{\text{mid}}}$, $\text{P}_{\text{mid}}$, and $\kappa_{\text{mid}}$ define the  temperature, the volumetric gas density, the sound speed, the pressure, and the opacity at the disc mid-plane, respectively. The decreases of the opacities in the inner part of the disc are due to the transition between the metal grain regime and the evaporation of metal grain regime in the opacity regimes given by \citet{Bell.Lin.1994}.}
  \label{fig:fig1}
\end{figure*}

\section{Results}
\label{sec3}

The aim of this work is to analyse how the updated type I migration rates described in Sec.~\ref{sec2-4-2}, and especially the thermal torque, impacts the process of planet formation. To do so, we first construct migration maps and then model the formation of a planet at different initial locations and for different protoplanetary disc models.   

\subsection{Migration maps for the updated type I migration rates}
\label{sec3-1}

\subsubsection{Comparison between \citet{paardekooper.etal2011} and \citet{jm2017} type I migration rates}
\label{sec3-1-1}

Here we compare the type I migration rates derived by
\citet{paardekooper.etal2011} and \citet{jm2017} constructing migration maps for both cases.  For the calculation of the torques of the disc on the planet we require the following quantities: the gas surface density, the mean viscosity, the scale height of the disc, the temperature, the volumetric gas density, the sound speed, the pressure, and the opacity at the mid-plane of the disc. In Fig.~\ref{fig:fig1} we plot the radial profiles of the aforementioned quantities from our fiducial simulation\footnote{Recall from Sect. \ref{Sec2-1} that for our fiducial model considers $R_{\text{c}}=39$~au, $\gamma=1$, $\text{M}_{\text{d}}=0.05~\text{M}_\odot$ and $\alpha=10^{-3}$.}, at four different stages of the disc early evolution (50 Kyr, 250 Kyr, 500 Kyr and 1 Myr).

\begin{figure*}
    \centering
    \includegraphics[width= 0.43\textwidth]{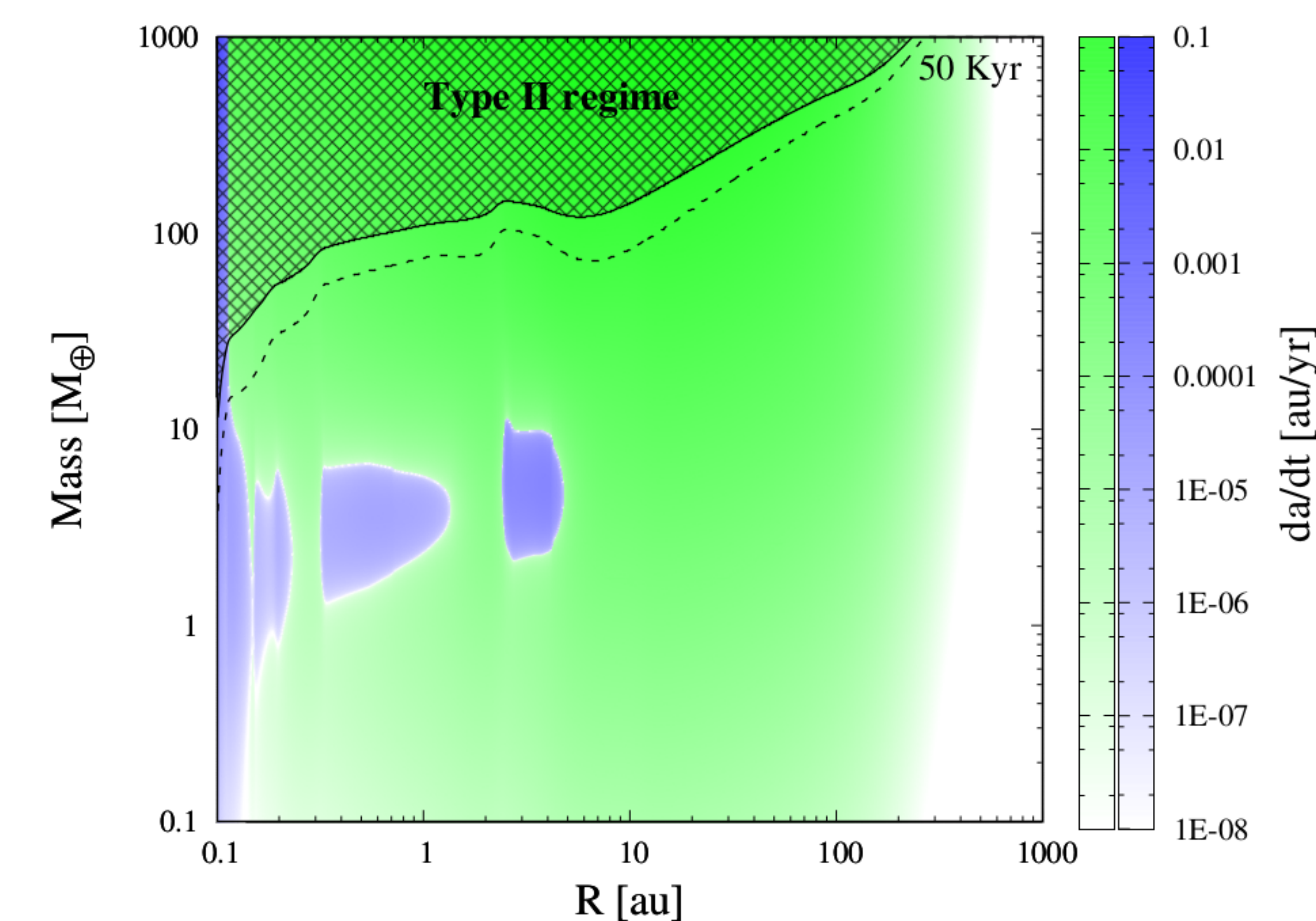} 
    \centering
    \includegraphics[width= 0.43\textwidth]{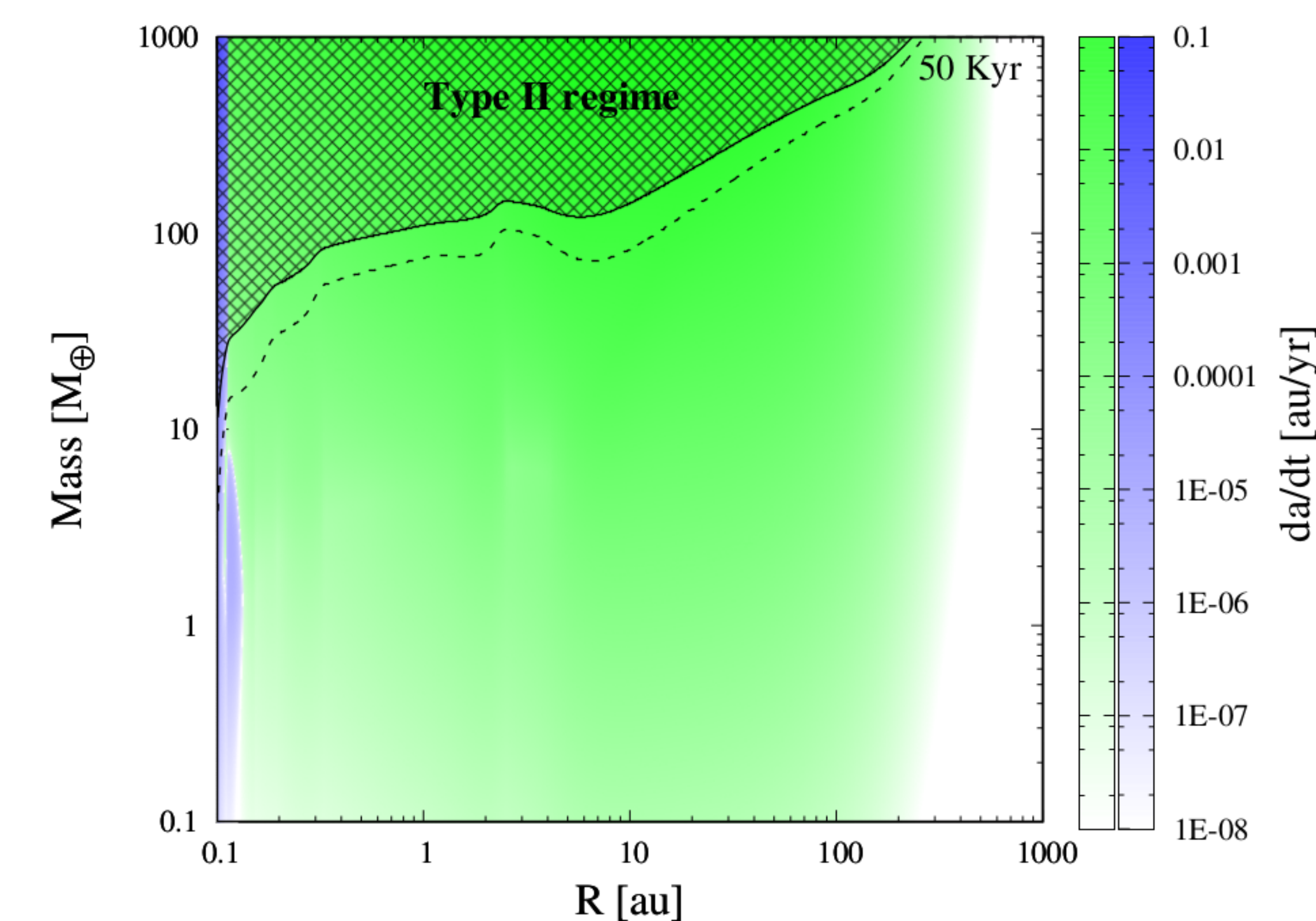} \\
    \centering
    \includegraphics[width= 0.43\textwidth]{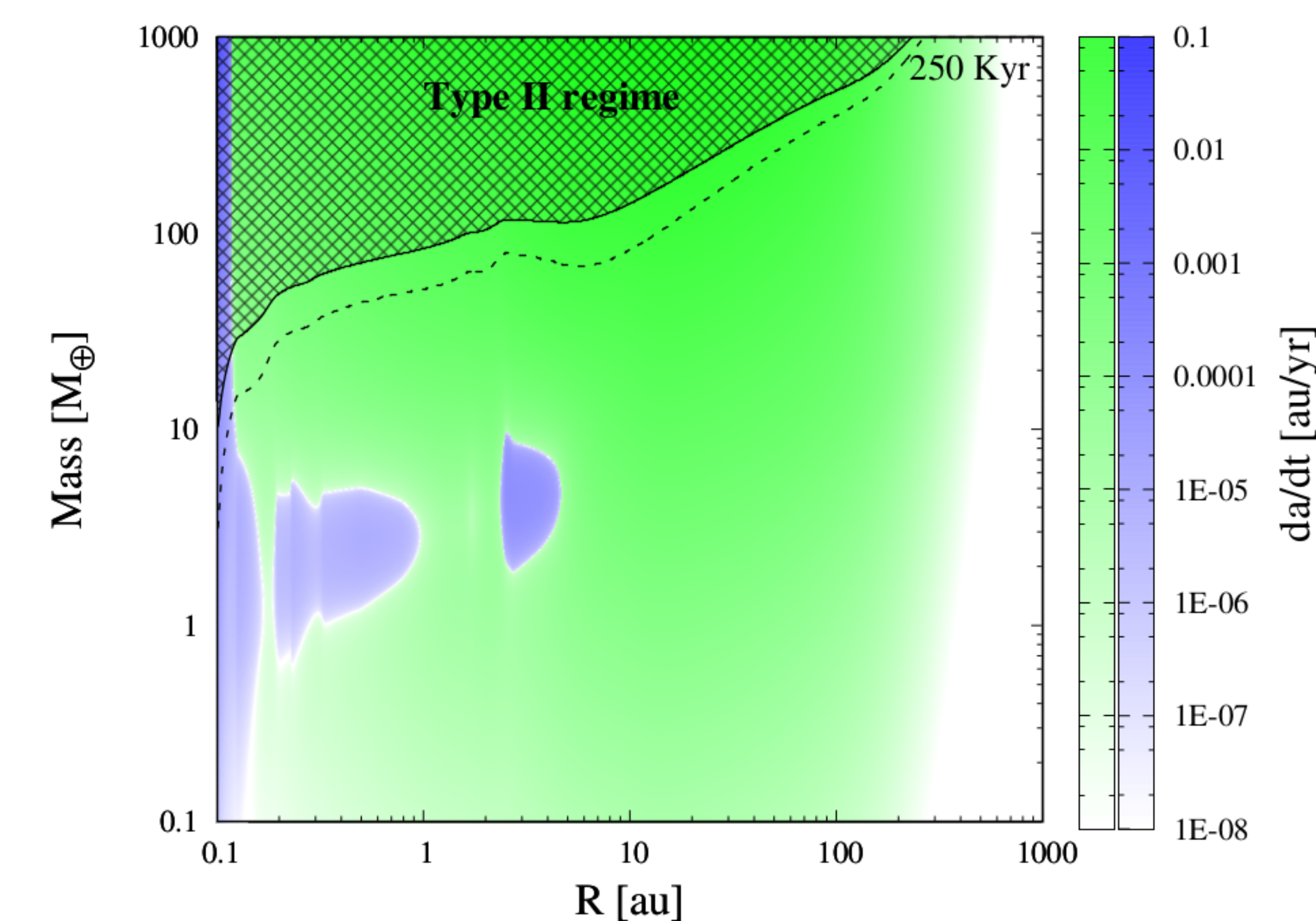} 
    \centering
    \includegraphics[width= 0.43\textwidth]{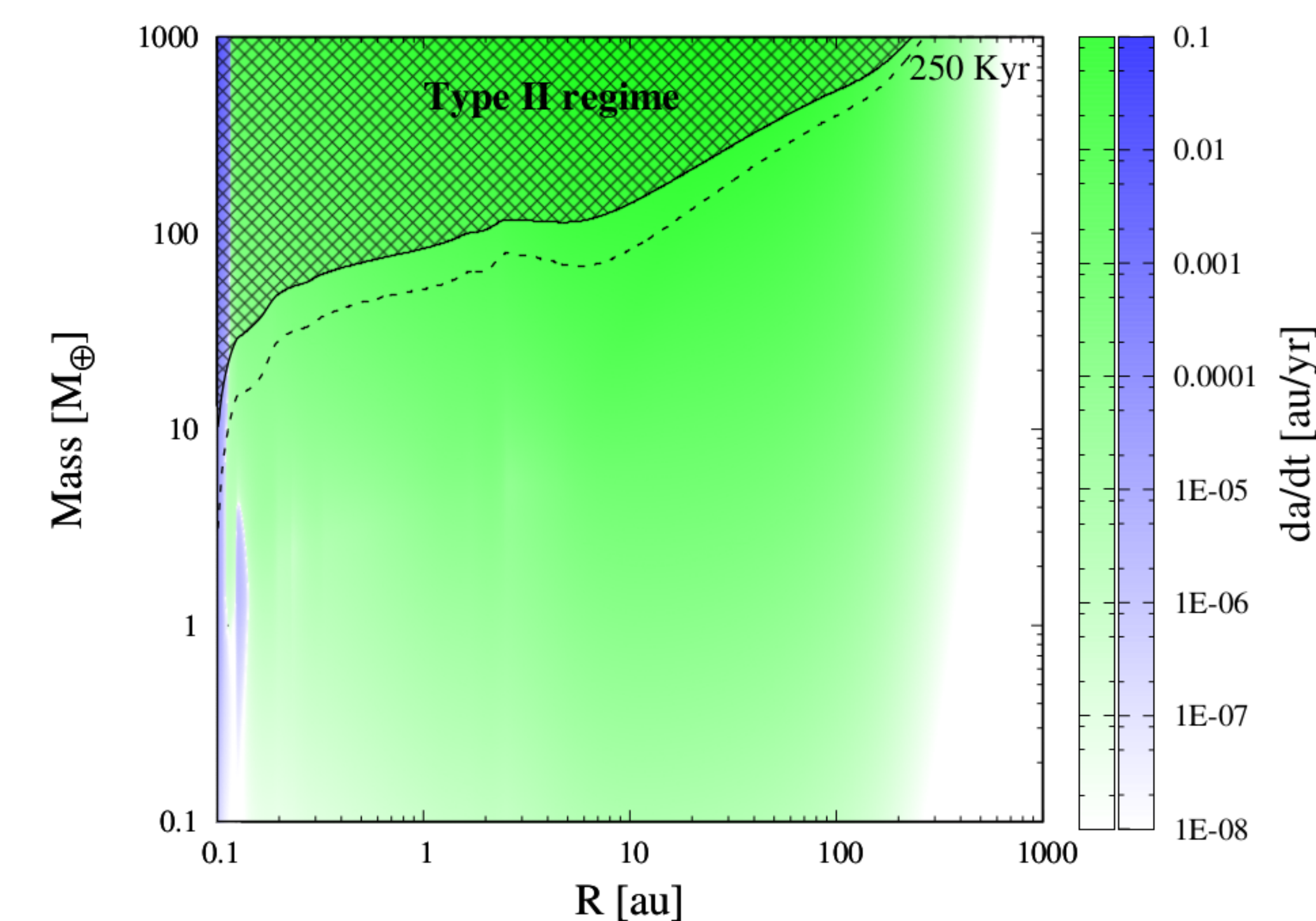} \\
    \centering
    \includegraphics[width= 0.43\textwidth]{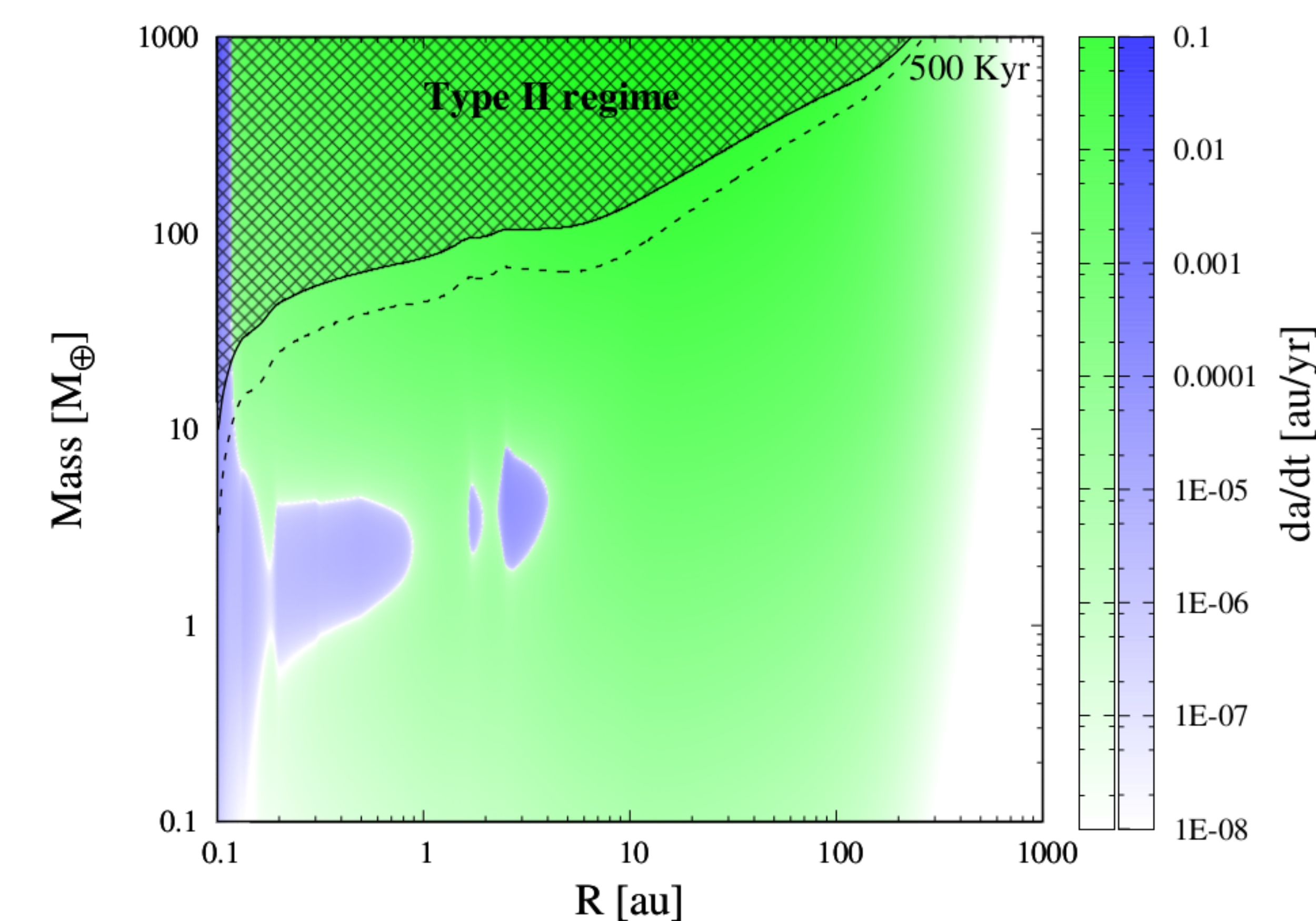} 
    \centering
    \includegraphics[width= 0.43\textwidth]{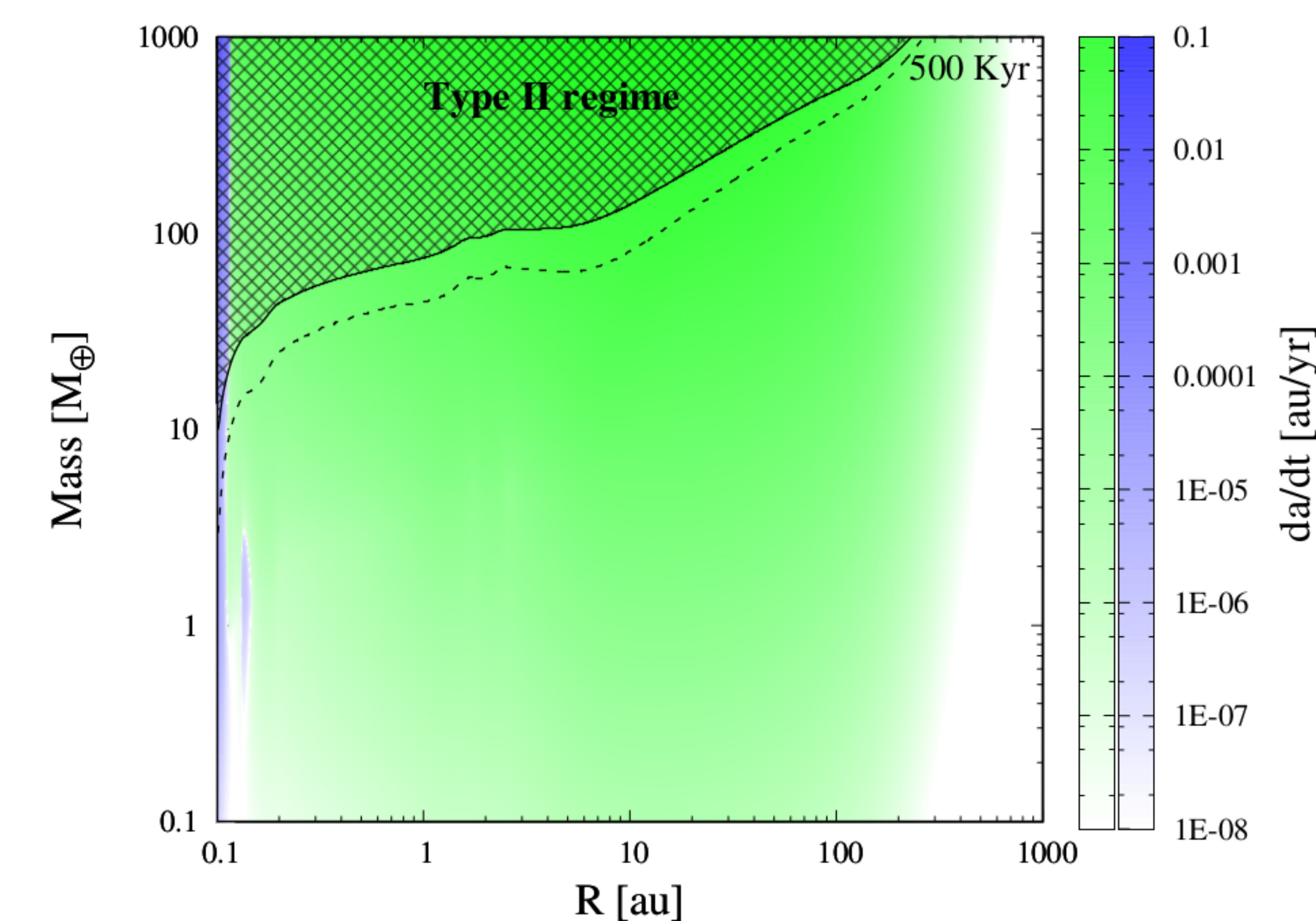} \\  
    \centering
    \includegraphics[width= 0.43\textwidth]{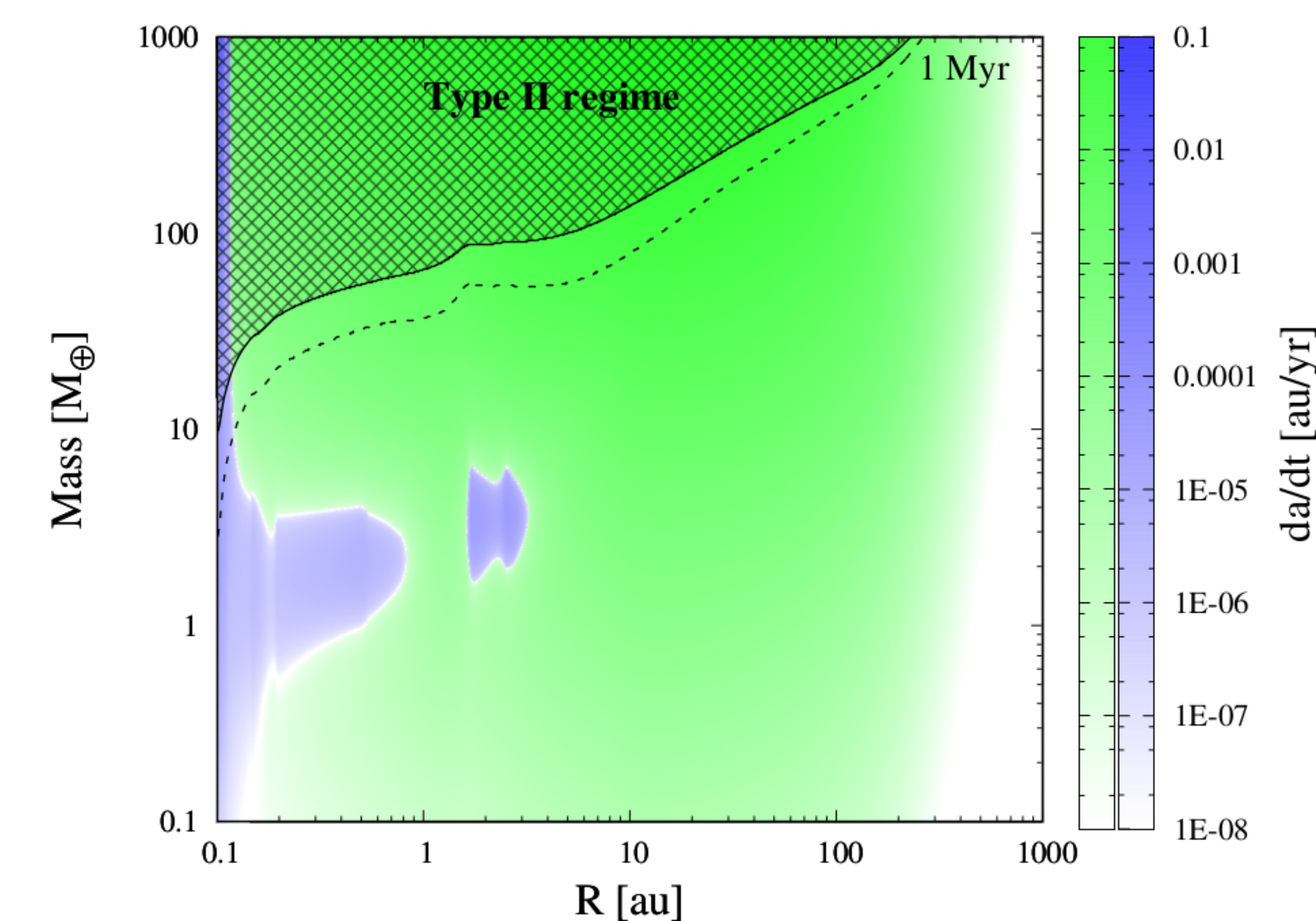} 
    \centering
    \includegraphics[width= 0.43\textwidth]{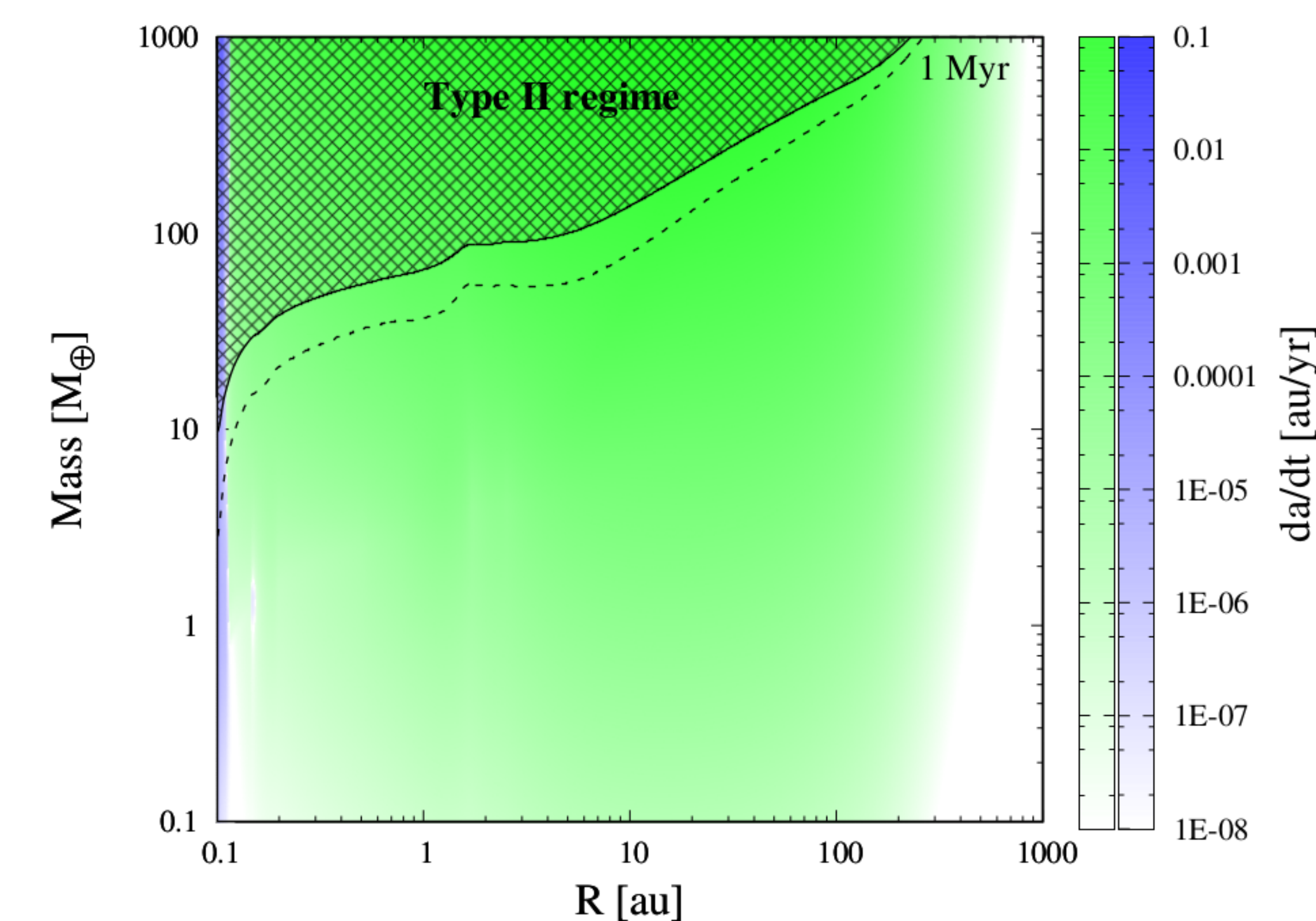}    
    \caption{Migration maps at the same four different times of Fig.~\ref{fig:fig1}. The green palette represents inward migration while the blue palette shows outward migration. The migration maps on the left and right columns are constructed with the type I migration rate prescriptions from \citet{paardekooper.etal2011} and from \citet{jm2017}, respectively. The black curve in all plots represents the planet mass needed to open a gap in the disc at a given distance, according to \citet{Crida2006}, and so the switch to the type II migration regime (shadowed region). In all plots, the dotted curve corresponds to the condition $q = 2h^3$ which defines the limit of an intermediate-mass planet.}
    \label{fig:fig2}
\end{figure*}

In Fig.~\ref{fig:fig2} we show the migration maps calculated using the type I migration rates by both \citet{paardekooper.etal2011} (left) and \citet{jm2017} (right panels). Each row corresponds to a different time, from 50~Kyr (top) to 1~Myr (bottom). These maps were constructed using $1000\times1000$ radial and mass bins, both logarithmic equally spaced. The disc extension and the mass range are comprised between 0.1~au and 1000~au and between $0.1~\text{M}_{\oplus}$ and $1000~\text{M}_{\oplus}$, respectively. The green and blue regions of the maps indicate inward and outward migration of the planet, respectively. The white region in the migration maps indicates that the migration rate is very low, reaching the bottom of the colour scale. The shadowed region for each panel at high masses represents the planet mass at which a gap is opened in the disc, following \citet{Crida2006}. This implies that planets above this line migrate in our simulations according to type II migration rates. The dashed line across each panel represents the planet mass for which the condition $q=2h^3$ is satisfied. We observe that this curve is close to the transition between type I and type II migration regimes. We see in Fig.~\ref{fig:fig2} that the migration maps generated with the two prescriptions considered are quite different, especially the location of the outward migration (blue) regions. Using the migration recipes from \citet{paardekooper.etal2011}, two outward migration zones appear. One at $\sim 3$ au for planet masses between $1 \lesssim \text{M}_{\text{P}}/\text{M}_{\oplus} \lesssim 10$, and an inner one at $R<1$~au for low-mass planets ($1 \lesssim \text{M}_{\text{P}}/\text{M}_{\oplus} \lesssim 5$). Moreover, these regions of outward migration remain roughly constant over the first $1$~Myr of the disc evolution. On the other hand, the model by \citet{jm2017} does not present blue regions of outward migration during the first $1$~Myr. The comparison between the two sets of migration maps shows that the corotation torques predicted by the migration recipes from \citet{paardekooper.etal2011} are higher than the corresponding predicted by \citet{jm2017}.

\begin{figure*}
    \centering
    \includegraphics[width= 0.47\textwidth]{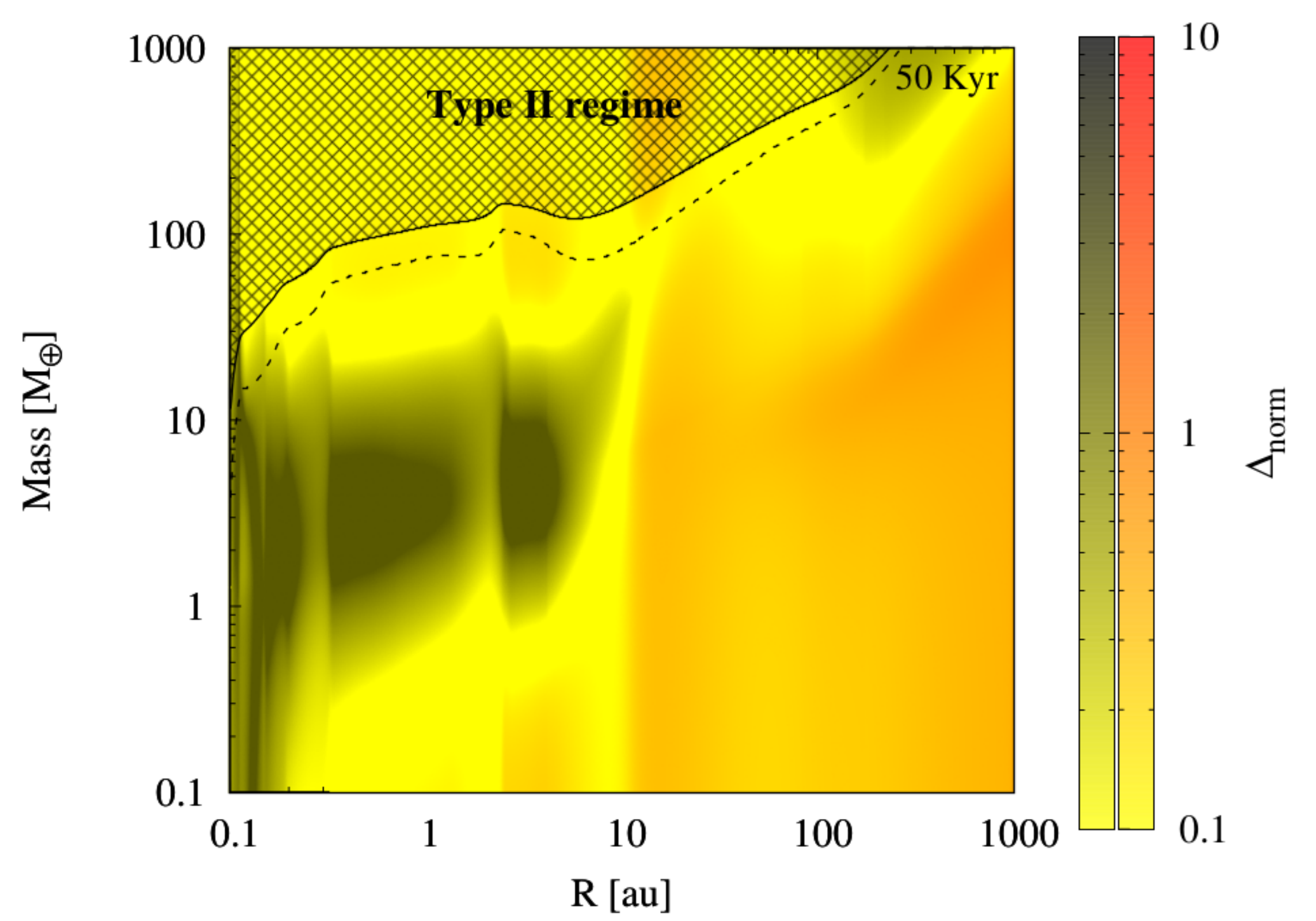} 
    \centering
    \includegraphics[width= 0.47\textwidth]{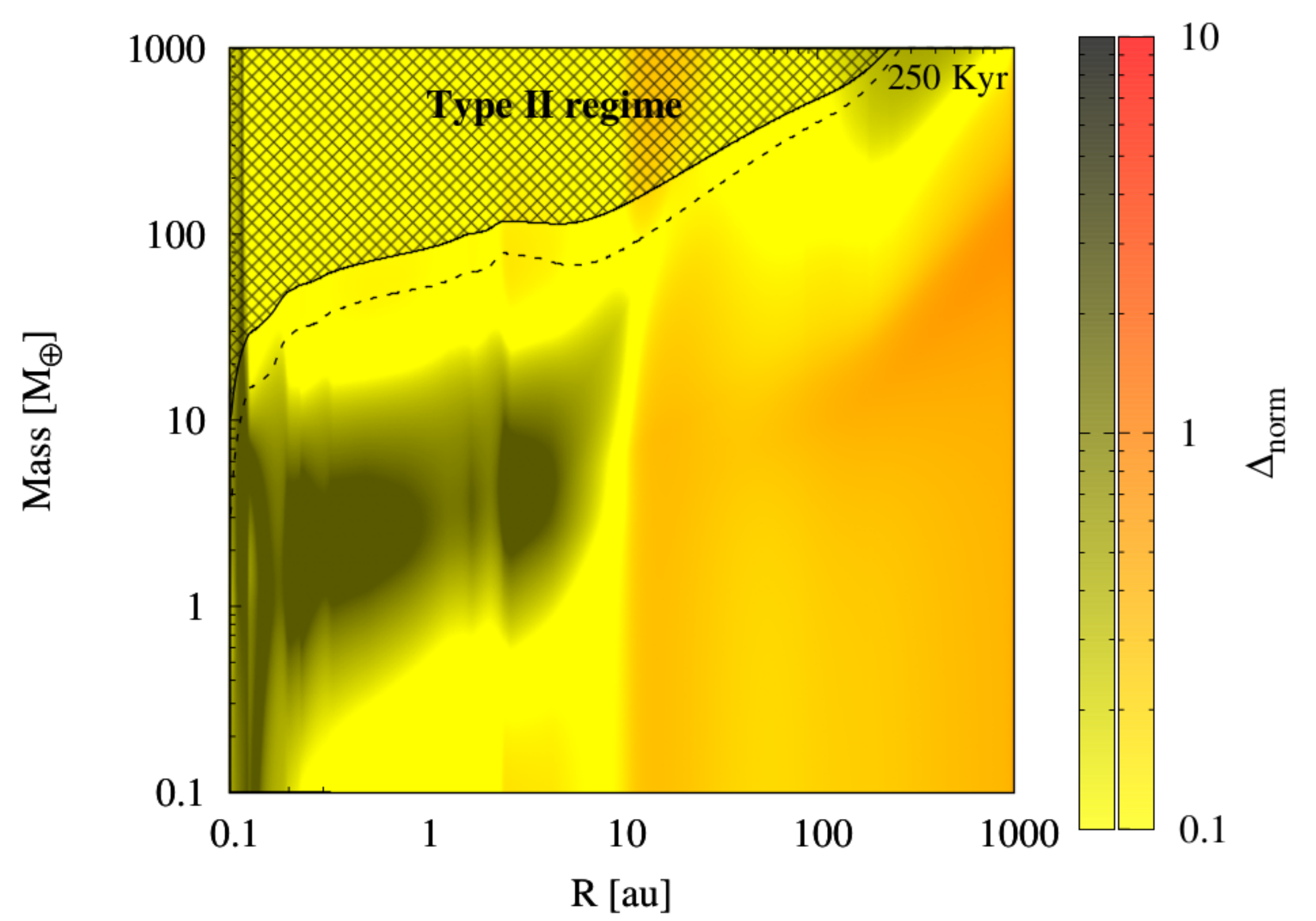} \\
    \centering
    \includegraphics[width= 0.47\textwidth]{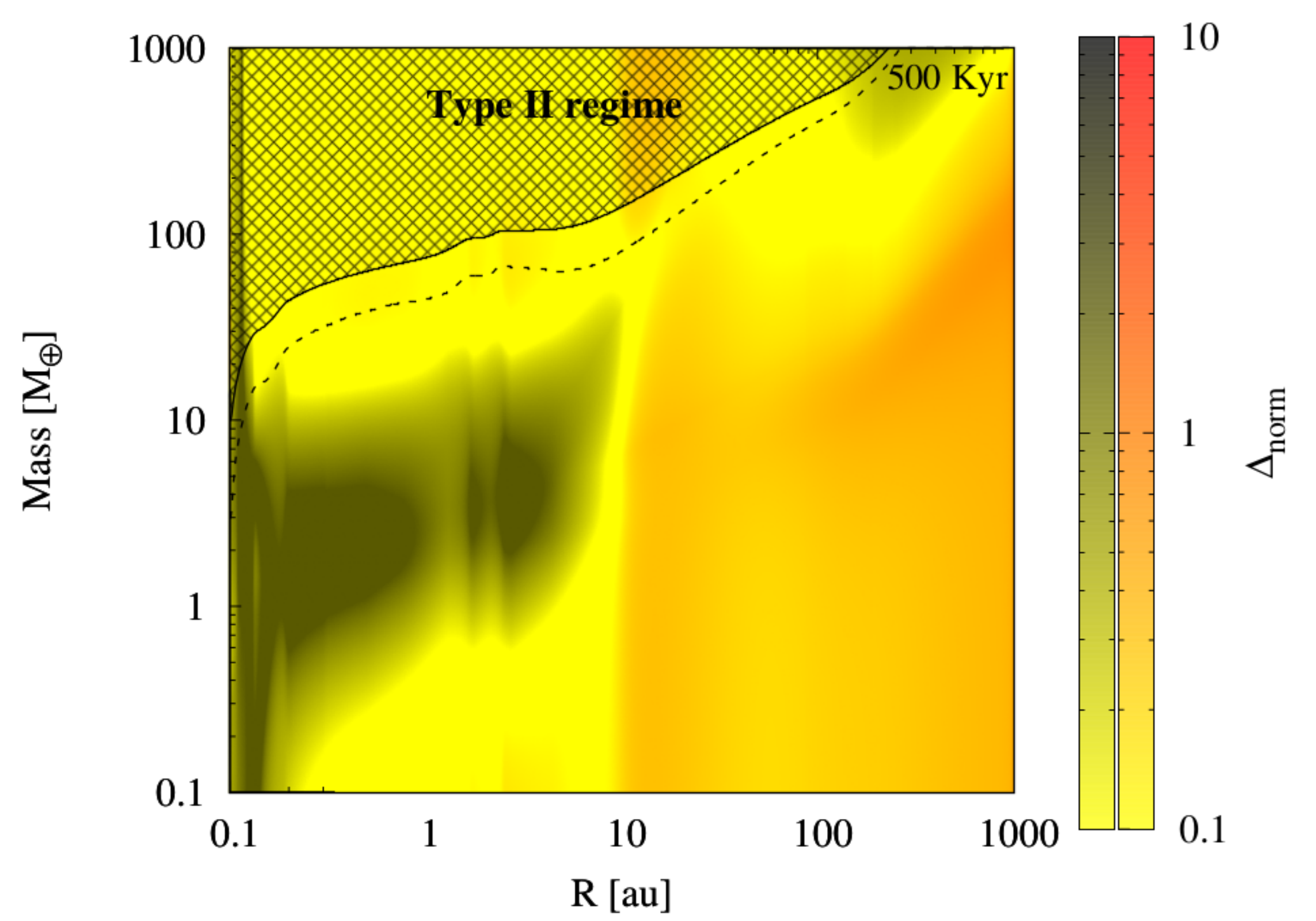} 
    \centering
    \includegraphics[width= 0.47\textwidth]{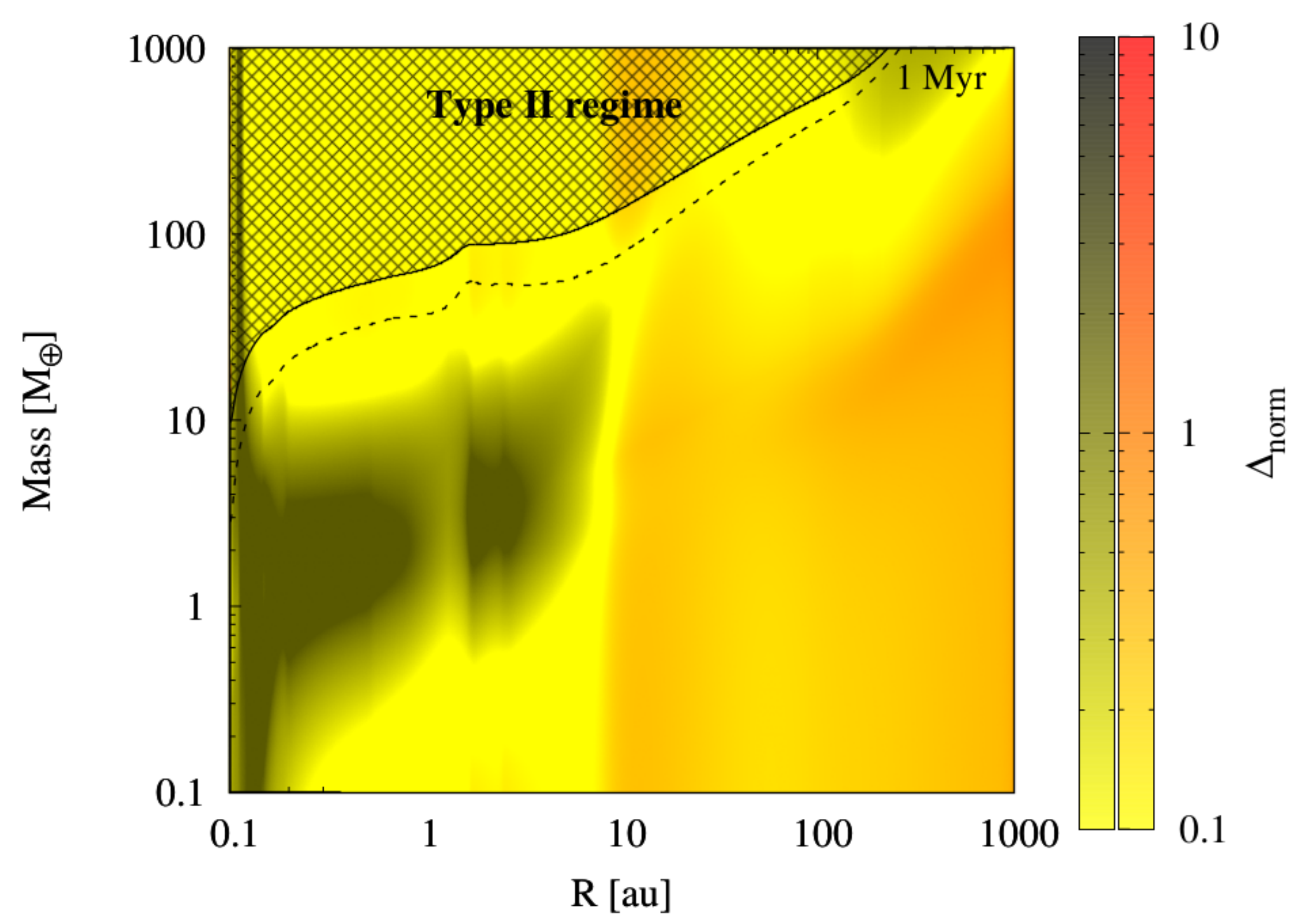}
    \caption{The normalised difference at the four different times of the disc evolution between the type I migration rate prescriptions from \citet{jm2017} and \citet{paardekooper.etal2011}. The yellow-black and the yellow-red palettes represent negative and positive differences, respectively. $\Delta_{\text{norm}}$ represents the normalised difference given by $\Delta_{\text{norm}}= (\dot{a}_{\text{JM2017}} - \dot{a}_{\text{P2011}}) / [(|\dot{a}_{\text{JM2017}}| + |\dot{a}_{\text{P2011}}|)/2]$, where $\dot{a}_{\text{JM2017}}$ and $\dot{a}_{\text{P2011}}$ represent the migration rates from \citet{jm2017} and \citet{paardekooper.etal2011}, respectively.}
  \label{fig:fig3}
\end{figure*}

In Fig.~\ref{fig:fig3}, we plot the normalised difference between the migration maps for the four different times, defined as:
\begin{eqnarray}
  \Delta_{\text{norm}}= \dfrac{\dot{a}_{\text{JM2017}} - \dot{a}_{\text{P2011}}}{(|\dot{a}_{\text{JM2017}}| + |\dot{a}_{\text{P2011}}|)/2}, 
  \label{eq:difnorm1}
\end{eqnarray}
where $\dot{a}_{\text{JM2017}}$ and $\dot{a}_{\text{P2011}}$ represent the migration rates from \citet{jm2017} and \citet{paardekooper.etal2011}, respectively. In these panels, the yellow-black (yellow-red) colour palette represents a negative (positive) difference. We clearly see that there are significant differences between both migration prescriptions for the four different times. These are particularly important in the inner disc regions at $R\lesssim20$~au, where planet formation is expected to take place. These differences also correspond to a wide range of planetary masses, from $0.1~\text{M}_{\oplus}$ up to tens of Earth masses. Moreover, the normalised difference becomes extremely high in the regions where both migration recipes predict opposite migration directions.  As a consequence, each prescription is expected to produce a different early planetary evolution. As we will show in next sections, these differences in the migration rates have a dramatic impact on planet formation.

\begin{figure*}
\centering
\includegraphics[width= 1.0\textwidth]{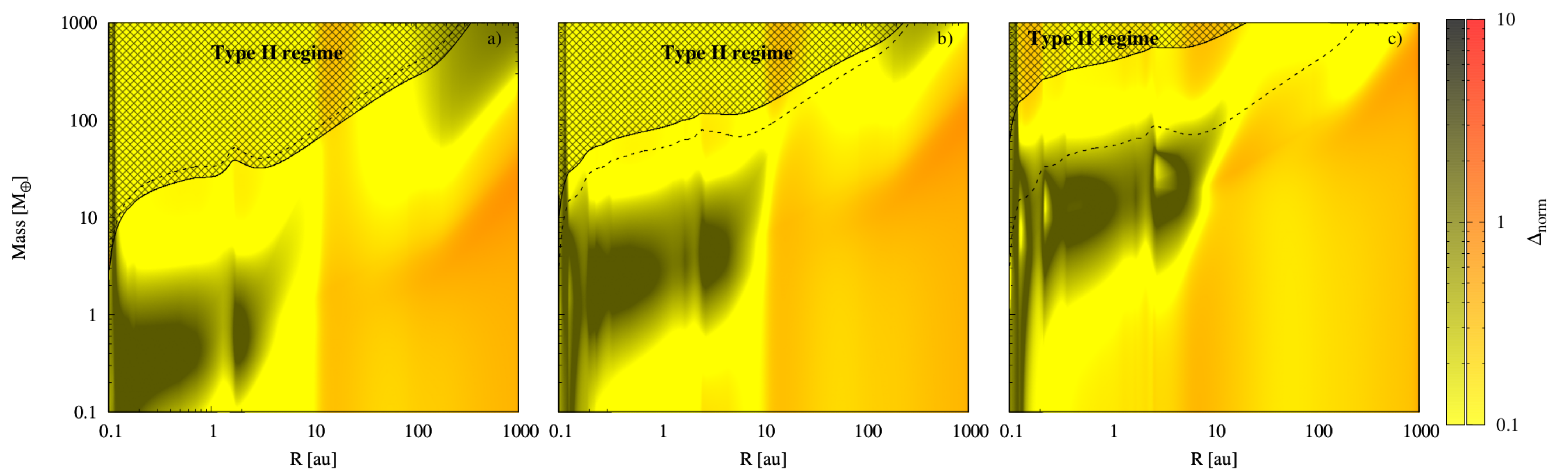}
\caption{Dependence of the normalised difference of the type I migration rate prescriptions at 250 Kyr with the Shakura-Sunyaev $\alpha$-viscosity parameter. The panel a) represents the fiducial case with $\alpha$= 1.e-4, panel b) is the fiducial case ($\alpha$= 1.e-3), and panel c) represents the fiducial case with $\alpha$= 1.e-2. $\Delta_{\text{norm}}$ is defined as in Fig.~\ref{fig:fig3}.}
\label{fig:fig4}
\end{figure*}

\begin{figure*}
\centering
\includegraphics[width= 1.0\textwidth]{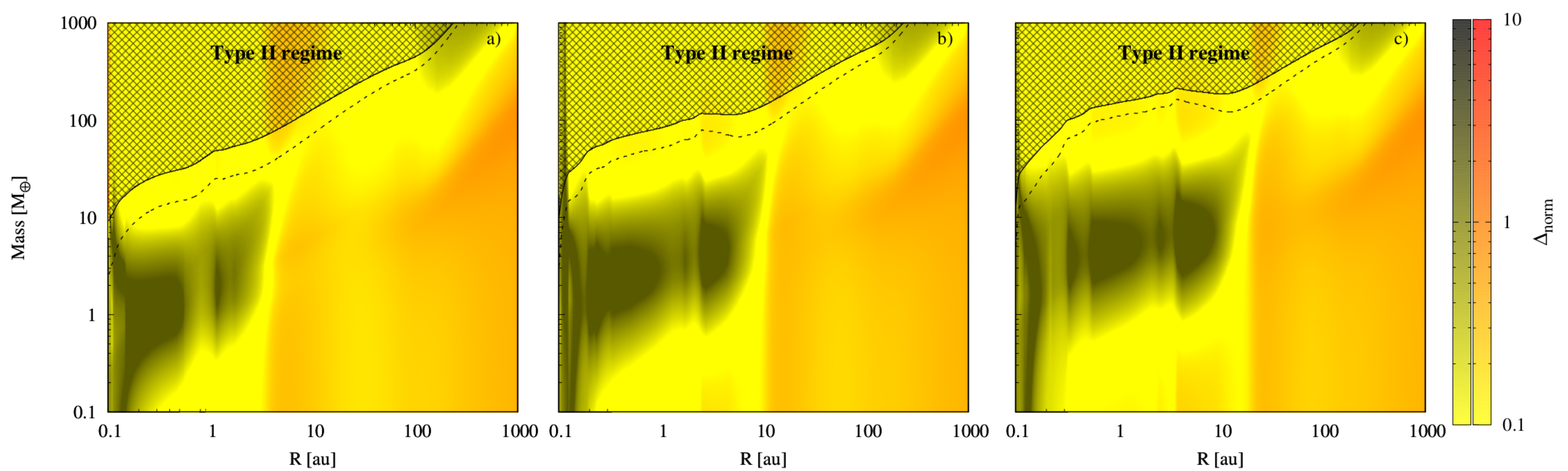}
\caption{Dependence of the normalised difference of the type I migration rate prescriptions at 250 Kyr with the mass of the disc. The panel a) represents the fiducial case for a low-mass disc of $\text{M}_{\text{d}}= 0.01~\text{M}_{\odot}$, panel b) is the fiducial case ($\text{M}_{\text{d}}= 0.05~\text{M}_{\odot}$), and panel c) represents the fiducial case for a massive disc of $\text{M}_{\text{d}}= 0.15~\text{M}_{\odot}$. $\Delta_{\text{norm}}$ is defined as in Fig.~\ref{fig:fig3}.}
  \label{fig:fig5}
\end{figure*}

It is well known that the Shakura-Sunyaev $\alpha$-viscosity parameter and the disc mass (which defines the initial gas surface density) play an important role in the time and radial disc evolution. This in turn affects the torques exerted on the planet. In Fig.~\ref{fig:fig4}, we plot the dependence of the normalised difference between the migration maps at 250 Kyr with the $\alpha$-viscosity parameter. We compare three different migration maps, one using the fiducial disc (panel b), and the other two using the same parameters as the fiducial disc except for the $\alpha$-viscosity parameter. In particular, we consider lower ($10^{-4}$, panel a) and higher ($10^{-2}$, panel c) values for $\alpha$. In Fig.~\ref{fig:fig5} we plot, also at 250 Kyr, the dependence of the normalised difference with the disc mass. The panel b) represents The fiducial case is shown in panel b), while the disc model with lower ($\text{M}_{\text{d}}= 0.01~\text{M}_{\odot}$) and higher ($\text{M}_{\text{d}}= 0.15~\text{M}_{\odot}$) masses are shown in panels a) and c), respectively. Figures~\ref{fig:fig4} and \ref{fig:fig5} show that --independently of the values of $\alpha$ and $\text{M}_{\text{d}}$-- the migration maps are substantially different for the two recipes considered. This result is robust since it applies for a wide range of disc parameters. Additionally, in Fig.~\ref{fig:fig4} and Fig.~\ref{fig:fig5} we see that the black region moves to higher masses as the $\alpha$-viscosity parameter and the disc mass increase (i.e. as the viscosity increases), being more evident the dependence with the $\alpha$-viscosity parameter. This black region presents the major discrepancy between both models and it is related with the unsaturated horseshoe component of the corotation torque. For low masses, the corotation torque is linear and its value is similar from both recipes, while at high masses the corotation torque is saturated and it tends to vanish, also for both recipes (see Fig~\ref{fig:fig12}). The transition mass between the linear regime and the horseshoe drag regime of the corotation torque depends on the viscosity (the transition mass increases as the viscosity increases). In the same way, the mass for the saturation of the horseshoe drag also increases as the viscosity increases. This explains why the black region moves to higher masses as the viscosity increases.

\subsubsection{Comparison between \citet{jm2017} type I migration rates with and without \citet{masset2017} thermal torque}
\label{sec3-1-2}

\begin{figure*}
  \centering
  \includegraphics[width= 0.49\textwidth]{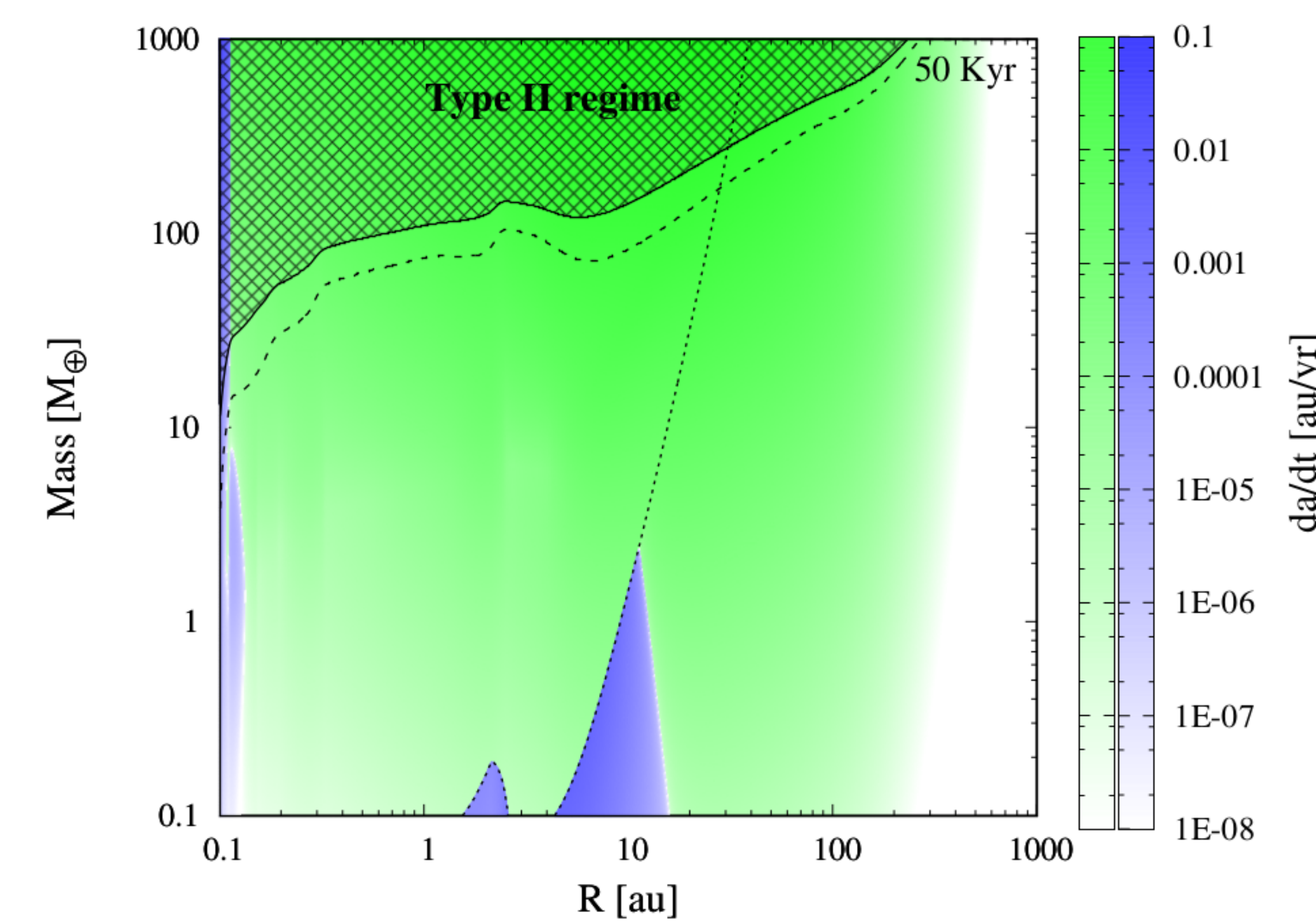} 
  \centering
  \includegraphics[width= 0.49\textwidth]{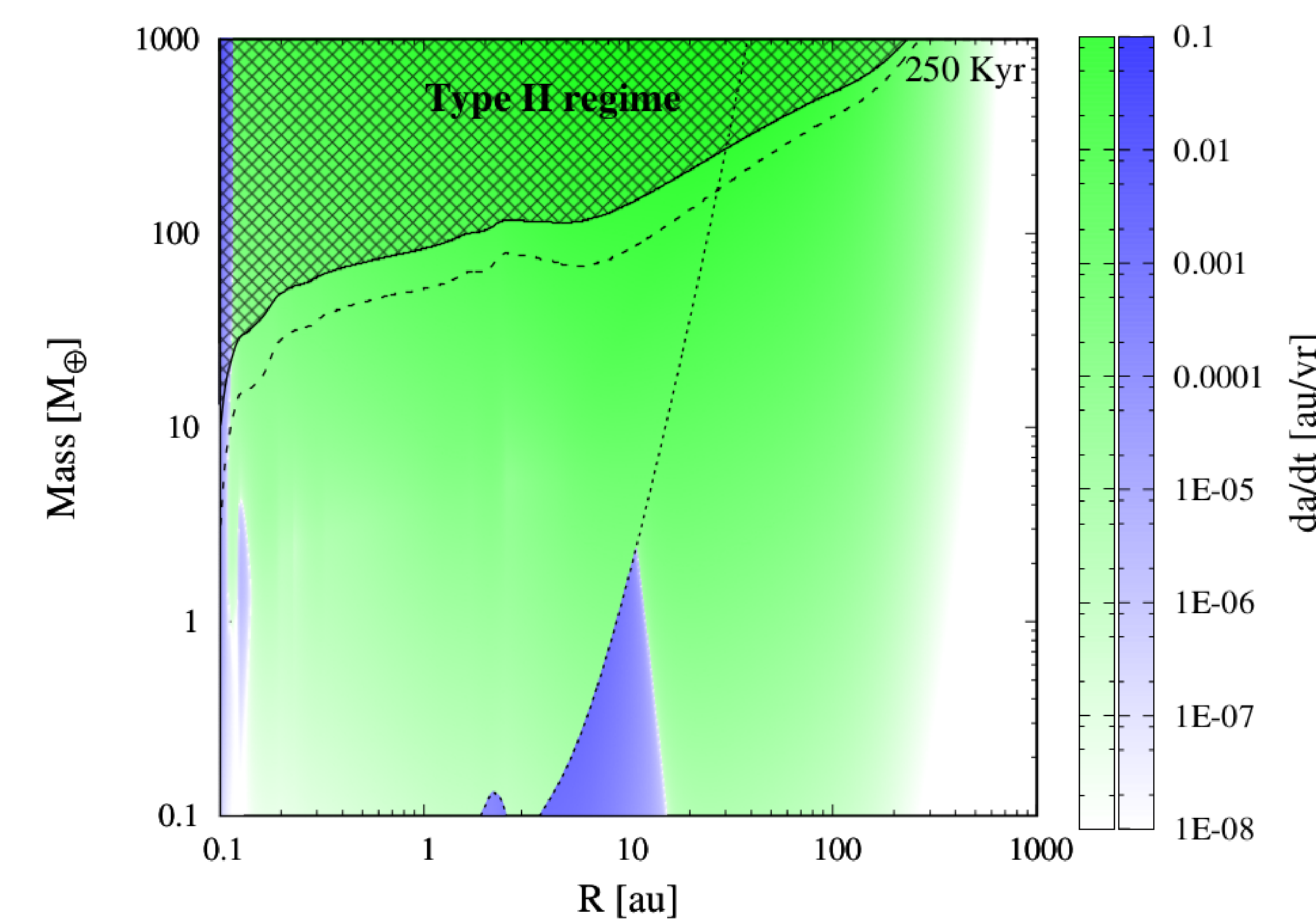} \\
  \includegraphics[width= 0.49\textwidth]{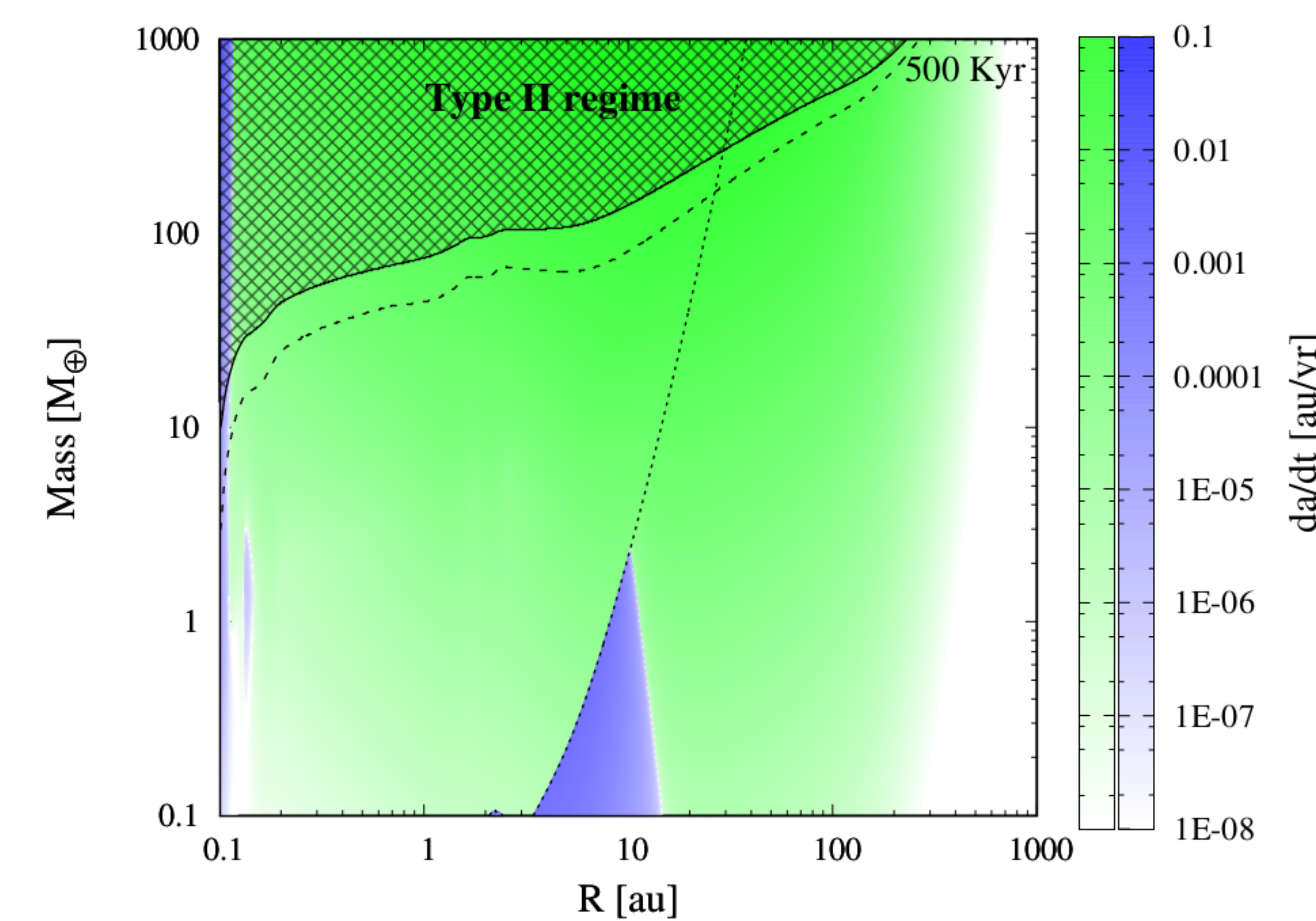}   
  \centering
  \includegraphics[width= 0.49\textwidth]{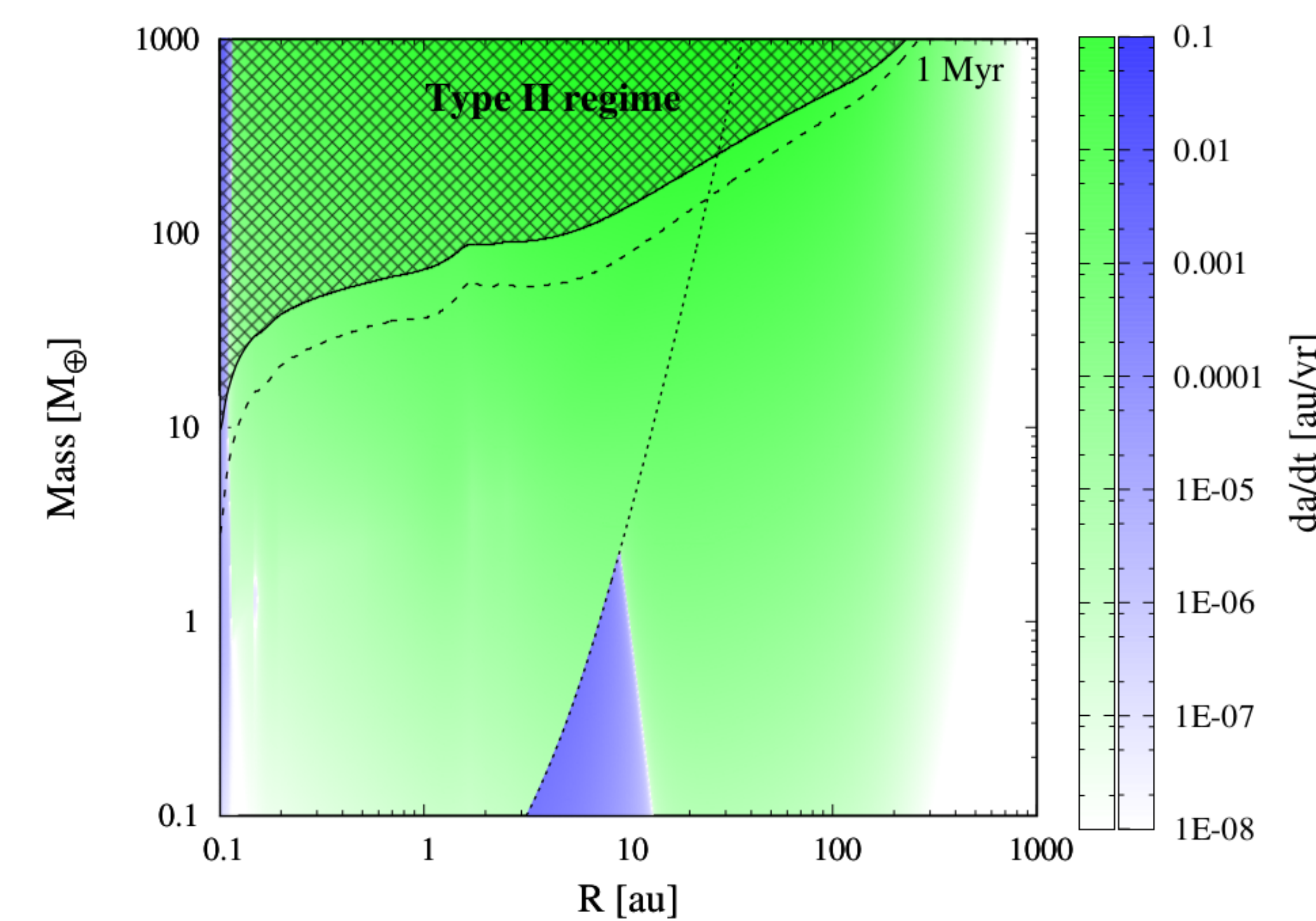}    
  \caption{Migration maps, at the same four different times of Fig.~\ref{fig:fig2}, using the type I migration rate prescriptions combined from \citet{jm2017} and \citet{masset2017}. For the thermal torque derived by \citet{masset2017}, we adopt a constant planet luminosity, corresponding to a constant solid accretion rate of $10^{-5}~\text{M}_{\oplus}/{\rm yr}$ (which is a typical value for small planetesimals in this work). The green palette represents inward migration while the blue palette represents outward migration. The black dotted curve in all plots represents the critical thermal mass above which the thermal torque is suppressed.}
  \label{fig:fig8}
\end{figure*}

Here we compare the migration maps obtained for 3D adiabatic discs \citep{jm2017}, with and without the inclusion of the thermal torque derived by \citep{masset2017}. It is important to note that we do not add the thermal torque to the type I migration prescriptions derived by \citet{paardekooper.etal2011}. As \citet{masset2017} pointed out, the thermal torque is derived taking into account how the disc thermal diffusion affects the torques over the planet. \citet{paardekooper.etal2011}, derived their migration prescriptions based on 2D radiative hydrodynamics simulations wherein thermal diffusion was already included. Thus, the migration recipes derived by \citet{paardekooper.etal2011} could have partially incorporated the effect of the cold torque, although we can anticipate that it would have a very small value in their two-dimensional experiments with a finite smoothing length. This does not occur with the heating torque because \citet{paardekooper.etal2011} did not considered the heat released by the planet due to solid accretion. As mentioned in Section~\ref{Intro}, we stress that the migration recipes derived by \citet{jm2017} do not take into account the effects of the cold and heating torques.   

In Fig.~\ref{fig:fig8}, we plot the migration maps constructed employing the type I migration rates combined from \citet{jm2017} with the ones corresponding to the thermal torque from \citet{masset2017}. As in Fig.~\ref{fig:fig2}, each panel corresponds to a different time, from 50~Kyr to 1~Myr. The green (blue) regions of the maps indicate an inward (outward) planet migration. Again, the shadowed region in each panel corresponds to the transition to the type II migration regime, and the dashed line corresponds to the condition $q=2h^3$. The dotted line in each panel corresponds to the critical thermal mass. These migration maps are constructed considering a constant solid accretion rate of $10^{-5}~\text{M}_{\oplus}/{\rm yr}$ for the thermal torque computation. We observe that the inclusion of the thermal torque introduces two regions of outward migration at low masses up to $\sim 2.5~\text{M}_{\oplus}$. The small blue island between 1~au and 3~au disappears as the disc evolves. However, the larger one between 4~au and 15~au, remains practically unchanged during the first Myr of the disc evolution. We note that in this region of outward migration, the transition to inward migration is smooth at the right side of the island. This means that in this specific region the Lindblad and cold torques dominate over the heating torque. On the other hand, at the left side of the blue island, the transition is more abrupt. This is due to the fact that in this case, as opposed to what happens on the right side, the planet reaches the critical mass for the thermal torque. The presence of this region of outward migration plays an important role in planetary evolution (see formation tracks in Sect.~\ref{sec3-2}).

\begin{figure*}
  \centering
  \includegraphics[width= 0.49\textwidth]{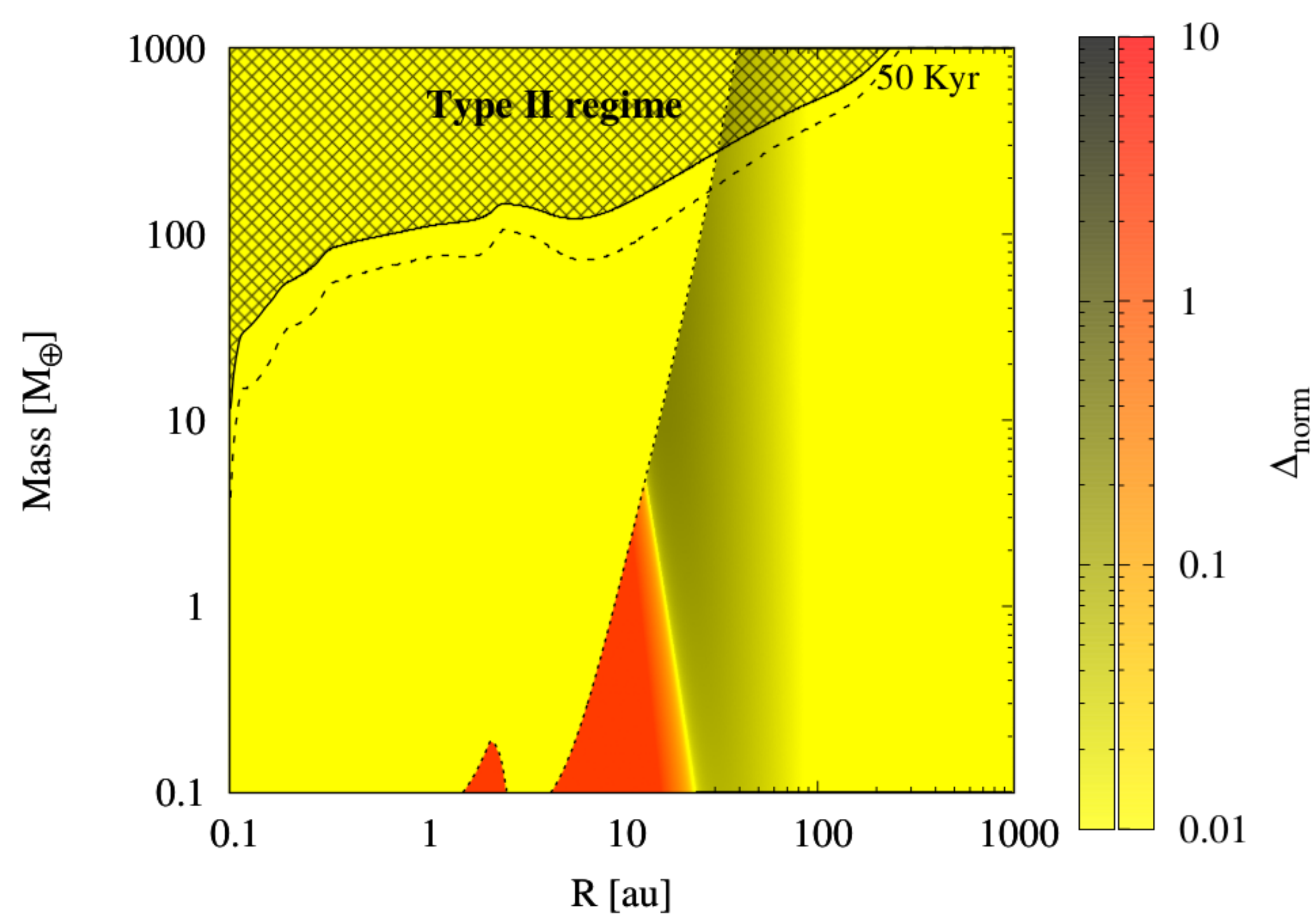} 
  \centering
  \includegraphics[width= 0.49\textwidth]{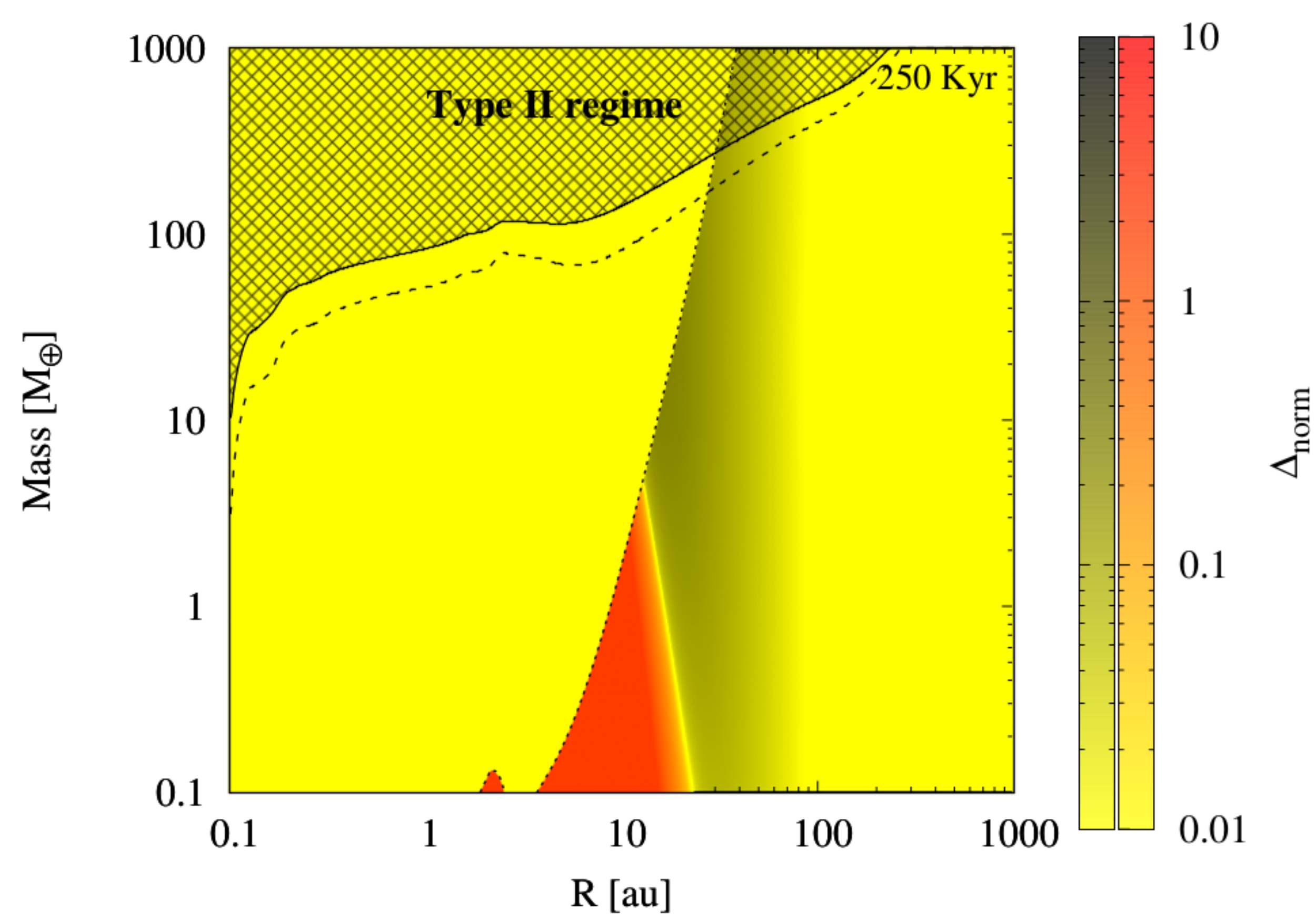} \\
  \includegraphics[width= 0.49\textwidth]{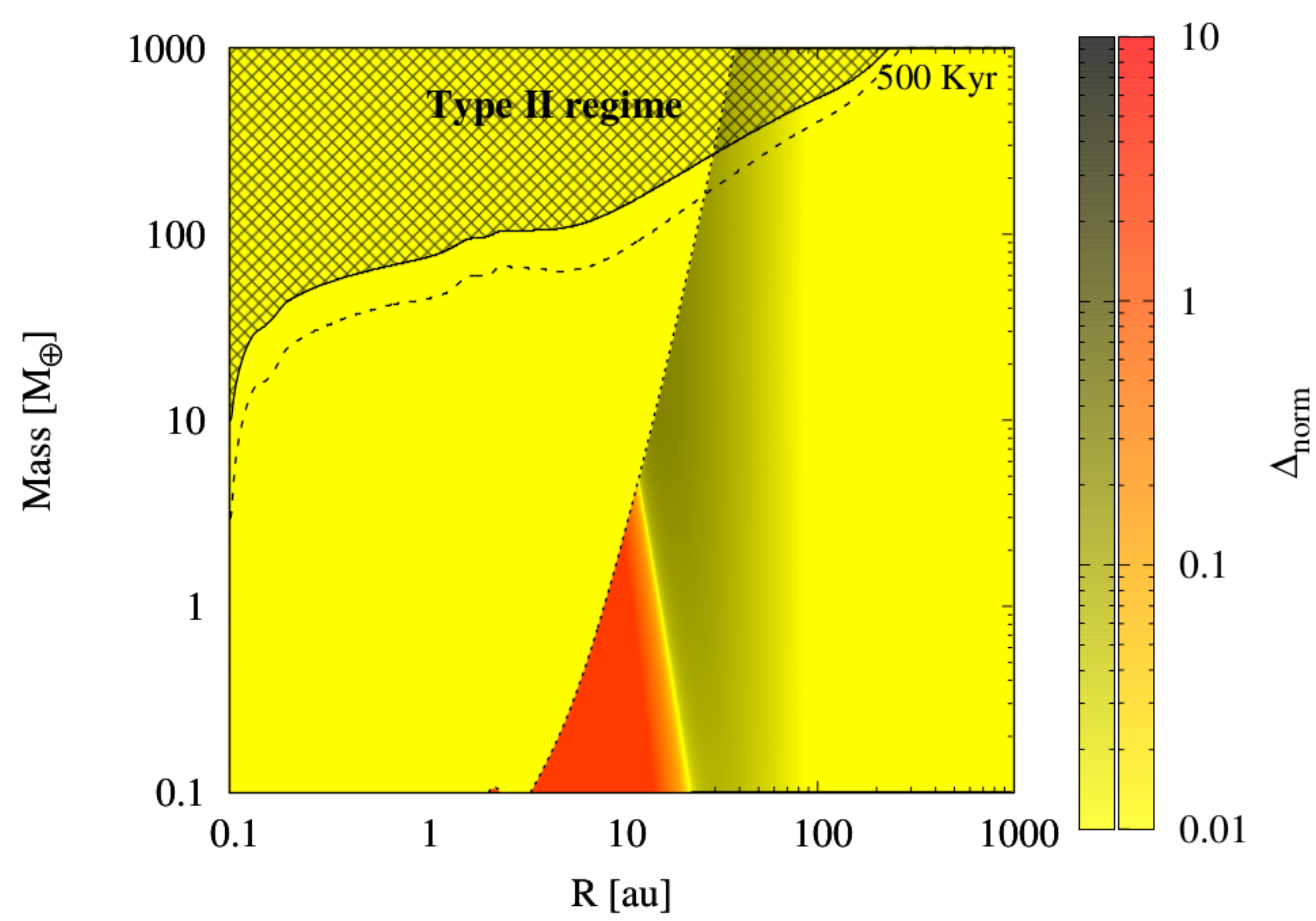}   
  \centering
  \includegraphics[width= 0.49\textwidth]{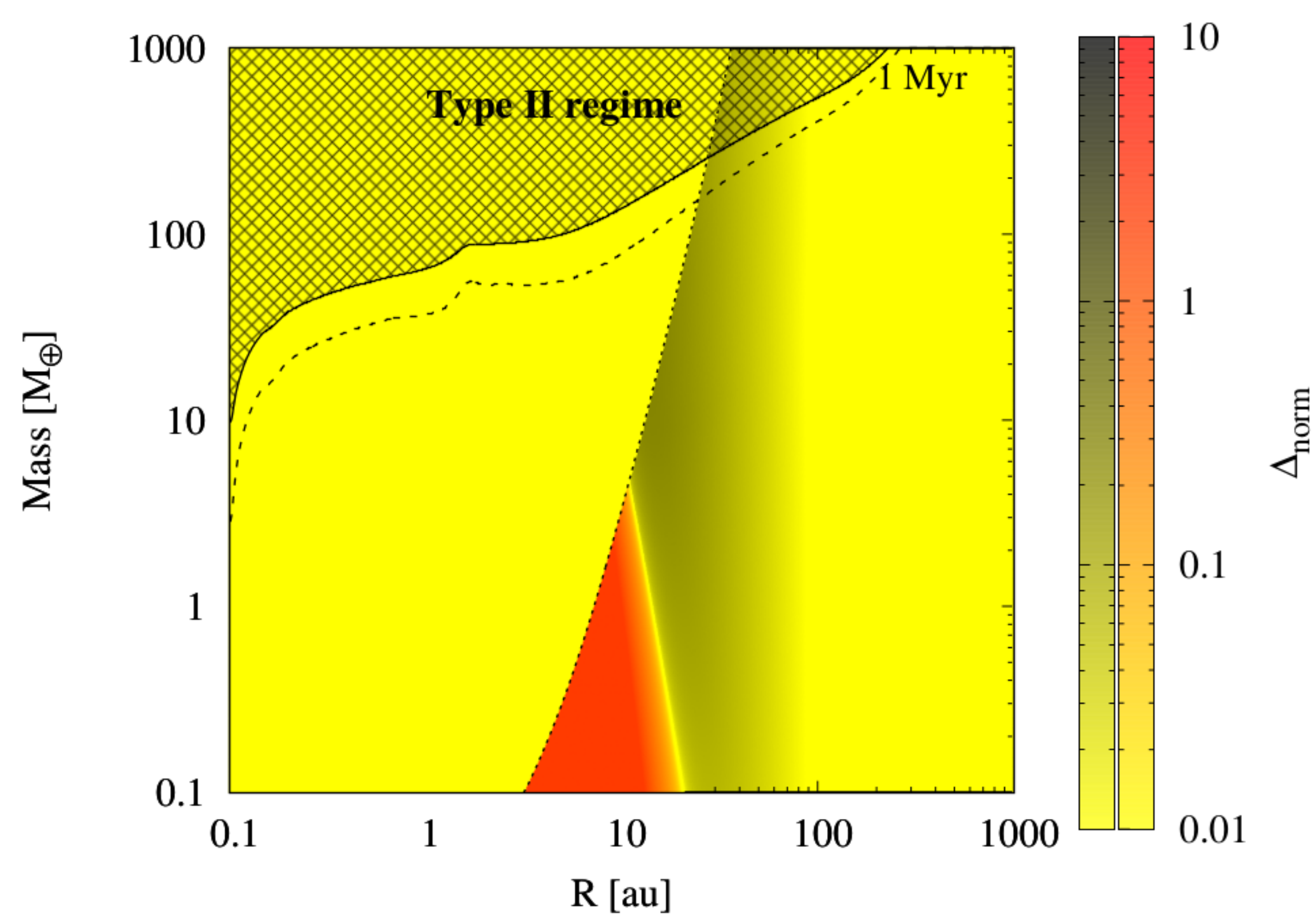}    
  \caption{The normalised difference at the four different times of the disc evolution between the type I migration rate prescriptions combined from \citet{jm2017} and \citet{masset2017}, and the ones from \citet{jm2017}. The yellow-black and yellow-red palettes represent negative and positive differences, respectively. $\Delta_{\text{norm}}$ represents here the normalised difference given by $\Delta_{\text{norm}}= (\dot{a}_{\text{JM\&M2017}} - \dot{a}_{\text{JM2017}}) / [(|\dot{a}_{\text{JM\&M2017}}| + |\dot{a}_{\text{JM2017}}|)/2]$, where $\dot{a}_{\text{JM\&M2017}}$ represents the migration rates combined from \citet{jm2017} and \citet{masset2017}, and $\dot{a}_{\text{JM2017}}$ represents the migration rates form \citet{jm2017}.}
  \label{fig:fig8bis}
\end{figure*}

In Fig.~\ref{fig:fig8bis}, we plot the normalised difference between the type I migration rate prescriptions combined from \citet{jm2017} and \citet{masset2017} minus the ones derived only from \citet{jm2017}. In the region where the planet mass is greater than the critical thermal mass (in bright yellow at the left of each panel), the type I migration prescriptions for both models are the same. The regions in red represent a positive normalised difference. These are regions of outward migration where the heating torque dominates the total torque. At some point, the combination between the Lindblad and the cold torque compensates the heating torque and an interface of small or null torque is generated (yellow transition between the red and black regions). This situation occurs due to the fact that, for a fixed solid accretion rate and a fixed mass of the planet, the heating torque decreases with distance to the central star  (see Eqs. \ref{eq19-sec2-4-1} and \ref{eq20-sec2-4-1}). The black region represents a negative normalised difference. This means that the total torque derived from the sum of \citet{jm2017} and \citet{masset2017} is more negative than the one derived only from \citet{jm2017}. Here the cold thermal torque dominates over the heating torque. In other words, the planet's luminosity is smaller than the critical value given by Eq.~\ref{eq20-sec2-4-1} and hence a more negative total torque is obtained. As we move far away from the central star, the contribution of the cold torque decreases (see Eq.~\ref{eq18-sec2-4-1}) and the planet enters the Type~I migration regime again (strong yellow).

\begin{figure*}
\centering
\includegraphics[width= 0.99\textwidth]{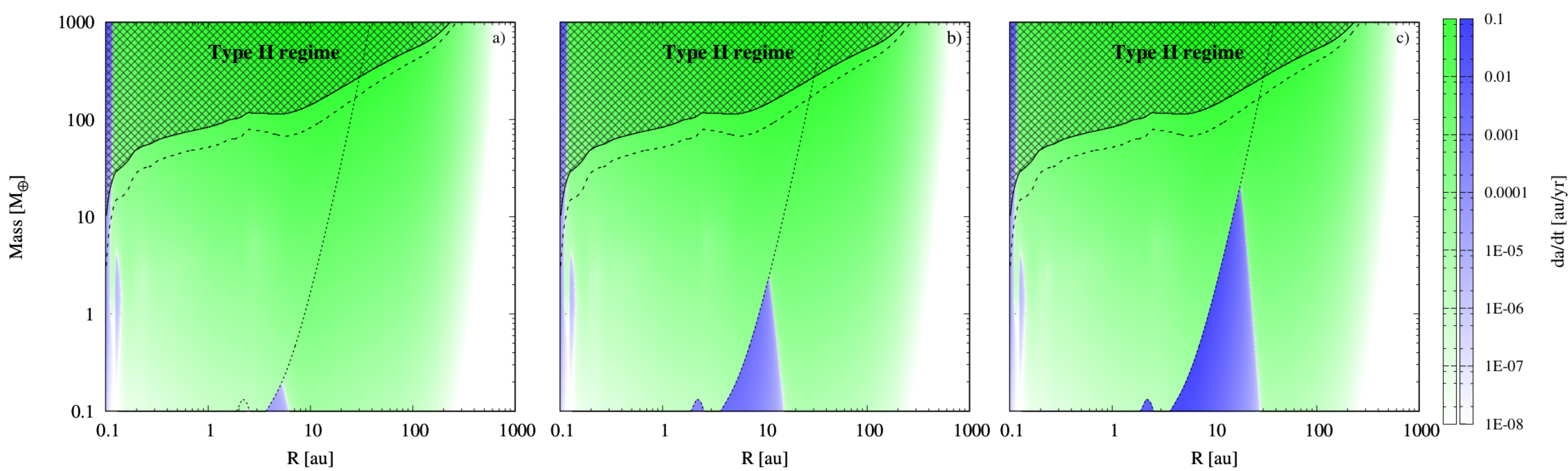}
\caption{Migration maps at 250 Kyr incorporating the thermal torques for three different constant solid accretion rates: $10^{-6}~\text{M}_{\oplus}/{\rm yr}$ (a), $10^{-5}~\text{M}_{\oplus}/{\rm yr}$ (b) and $10^{-4}~\text{M}_{\oplus}/{\rm yr}$ (c).}
\label{fig:fig9}
\end{figure*}

In Fig.~\ref{fig:fig9}, we plot the migration maps at 250 Kyr for three different constant solid accretion rates, $10^{-6}~\text{M}_{\oplus}/{\rm yr}$ (left), $10^{-5}~\text{M}_{\oplus}/{\rm yr}$ (centre), and $10^{-4}~\text{M}_{\oplus}/{\rm yr}$ (right). As the heating torque is proportional to the luminosity of the planet generated by the accretion of the pebbles or planetesimals, the blue region of outward migration increases as the solid accretion grows. For a constant solid accretion rate of $10^{-4}~\text{M}_{\oplus}/{\rm yr}$, the blue outward migration region extends up to ~30~au and up to a planet mass greater than $\sim 10~\text{M}_{\oplus}$. As before, at the right side of this region the transition to the inward migration regime is soft, but at the left side is abrupt because the planet reaches the critical mass for the thermal torque. As a planet grows, the solid accretion is expected to increase as well as long as there is enough available material in the planet feeding zone.

\subsection{Effects on planet formation}
\label{sec3-2}

In the previous sections, we constructed migration maps and showed that the migration rates predicted by different authors have significant differences, which can be of more than an order of magnitude for low-mass and intermediate-mass planets. In this section, we study the differences that arise from the adoption of either migration prescription in our planet formation model.

\subsubsection{Planet formation affected by \citet{paardekooper.etal2011} and \citet{jm2017} type I migration rates}
\label{sec3-2-1}

\begin{figure*}
    \centering
    \includegraphics[width= 0.325\textwidth, angle= 270]{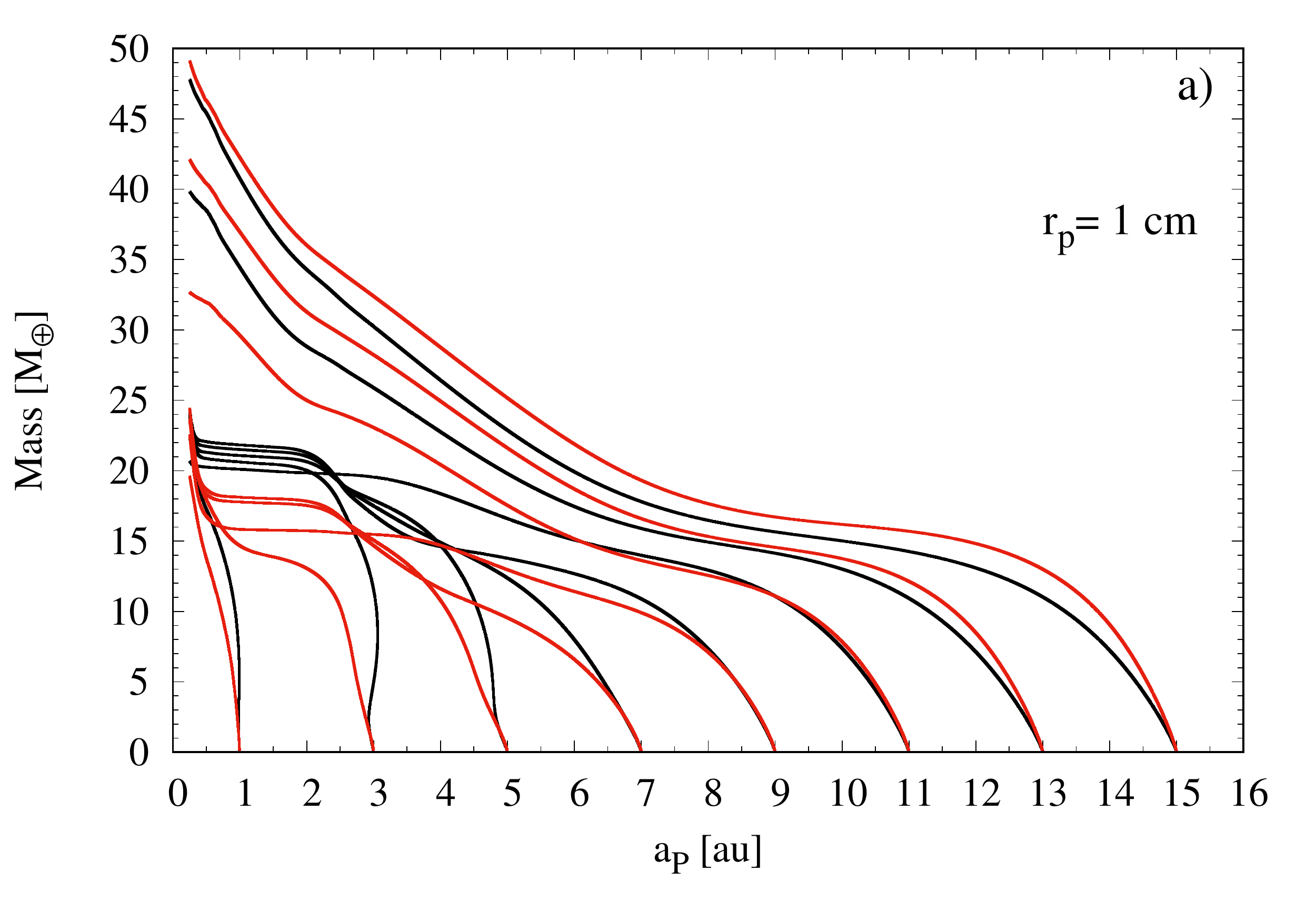} 
    \centering
    \includegraphics[width= 0.325\textwidth, angle= 270]{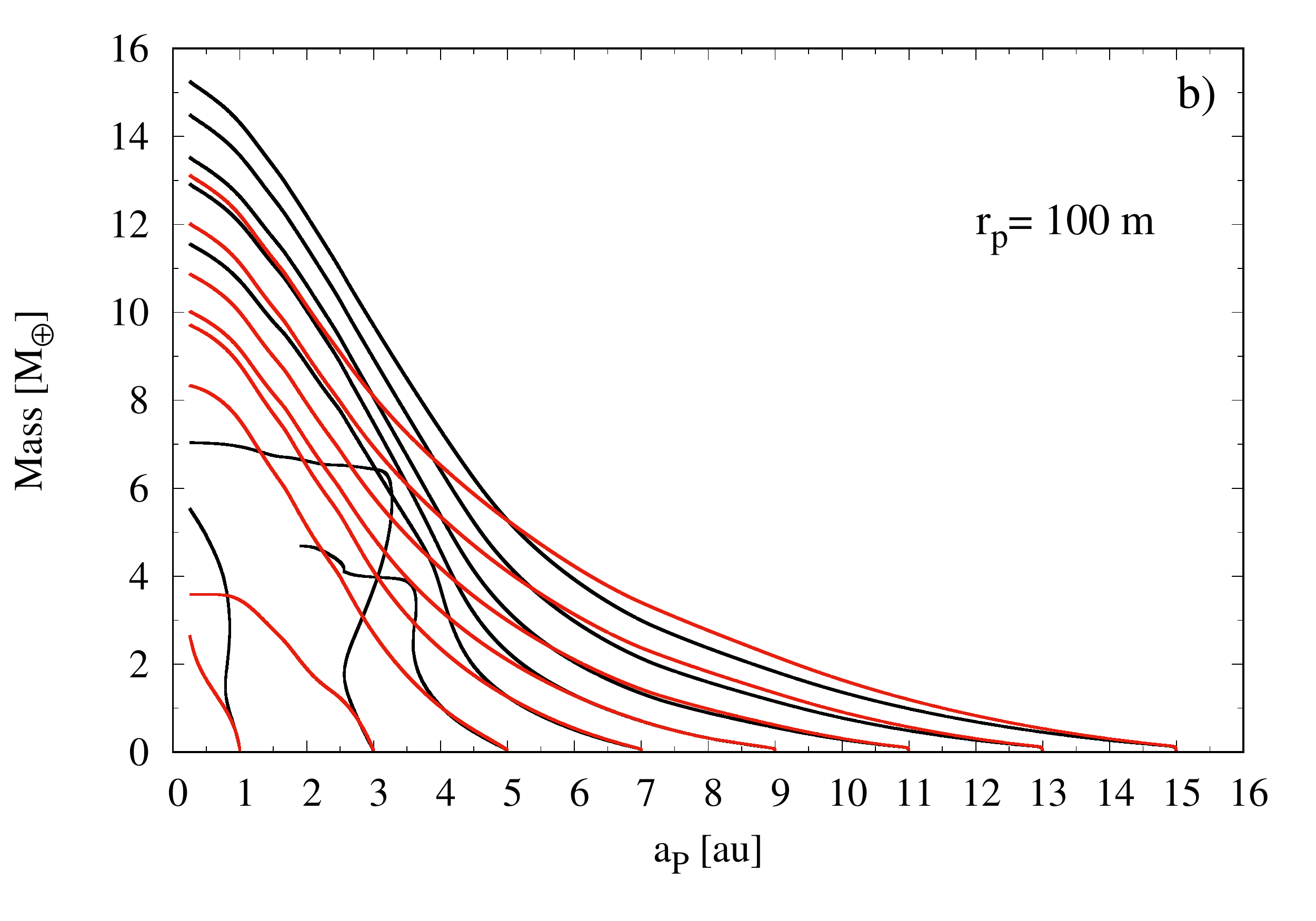} \\
    \centering
    \includegraphics[width= 0.325\textwidth, angle= 270]{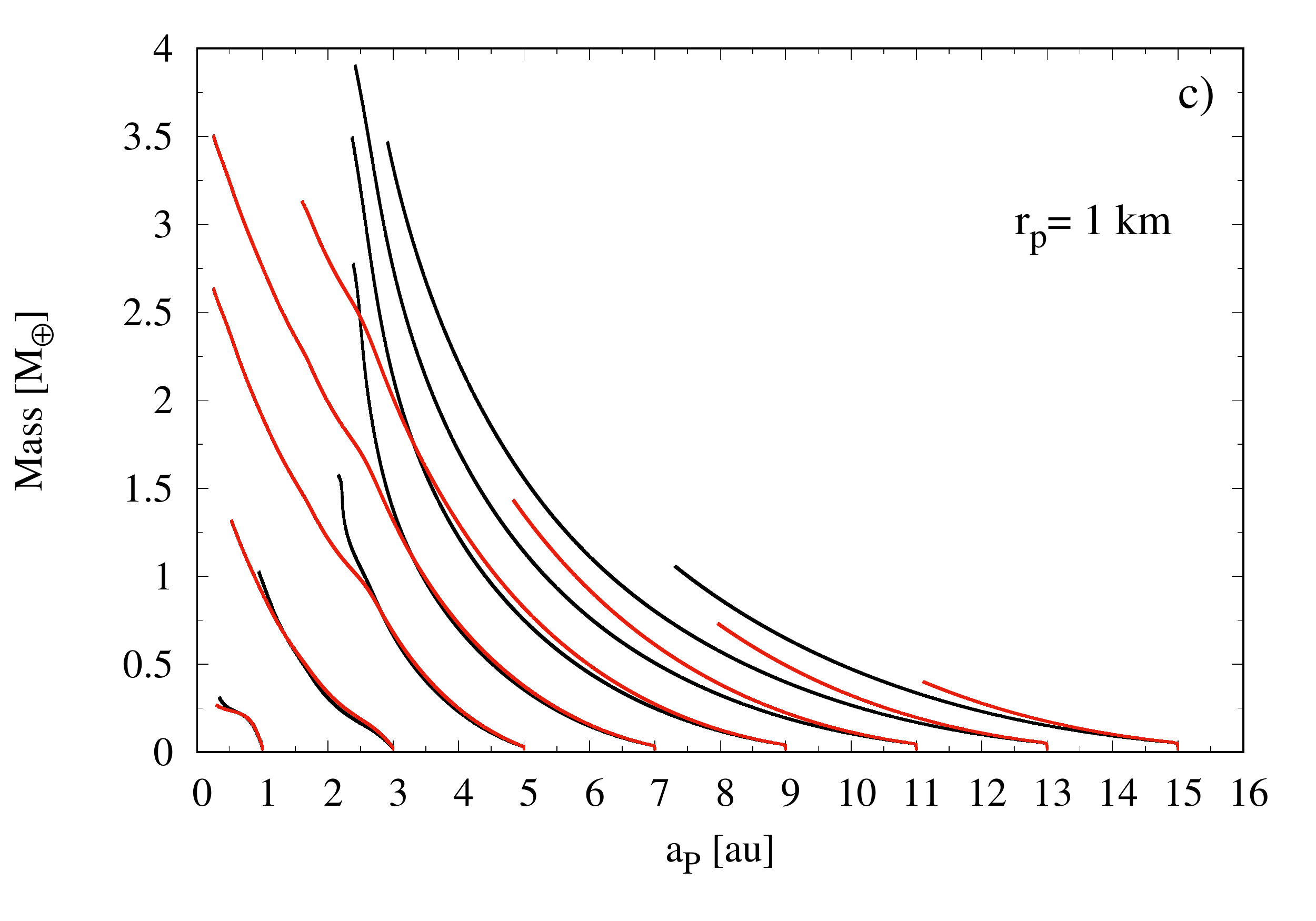} 
    \centering
    \includegraphics[width= 0.325\textwidth, angle= 270]{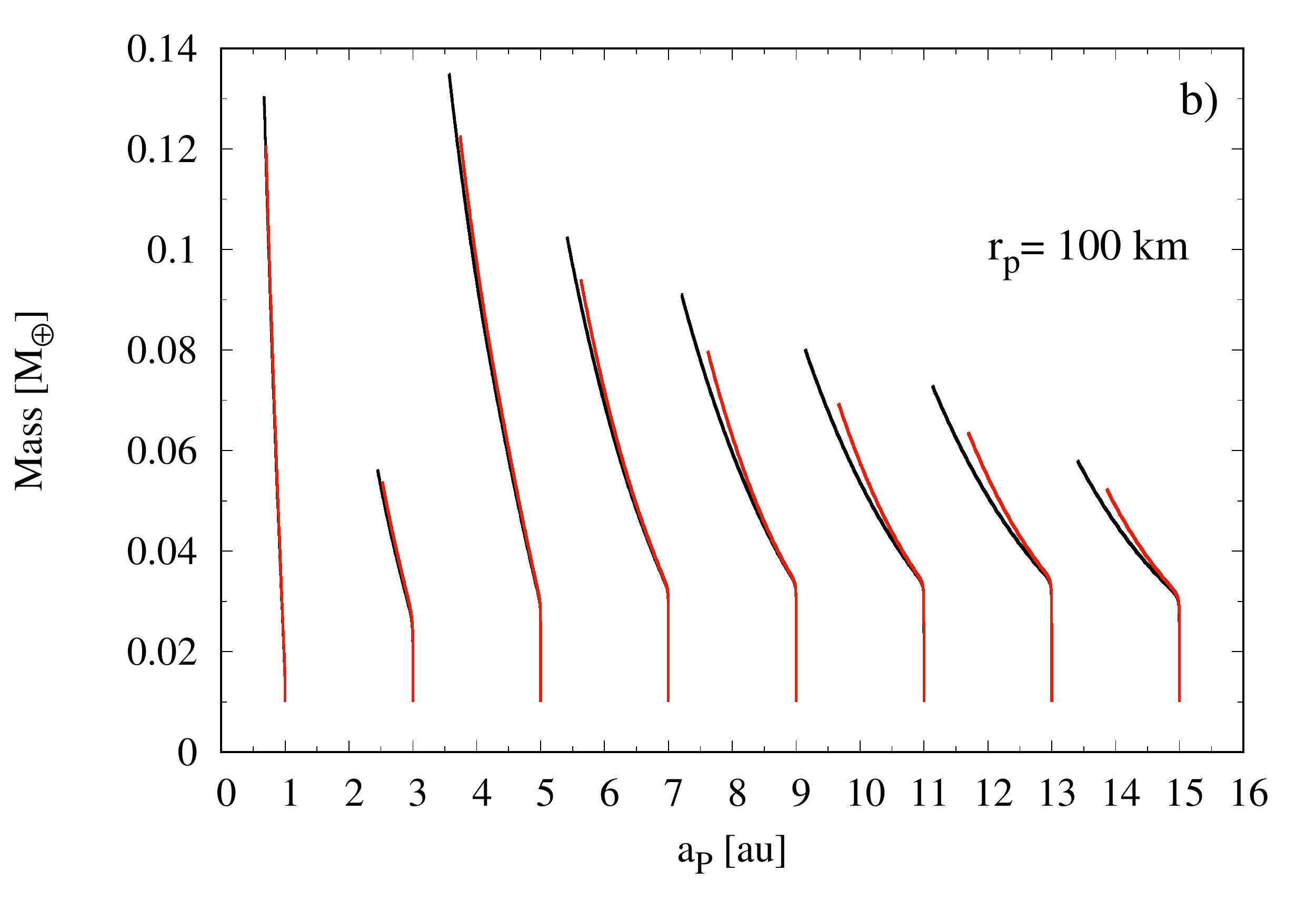} 
  \caption{Formation tracks of a planet initially located at different positions in the disc, from 1~au to 15~au, for the fiducial case. Each track corresponds to one simulation, i.e., we simulate the formation of only one planet per disc. The black and red curves correspond to the simulations using the type I migration rates from \citet{paardekooper.etal2011} and  \citet{jm2017}, respectively. The panels a, b, c, and d correspond to the case of pebbles of 1~cm of radii, and for the cases of planetesimals of 100~m of radii, 1~km of radii, and 100~km of radii, respectively. For the bigger planetesimals, the planet formation tracks for both models are practically identical. These planetesimals have the smaller accretion rates, so embryos only grow up to Mars masses. As type~I migration rates are proportional to the mass of the planet, both models do not exhibit remarkable differences.}
  \label{fig:fig6}
\end{figure*}

In Fig.~\ref{fig:fig6}, we plot the planet formation tracks considering different initial locations and different initial pebble and planetesimal sizes. The black lines represent the formation tracks using the migration prescriptions from \citet{paardekooper.etal2011}, while the red lines represent the tracks using the migration prescriptions from \citet{jm2017}. It is important to note that we performed one simulation per formation track, i.e. we do not calculate the simultaneous formation of several planets. In our simulations, the proto-planet begins as a rocky core of $0.01~\text{M}_{\oplus}$ immersed in our fiducial disc. The simulations end when the planet either reaches the disc inner radius or after 3~Myr of evolution.

When we considered pebbles of 1~cm radius (panel a of Fig.~\ref{fig:fig6}), the planets always reach the inner part of the disc experiencing both type I migration recipes. For planets with initial locations lower than 9~au, the planet formation tracks are very different despite the fact that all of them end with a similar mass (of $\sim 25~\text{M}_{\oplus}$). We note that if the planet is initially located at 11~au, the planet formation tracks and masses are dramatically different. For planets at further distances, the formation tracks and final masses remain similar.

For the case of initial planetesimals of 100~m of radius (panel b of Fig.~\ref{fig:fig6}), the formation tracks and final masses are very different using both type I migration rates (in this case for all the initial planets locations). For initial planet locations beyond 7~au, the type I migration rates derived using the prescription of \citet{jm2017} seem to be higher than the corresponding rates derived using the prescription of \citet{paardekooper.etal2011}. In all the simulations, the planets reach the inner radius of the disc, except for the planet initially located at 5~au using the migration recipes from \citep{paardekooper.etal2011}.

For the case of initial planetesimals of 1~km (panels c of Fig.~\ref{fig:fig6}), planet formation tracks present significant differences between both cases when planets grow up to a a few Earth masses.

Finally, for 100~km planetesimals (panels d of Fig.~\ref{fig:fig6}), the formation tracks are similar using the two migration prescriptions considered. This is due to the fact that planets only grow up to about the mass of Mars at most.

\begin{figure}
    \centering
    \includegraphics[width= 0.345\textwidth,angle= -90]{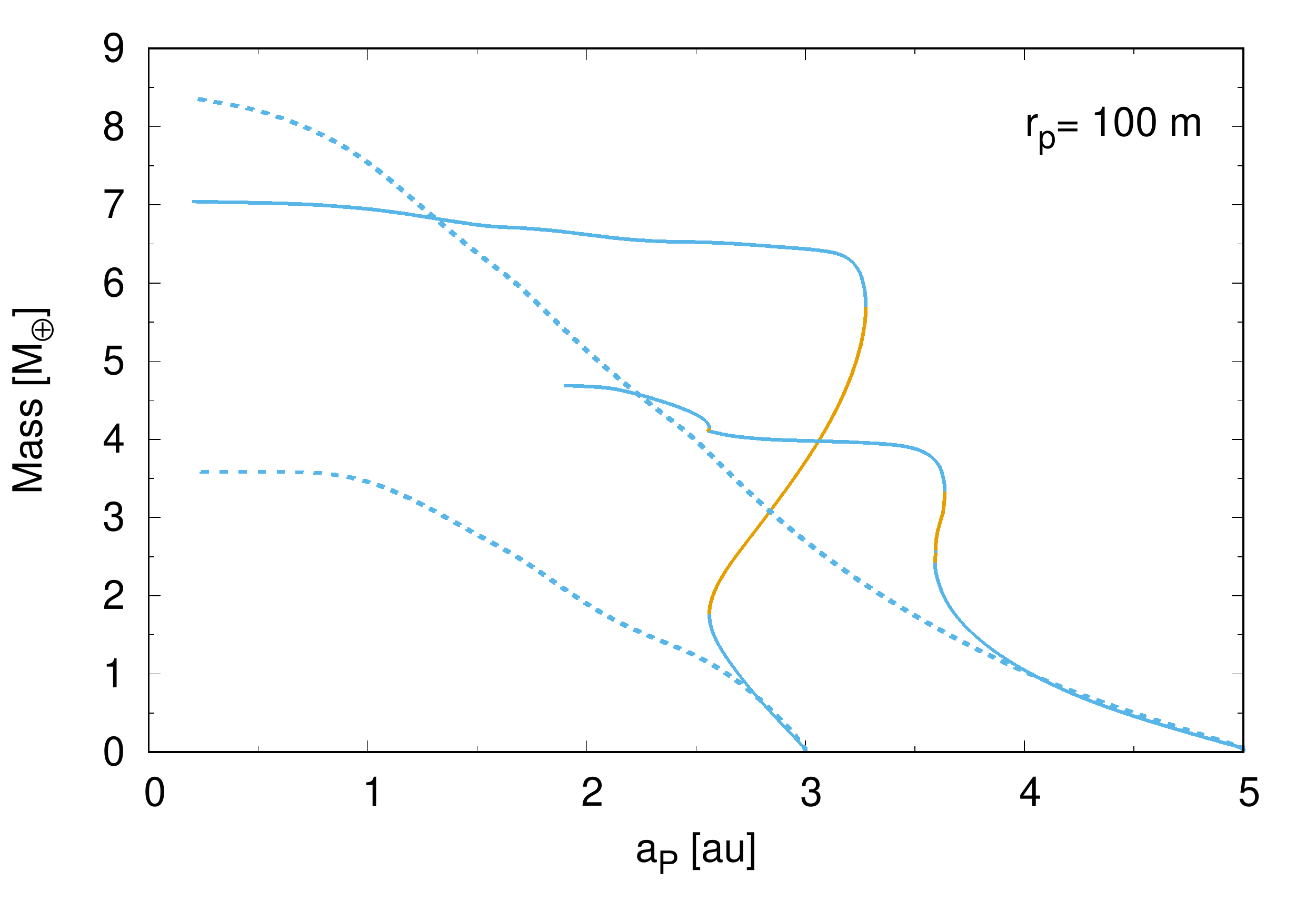} 
    \caption{Planet formation tracks for the fiducial model and initial planetesimals of 100~m. The solid lines correspond to the \citet{paardekooper.etal2011} type~I migration rates, and the dashed lines to the \citet{jm2017} rates. The light-blue (orange) curves indicate the part of the track for which the Lindblad (corotation) torque becomes dominant.}
  \label{fig:fig7}
\end{figure}

In Fig.~\ref{fig:fig7}, we plot the planet formation tracks using both migration prescriptions for initial planetesimals of 100~m of radius and the planet initial locations at 3~au and 5~au. These two cases are the ones showing the largest differences in the planet formation tracks presented in the panel b) of Fig.~\ref{fig:fig6}. We emphasise the role of the Lindblad and corotation torques for both migration planetary tracks. The light-blue and orange curves correspond to parts of the evolution in which either the Lindblad or the corotation torques dominate, respectively. It is clear that both prescriptions predict very different torques over the evolution of the planet. Adopting the migration recipes from \citet{paardekooper.etal2011}, the corotation torque becomes dominant at planet masses between $\sim 2~\text{M}_{\oplus}$ -- $\sim 5.5~\text{M}_{\oplus}$, and between $\sim 2.5~\text{M}_{\oplus}$ -- $\sim 3.5~\text{M}_{\oplus}$ for planets initially located at 3~au and 5~au, respectively. This is a natural consequence of the blue islands in the left column of Fig.~\ref{fig:fig2}, located at $R\sim 3$~au. On the other hand, adopting the migration recipes from \citet{jm2017}, the corotation torque never becomes dominant, and planets always migrate inward. That is the reason of the radically different formation tracks.

\subsubsection{Planet formation including \citet{masset2017} thermal torque} 
\label{sec3-2-2}

\begin{figure*}
    \centering
    \includegraphics[width= 0.325\textwidth, angle= 270]{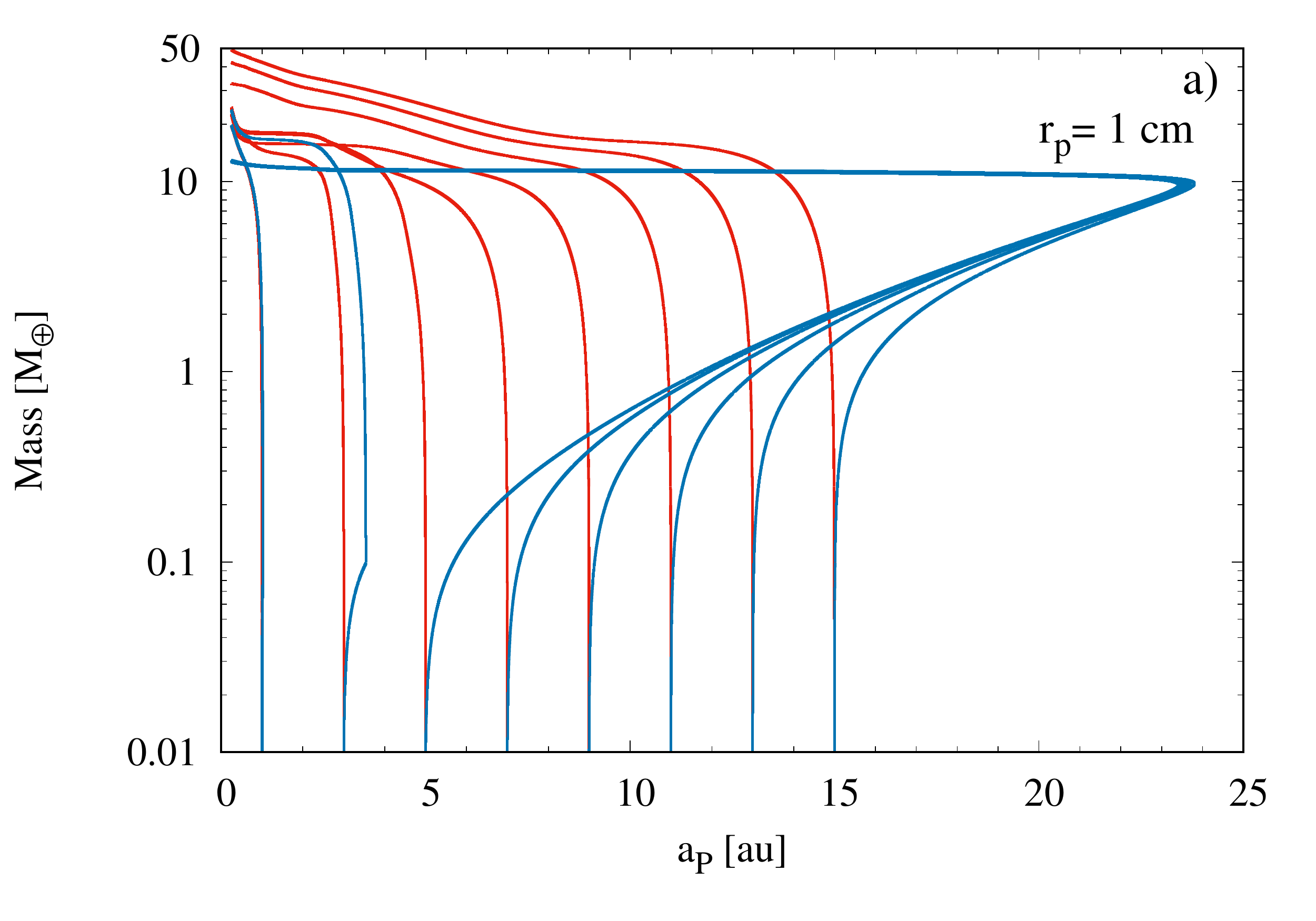} 
    \centering
    \includegraphics[width= 0.325\textwidth, angle= 270]{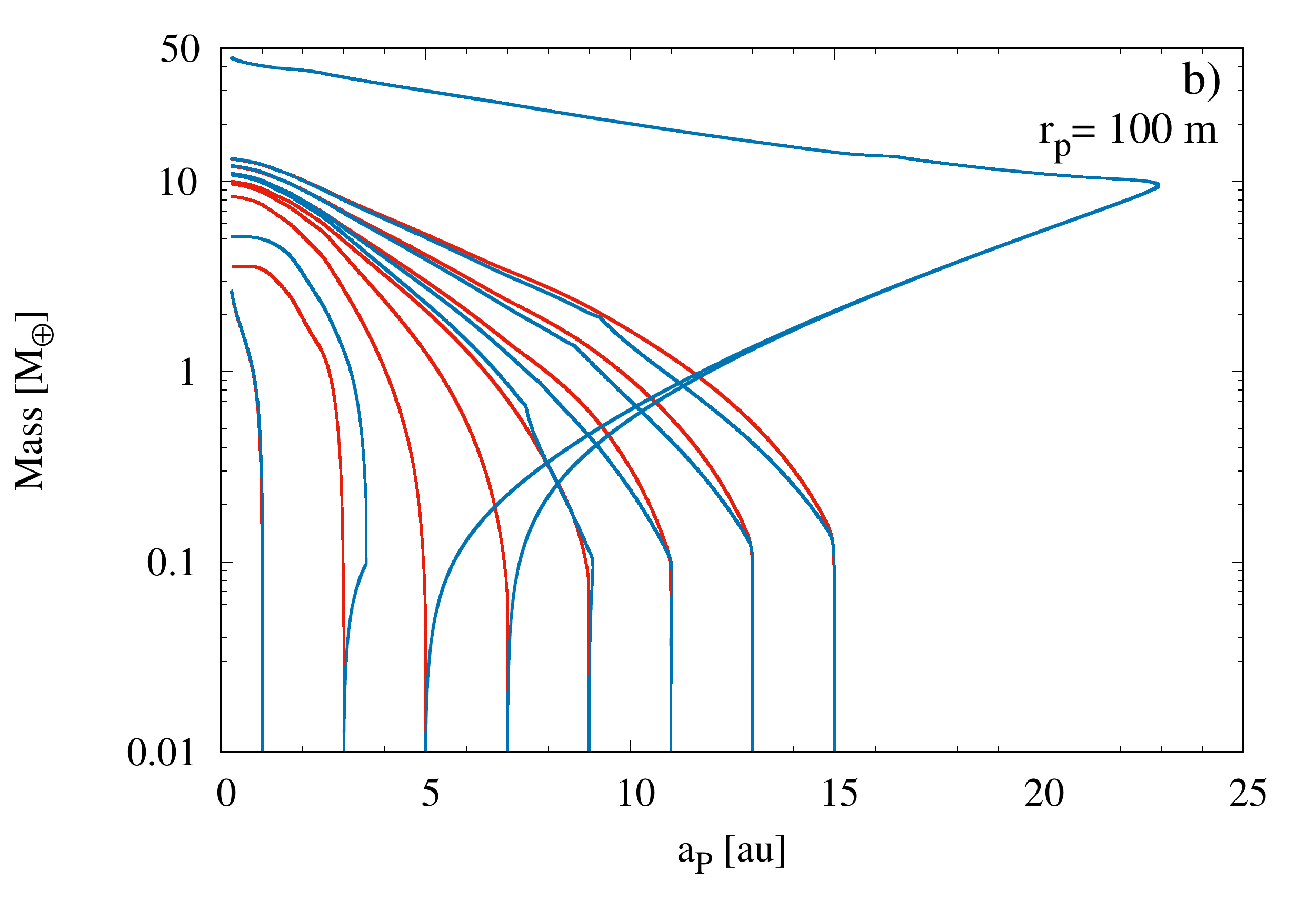} \\
    \centering
    \includegraphics[width= 0.325\textwidth, angle= 270]{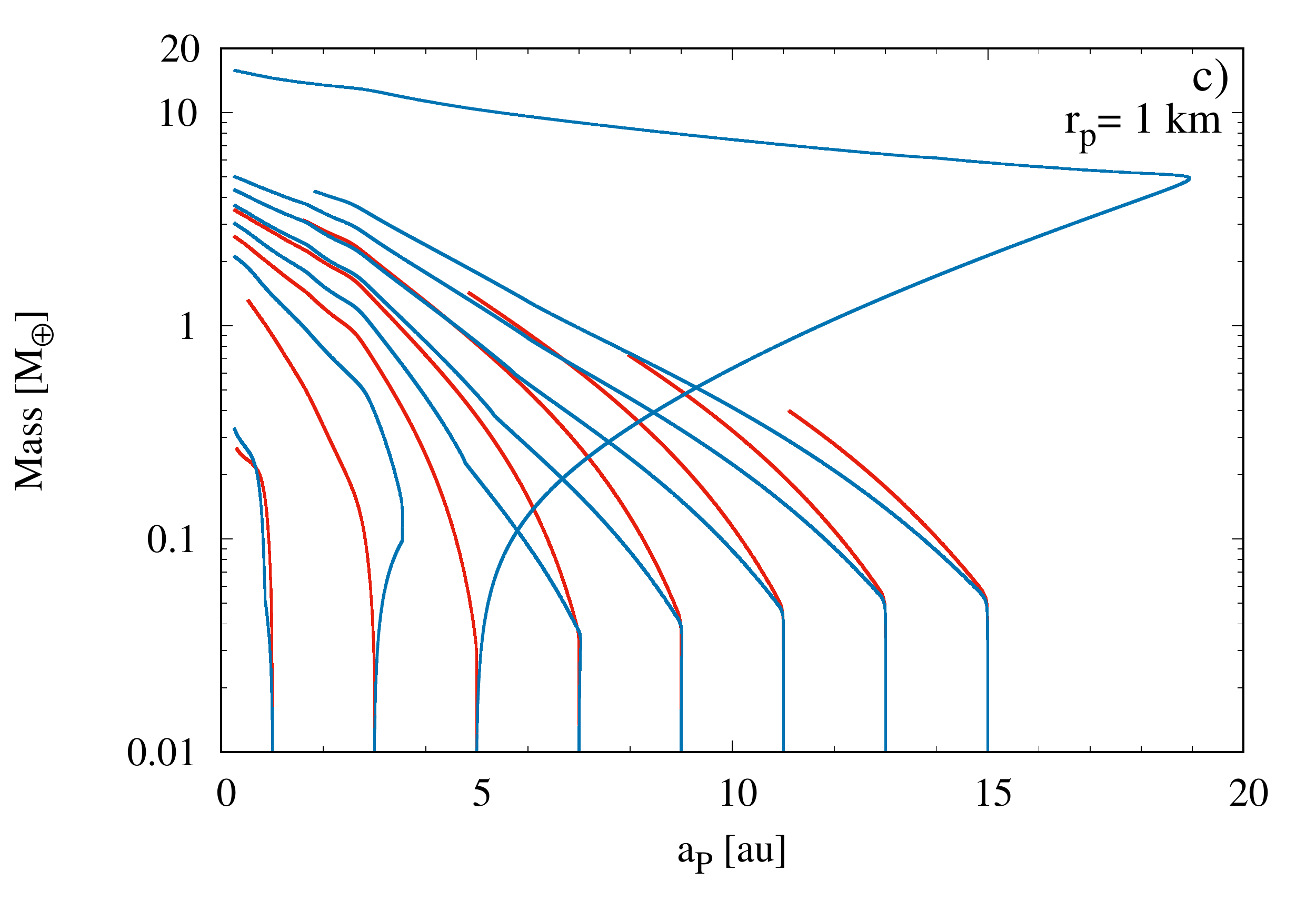} 
    \centering
    \includegraphics[width= 0.325\textwidth, angle= 270]{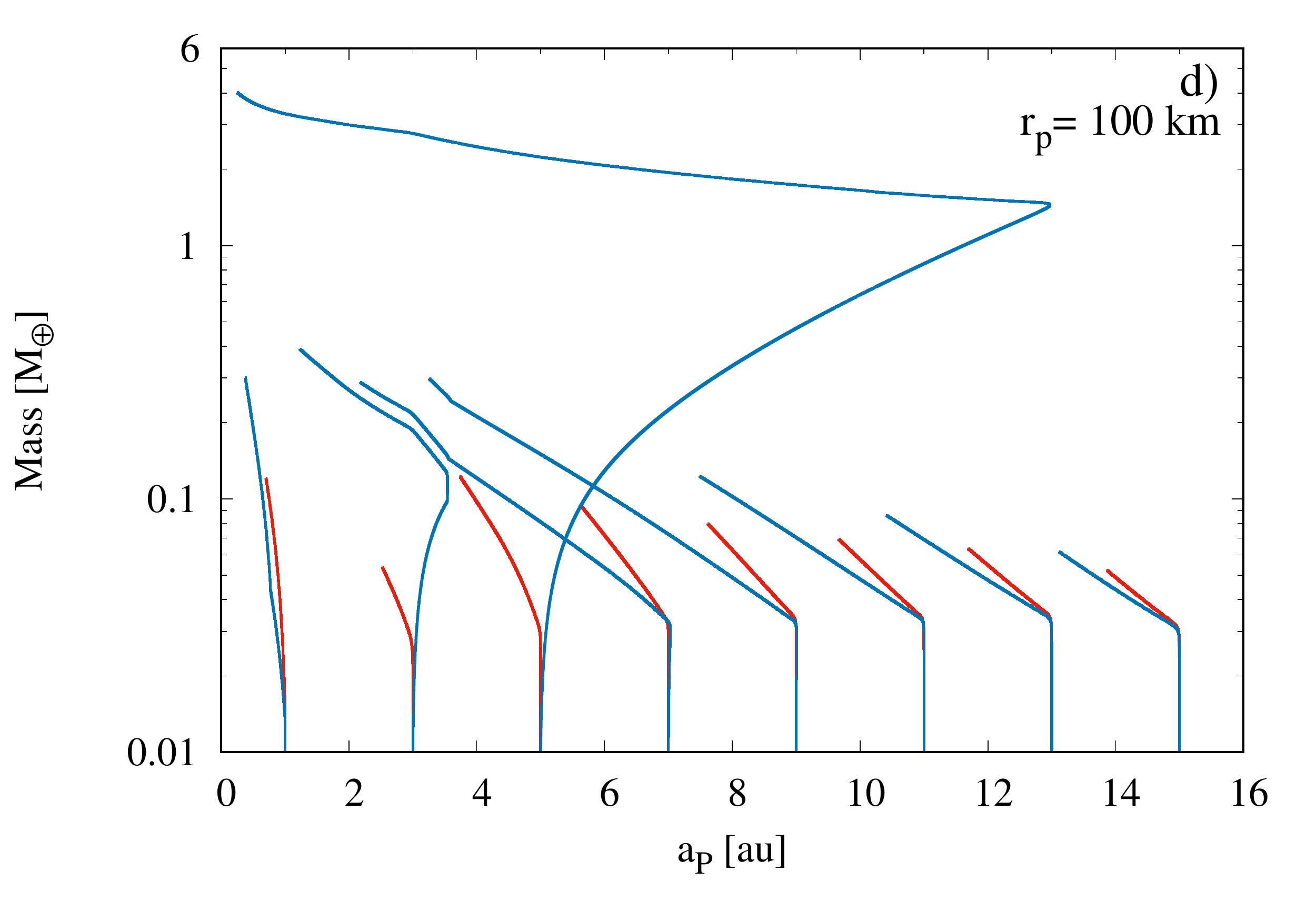} 
  \caption{Formation tracks of a planet initially located at different positions in the disc, from 1 au to 15 au, for the fiducial case. Each track corresponds to one simulation, i.e., we simulate the formation of only one planet per disc. As in Fig.~\ref{fig:fig6}, the red curves correspond to the simulations using the type I migration rates from \citet{jm2017}, while the blue curves correspond to the simulations wherein the thermal torques are included, this means that for these simulations we consider the type~I migration rates combined from \citet{jm2017} and \citet{masset2017}. The panels a, b, c, and d correspond to the case of pebbles of 1~cm of radii, and for the cases of planetesimals of 100~m of radii, 1~km of radii, and 100~km of radii, respectively. }
  \label{fig:fig10}
\end{figure*}

In this section, we analyse the role of the thermal torque over the formation of a planet. In Fig.~\ref{fig:fig10}, we plot the formation tracks of the planets considering different initial locations and different initial planetesimal sizes. The blue lines represent the formation tracks using the migration prescriptions from \citet{jm2017} and \citet{masset2017}, while the red lines only the ones from \citet{jm2017} (i.e. without the thermal torque). Again, we perform one simulation per formation track, where the planets are initially rocky cores of $0.01~\text{M}_{\oplus}$. The disc is the fiducial one and the simulations end when the planets reach the inner radius of the disc or at 3~Myr of evolution. 

\begin{figure}
    \centering
    \includegraphics[width= 0.345\textwidth,angle= -90]{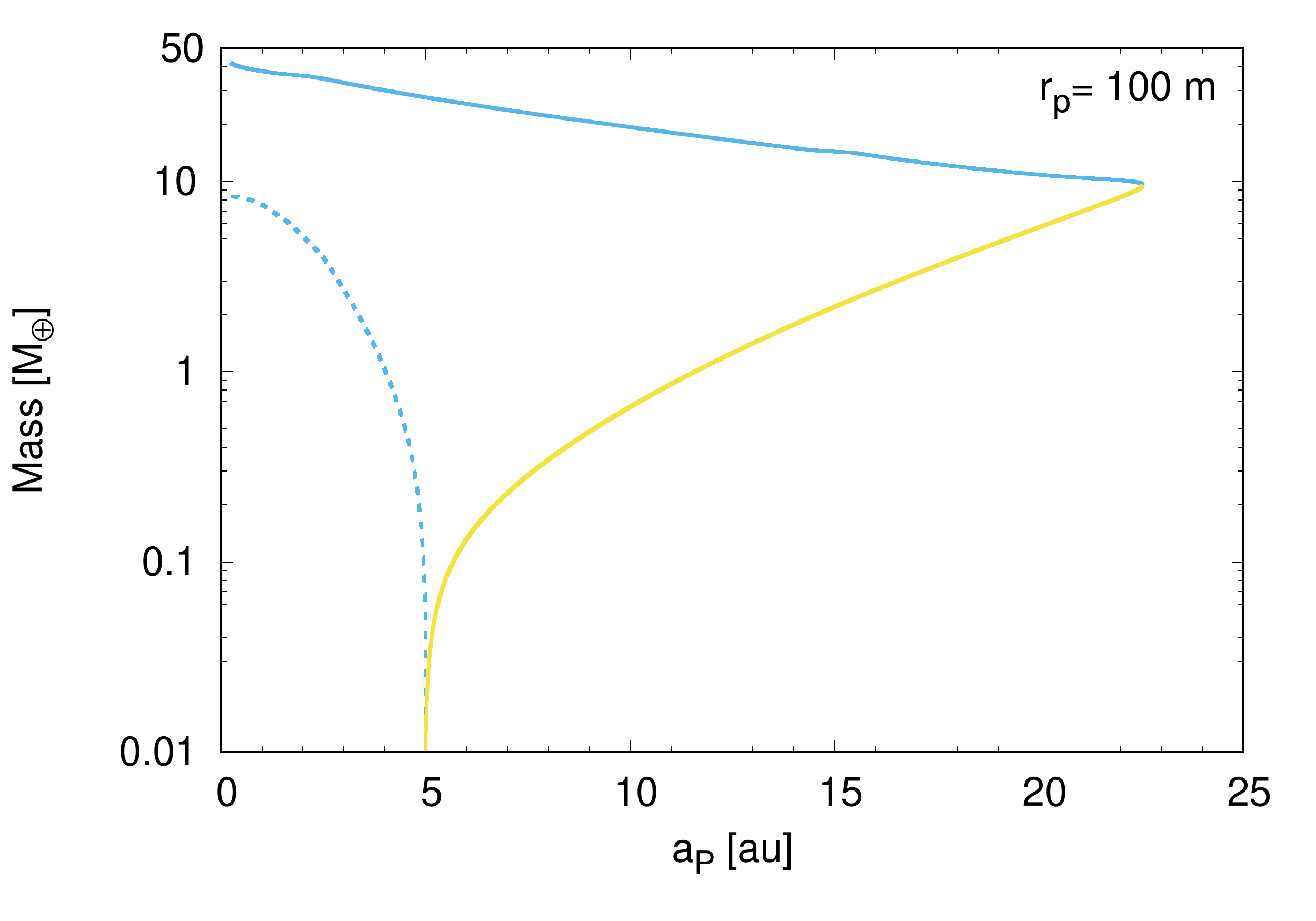} 
    \caption{Planet formation tracks for the fiducial model and initial planetesimals of 100~m. The yellow and light-blue curves represent whether the thermal or Lindblad torques, respectively, become dominant adopting the type I migration rates from \citet{jm2017} and \citet{masset2017}. The solid curve represents the case wherein the thermal torque is considered, while the dashed curve represents the case when not.}
 \label{fig:fig11}
\end{figure}

For the case where the solid disc is formed by pebbles of 1~cm (panel a of Fig.~\ref{fig:fig10}), and when the thermal torque is considered, the heating torque produces a significant outward migration. For the case of the planets initially located at $a_{\text{P}}\lesssim 3$~au, this migration is not significant, due to the fact that the planet reaches the critical thermal mass at a very low-mass and the thermal torque is then suppressed. Afterwards, the formation track is similar to the pure type~I case (as the case of the planet initially located at 1~au). However, for planets initially located beyond 3~au the heating torque becomes dominant (see Fig.~\ref{fig:fig11} that illustrates a similar situation) and the planets migrate outward until $\sim 24$~au. At this distance from the central star, the planets achieve the critical thermal mass (of about $10~\text{M}_{\oplus}$), and quickly migrate inward until reaching the disc inner radius. In these cases, the final planet masses are very different with respect to the case where the thermal torque is not considered. We note that not considering the pebble isolation mass does not affect the planet formation tracks before the planets reach the critical thermal mass since pebble isolation masses are normally higher than the critical thermal masses. However, this assumption could lead to an overestimation of pebble accretion rates at large masses.

When planetesimals of 100~m are considered (panel b of Fig.~\ref{fig:fig10}), only for planets initially located at 5~au and 7~au the heating torque becomes dominant\footnote{This happens because pebble accretion rates are higher than the planetesimal accretion rates, even for such small planetesimals.} and planets migrate outward until they reach the critical thermal mass of about $10~\text{M}_{\oplus}$ at $\sim 23$~au. Then, the planets migrate inward and they reach the disc inner radius with final masses significantly greater than the ones corresponding to the case where the thermal torque is not included. For the planets initially located at 3~au and 9~au, the heating torque only acts for very low-masses, so planets do not migrate outward efficiently. In the former case, the planet quickly achieves the critical thermal mass and the thermal torque is suppressed. In the latter, the planetesimal accretion rate is not high enough to produce a heating torque greater than the combination of the cold and the Lindbland torques. This results into inward migration. This is because in the fiducial case, as we mentioned before, the snowline is located near 3~au. Thus, as we move far away from the central star the planetesimal surface density decreases and then the planetesimal accretion rate also decreases. For planets located beyond 9~au we obtain a faster inward migration compared to the case where the thermal torque is not considered. This is due to the fact that, as we mentioned before, planetesimal accretion rates are lower as we move far way from the central star, and thus the heating torque is negligible. 
So, the cold torque dominates the thermal torque. When added to the Lindbland one, the net torque is then more negative. Similar situations occur for the cases where planetesimals of 1~km and 100~km are considered (panels c and d of Fig.~\ref{fig:fig10}, respectively). However, in these cases the planets that efficiently migrate outward only migrated until $\sim 19$~au for 1~km planetesimals and $\sim 13$~au for 100~km planetesimals. In these cases, since planetesimal accretion rates decrease as planetesimal sizes increase, the cold torque plus the Lindblad torque reverse the migration direction of the planet at lower distances with respect to the case of smaller planetesimals or pebbles.

Finally, in Fig.~\ref{fig:fig11} we plot the formation tracks for both migration models, for 100~m planetesimals and for a planet initially at 5~au. In this case, we compare the Lindblad (light-blue) and the thermal torques (yellow) by considering their absolute magnitudes. In this case, the corotation torque never becomes dominant. We observe that initially the thermal torque, and specifically the heating torque, dominates the migration track of the planet. This is due to the fact that planetesimal accretion rates are high for small planetesimals of 100~m at 5~au, while the cold and Lindblad torques are proportional to the mass of the planet, which is low initially. Thus, the planet migrates outward until it reaches $\sim 10~\text{M}_{\oplus}$ at $\sim 23$~au. When this happens, the Lindblad torque becomes dominant and the migration direction of the planet is reversed until it reaches the disc inner radius with a final mass of $\sim 50~\text{M}_{\oplus}$. For the case where the thermal torque is not considered, as we mentioned before, the Lindblad torques always dominate and the planet always migrates inward until reaching the inner disc radius with a final mass of $\sim 10~\text{M}_{\oplus}$. 

\section{Summary and discussion}
\label{sec4}

In this work, we incorporated the most updated type I migration rate prescriptions to analyse their role in the process of planet formation. These prescriptions have two components: one related to the classical type~I torque, in which the role of thermal diffusion within the disc is limited to its non-linear effect on the saturation of the corotation torque, and another related to the thermal torque. The latter takes into account the perturbation of the disc's flow arising from thermal diffusion and the heat released by the planet due to the accretion of solid material, in the linear limit. The first ones are derived from 3D hydrodynamical simulations \citep{jm2017}, improving the calculation of the corotation torque respect to previous 2D calculations \citep{MC2010,paardekooper.etal2011}. The thermal torques are derived from the analytical study by \cite{masset2017}, motivated by previous hydrodynamical simulations \citep{Lega.et.al.2014, Benitez-llambay.et.al.2015}. 

We first compared the migration maps constructed using type I migration prescriptions from 2D and 3D hydrodynamical simulations for a fiducial case. For the 2D hydrodynamical simulations we adopted the commonly used type I prescriptions of \citet{paardekooper.etal2011}, while for the 3D hydrodynamical simulations we adopted the new prescriptions derived by \citet{jm2017}. We showed that the corresponding maps present significant differences for the regions of outward and inward migration, which are related to the corotation torques. We also analysed the dependence of the migration maps with the $\alpha$-viscosity parameter and the disc mass finding large differences between both type I migration prescriptions in a wide range of such parameters. This is particularly striking in the disc inner regions, at $R \lesssim 20$~au, where planet formation is expected to take place.

Then, we added the calculation of the thermal torque derived by \citet{masset2017} to the updated prescriptions by \citet{jm2017}, for the first time to the best of our knowledge. We find that the inclusion of the thermal torque introduces a new region of outward migration, whose size directly depends on the planet solid accretion rate because the heating torque is proportional to this quantity. We also showed that at some point the sum of the Lindbland torque and the cold torque counteracts the total positive torque generated by the heating torque. Then, due to the cold torque, the total torque becomes more negative compared to the pure type I migration case. It is important to stress that, as in \citet{masset2017}, the critical thermal mass given by Eq.~\ref{eq23-sec2-4-1} is only an order of magnitude estimate. More detailed studies on the mass limit for which the planet stops to efficiently release heat on the disc are needed. In addition, the heating torque is calculated assuming that the heat released in the disc is only due to the solid accretion. This is certainly true for a low-mass or intermediate-mass planet. However, if the planet has a significant envelope, the heat released due to its contraction could also play an important role. We note that in this version of {\scriptsize PLANETALP}, the equations of transport and structure of the envelope are not calculated, and we implemented a parametric gas accretion model as in \citet{Ronco.et.al.2017}. Thus, we are not able to calculate in an accurate way such additional production of heat. This will be studied in detail in a future work.  

After the comparison between the migration maps, we studied the planet formation affected by the different type I migration rates. We first analysed the planet formation employing \citet{paardekooper.etal2011} and \citet{jm2017} prescriptions. To do so, we calculated the planet formation tracks adopting the fiducial disc for planets located at different initial locations, and for different pebble and planetesimal sizes. We showed that the planet formation tracks are systematically different, especially in the models where the planets grow up to several Earth masses or more. This is precisely when the corotation torque becomes significant. 

Then, we calculated the formation tracks considering the inclusion of the thermal torque. For the case of pebbles, we found that all the planets initially located beyond 3~au suffered a significant outward migration due to the thermal torque, and specifically to the heating torque, because of the high accretion rates onto them. In contrast to \citet{Bitsch.et.al.2015}, we found that the planet formation tracks are remarkably different compared to the case where the thermal torque is not considered. As a result, the planets end with significantly different final masses. We associate this discrepancy with the fact that, in this early work published before the release of an analytic expression for the thermal torque, they used an empirical fit of the heating torque measured by \citet{Benitez-llambay.et.al.2015}, which was obtained for planets with masses comparable to or larger than the critical mass. These could had been therefore subjected to thermal torques smaller than the ones predicted by \citet{masset2017}. Other discrepancies can arise from the solid accretion rates involved in both models: the higher the accrection rate, the stronger the heating torque. More recently, \citet{Chrenko2017} computed the simultaneous formation of four super-Earths (initially located between 5~au and 10~au) by the accretion of pebbles through 2D radiative hydrodynamical simulations. These calculations also included the heat released by the planets due to the pebble accretion onto the nearby gas disc. They analysed the orbital evolution and growth in three different scenarios: without pebble accretion, with pebble accretion but without heating, and with accretion plus the heating torque. For the latter case, their results are different with respect to the first two cases regarding for the outward migration due to heating torque.

For planetesimals of 100~m of radius, the significant planet outward migration only occurs for planets initially located at 5 and 7~au. This is because the snowline of the fiducial disc is slightly beyond 3~au, so at larger distances from the central star the planetesimal accretion rates become smaller. For planets with initial locations beyond 7~au, we note that these migrated faster compared to the case where the thermal torque was not considered. This is because the cold (negative) torque is added to the Lindblad torque resulting in a more negative total torque. Similar situations occur when planetesimals of 1~km and 100~km are considered, with the particularity that in these cases the planets did not reach high masses as in the previous cases. 

There are a number of shortcomings in our analysis. While we take into account the effects of the planet's luminosity on the time evolution of its semi-major axis, we neglect its impact on the eccentricity and inclination, which is potentially quite strong \citep{Eklund2017,Chrenko2017}. Sizeable eccentricities and inclinations may significantly alter the migration paths of planetary embryos, and may also affect the accretion rate on the embryos \citep{Chrenko2017} and the rate of collisions between embryos. As we consider only one planetary embryo at a time, such effect cannot be captured by this work. Our torque expressions assume that the torque is a function only of the underlying disc properties, but not of the planet's migration rate itself. The existence of a migration feedback on the corotation torque, initially considered for sub-giant planets \citep{2003ApJ...588..494M}, has been extended to lower mass planets \citep{2014MNRAS.444.2031P,2015MNRAS.454.2003P,2018MNRAS.477.4596M}. To the best of our knowledge, prescriptions for the dynamical corotation torque have not been used yet in planetary migration paths studies such as the one presented here. In spite of these shortcomings, our work shows that the new type I migration prescriptions by \citet{jm2017} and \citet{masset2017} are very different from the ones by \citet{MC2010} and \citet{paardekooper.etal2011} and that the planet formation tracks are highly sensitive to these differences.

This calls for a reanalysis of the predictions and conclusions of previous work based on previous type I migration rates. We also emphasise that the thermal torque, and especially the heating torque, can lead to very different planet formation tracks when the solid accretion rates are high. In this case, the planet can experience a significant outward migration. As a first step, we studied the formation of only one planet in a fiducial disc. The consequences of considering a population of embryos and the impact of these new type I migration prescriptions on a population synthesis will be explored in a future work. 

\section*{Acknowledgements}

We thanks the anonymous referee who help us to improve the work with his/her comments. OMG and M3B are supported by the PICT 2016-0053 from ANPCyT, Argentina. OMG also acknowledges the hosting as invited researcher from IA-PUC. NC acknowledges financial support provided by FONDECYT grant 3170680. MPR and JC acknowledge financial support from the Iniciativa Cient\'{\i}fica Milenio (ICM) via the N\'ucleo Milenio de Formaci\'on Planetaria Grant. NC, MPR and JC acknowledge support from CONICYT project Basal AFB-170002. MM acknowledge financial support from the Chinese Academy of Sciences (CAS) through a CAS-CONICYT Postdoctoral Fellowship administered by the CAS South America Center for Astronomy (CASSACA) in Santiago, Chile.


\bibliographystyle{mnras}
\bibliography{biblio}

\begin{thebibliography}{}
\makeatletter
\relax
\def\mn@urlcharsother{\let\do\@makeother \do\$\do\&\do\#\do\^\do\_\do\%\do\~}
\def\mn@doi{\begingroup\mn@urlcharsother \@ifnextchar [ {\mn@doi@}
  {\mn@doi@[]}}
\def\mn@doi@[#1]#2{\def\@tempa{#1}\ifx\@tempa\@empty \href
  {http://dx.doi.org/#2} {doi:#2}\else \href {http://dx.doi.org/#2} {#1}\fi
  \endgroup}
\def\mn@eprint#1#2{\mn@eprint@#1:#2::\@nil}
\def\mn@eprint@arXiv#1{\href {http://arxiv.org/abs/#1} {{\tt arXiv:#1}}}
\def\mn@eprint@dblp#1{\href {http://dblp.uni-trier.de/rec/bibtex/#1.xml}
  {dblp:#1}}
\def\mn@eprint@#1:#2:#3:#4\@nil{\def\@tempa {#1}\def\@tempb {#2}\def\@tempc
  {#3}\ifx \@tempc \@empty \let \@tempc \@tempb \let \@tempb \@tempa \fi \ifx
  \@tempb \@empty \def\@tempb {arXiv}\fi \@ifundefined
  {mn@eprint@\@tempb}{\@tempb:\@tempc}{\expandafter \expandafter \csname
  mn@eprint@\@tempb\endcsname \expandafter{\@tempc}}}

\bibitem[\protect\citeauthoryear{{Alibert}, {Mordasini}, {Benz}  \&
  {Winisdoerffer}}{{Alibert} et~al.}{2005}]{Alibert2005}
{Alibert} Y.,  {Mordasini} C.,  {Benz} W.,   {Winisdoerffer} C.,  2005, \mn@doi
  [\aap] {10.1051/0004-6361:20042032}, \href
  {http://adsabs.harvard.edu/abs/2005A%26A...434..343A} {434, 343}

\bibitem[\protect\citeauthoryear{{Alibert}, {Carron}, {Fortier}, {Pfyffer},
  {Benz}, {Mordasini}  \& {Swoboda}}{{Alibert}
  et~al.}{2013}]{Alibert.et.al.2013}
{Alibert} Y.,  {Carron} F.,  {Fortier} A.,  {Pfyffer} S.,  {Benz} W.,
  {Mordasini} C.,   {Swoboda} D.,  2013, \mn@doi [\aap]
  {10.1051/0004-6361/201321690}, \href
  {http://adsabs.harvard.edu/abs/2013A%26A...558A.109A} {558, A109}

\bibitem[\protect\citeauthoryear{{Andrews}, {Wilner}, {Hughes}, {Qi}  \&
  {Dullemond}}{{Andrews} et~al.}{2010}]{Andrews2010}
{Andrews} S.~M.,  {Wilner} D.~J.,  {Hughes} A.~M.,  {Qi} C.,   {Dullemond}
  C.~P.,  2010, \mn@doi [\apj] {10.1088/0004-637X/723/2/1241}, \href
  {http://adsabs.harvard.edu/abs/2010ApJ...723.1241A} {723, 1241}

\bibitem[\protect\citeauthoryear{{Bailli{\'e}}, {Charnoz}  \&
  {Pantin}}{{Bailli{\'e}} et~al.}{2015}]{Baillie2015}
{Bailli{\'e}} K.,  {Charnoz} S.,   {Pantin} E.,  2015, \mn@doi [\aap]
  {10.1051/0004-6361/201424987}, \href
  {http://adsabs.harvard.edu/abs/2015A%26A...577A..65B} {577, A65}

\bibitem[\protect\citeauthoryear{{Bailli{\'e}}, {Charnoz}  \&
  {Pantin}}{{Bailli{\'e}} et~al.}{2016}]{Baillie2016}
{Bailli{\'e}} K.,  {Charnoz} S.,   {Pantin} E.,  2016, \mn@doi [\aap]
  {10.1051/0004-6361/201528027}, \href
  {http://adsabs.harvard.edu/abs/2016A%26A...590A..60B} {590, A60}

\bibitem[\protect\citeauthoryear{{Baraffe}, {Homeier}, {Allard}  \&
  {Chabrier}}{{Baraffe} et~al.}{2015}]{Baraffe2015}
{Baraffe} I.,  {Homeier} D.,  {Allard} F.,   {Chabrier} G.,  2015, \mn@doi
  [\aap] {10.1051/0004-6361/201425481}, \href
  {http://adsabs.harvard.edu/abs/2015A%26A...577A..42B} {577, A42}

\bibitem[\protect\citeauthoryear{{Baruteau} \& {Masset}}{{Baruteau} \&
  {Masset}}{2008}]{BM2008}
{Baruteau} C.,  {Masset} F.,  2008, \mn@doi [\apj] {10.1086/523667}, \href
  {http://adsabs.harvard.edu/abs/2008ApJ...672.1054B} {672, 1054}

\bibitem[\protect\citeauthoryear{{Bell} \& {Lin}}{{Bell} \&
  {Lin}}{1994}]{Bell.Lin.1994}
{Bell} K.~R.,  {Lin} D.~N.~C.,  1994, \mn@doi [\apj] {10.1086/174206}, \href
  {http://adsabs.harvard.edu/abs/1994ApJ...427..987B} {427, 987}

\bibitem[\protect\citeauthoryear{{Ben{\'{\i}}tez-Llambay}, {Masset},
  {Koenigsberger}  \& {Szul{\'a}gyi}}{{Ben{\'{\i}}tez-Llambay}
  et~al.}{2015}]{Benitez-llambay.et.al.2015}
{Ben{\'{\i}}tez-Llambay} P.,  {Masset} F.,  {Koenigsberger} G.,
  {Szul{\'a}gyi} J.,  2015, \mn@doi [\nat] {10.1038/nature14277}, \href
  {http://adsabs.harvard.edu/abs/2015Natur.520...63B} {520, 63}

\bibitem[\protect\citeauthoryear{{Bitsch} \& {Kley}}{{Bitsch} \&
  {Kley}}{2011}]{BK2011}
{Bitsch} B.,  {Kley} W.,  2011, \mn@doi [\aap] {10.1051/0004-6361/201117202},
  \href {http://adsabs.harvard.edu/abs/2011A%26A...536A..77B} {536, A77}

\bibitem[\protect\citeauthoryear{{Bitsch}, {Johansen}, {Lambrechts}  \&
  {Morbidelli}}{{Bitsch} et~al.}{2015a}]{Bitsch.et.al.2015b}
{Bitsch} B.,  {Johansen} A.,  {Lambrechts} M.,   {Morbidelli} A.,  2015a,
  \mn@doi [\aap] {10.1051/0004-6361/201424964}, \href
  {http://adsabs.harvard.edu/abs/2015A%26A...575A..28B} {575, A28}

\bibitem[\protect\citeauthoryear{{Bitsch}, {Lambrechts}  \&
  {Johansen}}{{Bitsch} et~al.}{2015b}]{Bitsch.et.al.2015}
{Bitsch} B.,  {Lambrechts} M.,   {Johansen} A.,  2015b, \mn@doi [\aap]
  {10.1051/0004-6361/201526463}, \href
  {http://adsabs.harvard.edu/abs/2015A%26A...582A.112B} {582, A112}

\bibitem[\protect\citeauthoryear{{Bitsch}, {Morbidelli}, {Johansen}, {Lega},
  {Lambrechts}  \& {Crida}}{{Bitsch} et~al.}{2018}]{Bitsch-et-al-2018}
{Bitsch} B.,  {Morbidelli} A.,  {Johansen} A.,  {Lega} E.,  {Lambrechts} M.,
  {Crida} A.,  2018, \mn@doi [\aap] {10.1051/0004-6361/201731931}, \href
  {http://adsabs.harvard.edu/abs/2018A%26A...612A..30B} {612, A30}

\bibitem[\protect\citeauthoryear{{Blum}}{{Blum}}{2018}]{Blum2018}
{Blum} J.,  2018, \mn@doi [\ssr] {10.1007/s11214-018-0486-5}, \href
  {http://adsabs.harvard.edu/abs/2018SSRv..214...52B} {214, 52}

\bibitem[\protect\citeauthoryear{{Bodenheimer} \& {Pollack}}{{Bodenheimer} \&
  {Pollack}}{1986}]{bodenheimer_pollack1986}
{Bodenheimer} P.,  {Pollack} J.~B.,  1986, \mn@doi [\icarus]
  {10.1016/0019-1035(86)90122-3}, \href
  {http://adsabs.harvard.edu/abs/1986Icar...67..391B} {67, 391}

\bibitem[\protect\citeauthoryear{{Brouwers}, {Vazan}  \& {Ormel}}{{Brouwers}
  et~al.}{2018}]{Brouwers-et-al2018}
{Brouwers} M.~G.,  {Vazan} A.,   {Ormel} C.~W.,  2018, \mn@doi [\aap]
  {10.1051/0004-6361/201731824}, \href
  {http://adsabs.harvard.edu/abs/2018A%26A...611A..65B} {611, A65}

\bibitem[\protect\citeauthoryear{{Chambers}}{{Chambers}}{2008}]{Chambers2008}
{Chambers} J.,  2008, \mn@doi [\icarus] {10.1016/j.icarus.2008.06.011}, \href
  {http://adsabs.harvard.edu/abs/2008Icar..198..256C} {198, 256}

\bibitem[\protect\citeauthoryear{{Chambers}}{{Chambers}}{2016}]{Chambers2016}
{Chambers} J.~E.,  2016, \mn@doi [\apj] {10.3847/0004-637X/825/1/63}, \href
  {http://adsabs.harvard.edu/abs/2016ApJ...825...63C} {825, 63}

\bibitem[\protect\citeauthoryear{{Chiang} \& {Goldreich}}{{Chiang} \&
  {Goldreich}}{1997}]{CG1997}
{Chiang} E.~I.,  {Goldreich} P.,  1997, \mn@doi [\apj] {10.1086/304869}, \href
  {http://adsabs.harvard.edu/abs/1997ApJ...490..368C} {490, 368}

\bibitem[\protect\citeauthoryear{{Chrenko}, {Bro{\v z}}  \&
  {Lambrechts}}{{Chrenko} et~al.}{2017}]{Chrenko2017}
{Chrenko} O.,  {Bro{\v z}} M.,   {Lambrechts} M.,  2017, \mn@doi [\aap]
  {10.1051/0004-6361/201731033}, \href
  {http://adsabs.harvard.edu/abs/2017A%26A...606A.114C} {606, A114}

\bibitem[\protect\citeauthoryear{{Coleman} \& {Nelson}}{{Coleman} \&
  {Nelson}}{2014}]{CL2014}
{Coleman} G.~A.~L.,  {Nelson} R.~P.,  2014, \mn@doi [\mnras]
  {10.1093/mnras/stu1715}, \href
  {http://adsabs.harvard.edu/abs/2014MNRAS.445..479C} {445, 479}

\bibitem[\protect\citeauthoryear{{Coleman} \& {Nelson}}{{Coleman} \&
  {Nelson}}{2016}]{CL2016}
{Coleman} G.~A.~L.,  {Nelson} R.~P.,  2016, \mn@doi [\mnras]
  {10.1093/mnras/stw1177}, \href
  {http://adsabs.harvard.edu/abs/2016MNRAS.460.2779C} {460, 2779}

\bibitem[\protect\citeauthoryear{{Crida}, {Morbidelli}  \& {Masset}}{{Crida}
  et~al.}{2006}]{Crida2006}
{Crida} A.,  {Morbidelli} A.,   {Masset} F.,  2006, \mn@doi [\icarus]
  {10.1016/j.icarus.2005.10.007}, \href
  {http://adsabs.harvard.edu/abs/2006Icar..181..587C} {181, 587}

\bibitem[\protect\citeauthoryear{{Dittkrist}, {Mordasini}, {Klahr}, {Alibert}
  \& {Henning}}{{Dittkrist} et~al.}{2014}]{Dittkrist.et.al.2015}
{Dittkrist} K.-M.,  {Mordasini} C.,  {Klahr} H.,  {Alibert} Y.,   {Henning} T.,
   2014, \mn@doi [\aap] {10.1051/0004-6361/201322506}, \href
  {http://adsabs.harvard.edu/abs/2014A%26A...567A.121D} {567, A121}

\bibitem[\protect\citeauthoryear{{Eklund} \& {Masset}}{{Eklund} \&
  {Masset}}{2017}]{Eklund2017}
{Eklund} H.,  {Masset} F.~S.,  2017, \mn@doi [\mnras] {10.1093/mnras/stx856},
  \href {http://adsabs.harvard.edu/abs/2017MNRAS.469..206E} {469, 206}

\bibitem[\protect\citeauthoryear{{Fortier}, {Benvenuto}  \&
  {Brunini}}{{Fortier} et~al.}{2007}]{Fortier.et.al.2007}
{Fortier} A.,  {Benvenuto} O.~G.,   {Brunini} A.,  2007, \mn@doi [\aap]
  {10.1051/0004-6361:20066729}, \href
  {http://adsabs.harvard.edu/abs/2007A%26A...473..311F} {473, 311}

\bibitem[\protect\citeauthoryear{{Fortier}, {Alibert}, {Carron}, {Benz}  \&
  {Dittkrist}}{{Fortier} et~al.}{2013}]{Fortier2013}
{Fortier} A.,  {Alibert} Y.,  {Carron} F.,  {Benz} W.,   {Dittkrist} K.-M.,
  2013, \mn@doi [\aap] {10.1051/0004-6361/201220241}, \href
  {http://adsabs.harvard.edu/abs/2013A%26A...549A..44F} {549, A44}

\bibitem[\protect\citeauthoryear{{Greenzweig} \& {Lissauer}}{{Greenzweig} \&
  {Lissauer}}{1992}]{GL1992}
{Greenzweig} Y.,  {Lissauer} J.~J.,  1992, \mn@doi [\icarus]
  {10.1016/0019-1035(92)90110-S}, \href
  {http://adsabs.harvard.edu/abs/1992Icar..100..440G} {100, 440}

\bibitem[\protect\citeauthoryear{{Guilera} \& {S{\'a}ndor}}{{Guilera} \&
  {S{\'a}ndor}}{2017}]{Guilera.Sandor.2017}
{Guilera} O.~M.,  {S{\'a}ndor} Z.,  2017, \mn@doi [\aap]
  {10.1051/0004-6361/201629843}, \href
  {http://adsabs.harvard.edu/abs/2017A%26A...604A..10G} {604, A10}

\bibitem[\protect\citeauthoryear{{Guilera}, {Brunini}  \&
  {Benvenuto}}{{Guilera} et~al.}{2010}]{Guilera2010}
{Guilera} O.~M.,  {Brunini} A.,   {Benvenuto} O.~G.,  2010, \mn@doi [\aap]
  {10.1051/0004-6361/201014365}, \href
  {http://adsabs.harvard.edu/abs/2010A%26A...521A..50G} {521, A50}

\bibitem[\protect\citeauthoryear{{Guilera}, {Fortier}, {Brunini}  \&
  {Benvenuto}}{{Guilera} et~al.}{2011}]{Guilera2011}
{Guilera} O.~M.,  {Fortier} A.,  {Brunini} A.,   {Benvenuto} O.~G.,  2011,
  \mn@doi [\aap] {10.1051/0004-6361/201015731}, \href
  {http://adsabs.harvard.edu/abs/2011A%26A...532A.142G} {532, A142}

\bibitem[\protect\citeauthoryear{{Guilera}, {de El{\'{\i}}a}, {Brunini}  \&
  {Santamar{\'{\i}}a}}{{Guilera} et~al.}{2014}]{Guilera2014}
{Guilera} O.~M.,  {de El{\'{\i}}a} G.~C.,  {Brunini} A.,   {Santamar{\'{\i}}a}
  P.~J.,  2014, \mn@doi [\aap] {10.1051/0004-6361/201322061}, \href
  {http://adsabs.harvard.edu/abs/2014A%26A...565A..96G} {565, A96}

\bibitem[\protect\citeauthoryear{{Guilera}, {Miller Bertolami}  \&
  {Ronco}}{{Guilera} et~al.}{2017}]{Guilera.et.al.2017}
{Guilera} O.~M.,  {Miller Bertolami} M.~M.,   {Ronco} M.~P.,  2017, \mn@doi
  [\mnras] {10.1093/mnrasl/slx095}, \href
  {http://adsabs.harvard.edu/abs/2017MNRAS.471L..16G} {471, L16}

\bibitem[\protect\citeauthoryear{{Guilet}, {Baruteau}  \&
  {Papaloizou}}{{Guilet} et~al.}{2013}]{Guilet.et.al.2013}
{Guilet} J.,  {Baruteau} C.,   {Papaloizou} J.~C.~B.,  2013, \mn@doi [\mnras]
  {10.1093/mnras/sts720}, \href
  {http://adsabs.harvard.edu/abs/2013MNRAS.430.1764G} {430, 1764}

\bibitem[\protect\citeauthoryear{{Ida} \& {Lin}}{{Ida} \&
  {Lin}}{2004}]{Ida-Lin2004}
{Ida} S.,  {Lin} D.~N.~C.,  2004, \mn@doi [\apj] {10.1086/381724}, \href
  {http://adsabs.harvard.edu/abs/2004ApJ...604..388I} {604, 388}

\bibitem[\protect\citeauthoryear{{Ida} \& {Makino}}{{Ida} \&
  {Makino}}{1993}]{IdaMakino1993}
{Ida} S.,  {Makino} J.,  1993, \mn@doi [\icarus] {10.1006/icar.1993.1167},
  \href {http://adsabs.harvard.edu/abs/1993Icar..106..210I} {106, 210}

\bibitem[\protect\citeauthoryear{{Inaba} \& {Ikoma}}{{Inaba} \&
  {Ikoma}}{2003}]{Inaba&Ikoma-2003}
{Inaba} S.,  {Ikoma} M.,  2003, \mn@doi [\aap] {10.1051/0004-6361:20031248},
  \href {http://adsabs.harvard.edu/abs/2003A%26A...410..711I} {410, 711}

\bibitem[\protect\citeauthoryear{{Inaba}, {Tanaka}, {Nakazawa}, {Wetherill}  \&
  {Kokubo}}{{Inaba} et~al.}{2001}]{Inaba2001}
{Inaba} S.,  {Tanaka} H.,  {Nakazawa} K.,  {Wetherill} G.~W.,   {Kokubo} E.,
  2001, \mn@doi [\icarus] {10.1006/icar.2000.6533}, \href
  {http://adsabs.harvard.edu/abs/2001Icar..149..235I} {149, 235}

\bibitem[\protect\citeauthoryear{{Izidoro}, {Ogihara}, {Raymond}, {Morbidelli},
  {Pierens}, {Bitsch}, {Cossou}  \& {Hersant}}{{Izidoro}
  et~al.}{2017}]{Izidoro2017}
{Izidoro} A.,  {Ogihara} M.,  {Raymond} S.~N.,  {Morbidelli} A.,  {Pierens} A.,
   {Bitsch} B.,  {Cossou} C.,   {Hersant} F.,  2017, \mn@doi [\mnras]
  {10.1093/mnras/stx1232}, \href
  {http://adsabs.harvard.edu/abs/2017MNRAS.470.1750I} {470, 1750}

\bibitem[\protect\citeauthoryear{{Jim{\'e}nez} \& {Masset}}{{Jim{\'e}nez} \&
  {Masset}}{2017}]{jm2017}
{Jim{\'e}nez} M.~A.,  {Masset} F.~S.,  2017, \mn@doi [\mnras]
  {10.1093/mnras/stx1946}, \href
  {http://adsabs.harvard.edu/abs/2017MNRAS.471.4917J} {471, 4917}

\bibitem[\protect\citeauthoryear{{Johansen} \& {Lacerda}}{{Johansen} \&
  {Lacerda}}{2010}]{J&L2010}
{Johansen} A.,  {Lacerda} P.,  2010, \mn@doi [\mnras]
  {10.1111/j.1365-2966.2010.16309.x}, \href
  {http://adsabs.harvard.edu/abs/2010MNRAS.404..475J} {404, 475}

\bibitem[\protect\citeauthoryear{{Johansen}, {Oishi}, {Mac Low}, {Klahr},
  {Henning}  \& {Youdin}}{{Johansen} et~al.}{2007}]{Johansen.et.al.2007}
{Johansen} A.,  {Oishi} J.~S.,  {Mac Low} M.-M.,  {Klahr} H.,  {Henning} T.,
  {Youdin} A.,  2007, \mn@doi [\nat] {10.1038/nature06086}, \href
  {http://adsabs.harvard.edu/abs/2007Natur.448.1022J} {448, 1022}

\bibitem[\protect\citeauthoryear{{Kippenhahn}, {Weigert}  \&
  {Weiss}}{{Kippenhahn} et~al.}{2012}]{2012sse..book.....K}
{Kippenhahn} R.,  {Weigert} A.,   {Weiss} A.,  2012, {Stellar Structure and
  Evolution}, \mn@doi{10.1007/978-3-642-30304-3.
}

\bibitem[\protect\citeauthoryear{{Kley}, {Bitsch}  \& {Klahr}}{{Kley}
  et~al.}{2009}]{Kley.et.al.2009}
{Kley} W.,  {Bitsch} B.,   {Klahr} H.,  2009, \mn@doi [\aap]
  {10.1051/0004-6361/200912072}, \href
  {http://adsabs.harvard.edu/abs/2009A%26A...506..971K} {506, 971}

\bibitem[\protect\citeauthoryear{{Kobayashi} \& {Tanaka}}{{Kobayashi} \&
  {Tanaka}}{2018}]{KB2018}
{Kobayashi} H.,  {Tanaka} H.,  2018, preprint, \href
  {http://adsabs.harvard.edu/abs/2018arXiv180607354K} {} (\mn@eprint {arXiv}
  {1806.07354})

\bibitem[\protect\citeauthoryear{{Kokubo} \& {Ida}}{{Kokubo} \&
  {Ida}}{1998}]{KokuboIda1998}
{Kokubo} E.,  {Ida} S.,  1998, \mn@doi [\icarus] {10.1006/icar.1997.5840},
  \href {http://adsabs.harvard.edu/abs/1998Icar..131..171K} {131, 171}

\bibitem[\protect\citeauthoryear{{Lambrechts} \& {Johansen}}{{Lambrechts} \&
  {Johansen}}{2012}]{Lambrechts&Johansen2012}
{Lambrechts} M.,  {Johansen} A.,  2012, \mn@doi [\aap]
  {10.1051/0004-6361/201219127}, \href
  {http://adsabs.harvard.edu/abs/2012A%26A...544A..32L} {544, A32}

\bibitem[\protect\citeauthoryear{{Lambrechts}, {Johansen}  \&
  {Morbidelli}}{{Lambrechts} et~al.}{2014}]{Lambrechts.et.al.2014}
{Lambrechts} M.,  {Johansen} A.,   {Morbidelli} A.,  2014, \mn@doi [\aap]
  {10.1051/0004-6361/201423814}, \href
  {http://adsabs.harvard.edu/abs/2014A%26A...572A..35L} {572, A35}

\bibitem[\protect\citeauthoryear{{Lega}, {Crida}, {Bitsch}  \&
  {Morbidelli}}{{Lega} et~al.}{2014}]{Lega.et.al.2014}
{Lega} E.,  {Crida} A.,  {Bitsch} B.,   {Morbidelli} A.,  2014, \mn@doi
  [\mnras] {10.1093/mnras/stu304}, \href
  {http://adsabs.harvard.edu/abs/2014MNRAS.440..683L} {440, 683}

\bibitem[\protect\citeauthoryear{{Lega}, {Morbidelli}, {Bitsch}, {Crida}  \&
  {Szul{\'a}gyi}}{{Lega} et~al.}{2015}]{Lega2015}
{Lega} E.,  {Morbidelli} A.,  {Bitsch} B.,  {Crida} A.,   {Szul{\'a}gyi} J.,
  2015, \mn@doi [\mnras] {10.1093/mnras/stv1385}, \href
  {http://adsabs.harvard.edu/abs/2015MNRAS.452.1717L} {452, 1717}

\bibitem[\protect\citeauthoryear{{Lin} \& {Papaloizou}}{{Lin} \&
  {Papaloizou}}{1986}]{LinPapaloizou1986}
{Lin} D.~N.~C.,  {Papaloizou} J.,  1986, \mn@doi [\apj] {10.1086/164653}, \href
  {http://adsabs.harvard.edu/abs/1986ApJ...309..846L} {309, 846}

\bibitem[\protect\citeauthoryear{{Lodders}, {Palme}  \& {Gail}}{{Lodders}
  et~al.}{2009}]{Lodders.et.al.2009}
{Lodders} K.,  {Palme} H.,   {Gail} H.-P.,  2009, \mn@doi [Landolt
  B{\"o}rnstein] {10.1007/978-3-540-88055-4_34}, \href
  {http://adsabs.harvard.edu/abs/2009LanB...4B...44L} {}

\bibitem[\protect\citeauthoryear{{Masset}}{{Masset}}{2017}]{masset2017}
{Masset} F.~S.,  2017, \mn@doi [\mnras] {10.1093/mnras/stx2271}, \href
  {http://adsabs.harvard.edu/abs/2017MNRAS.472.4204M} {472, 4204}

\bibitem[\protect\citeauthoryear{{Masset} \& {Casoli}}{{Masset} \&
  {Casoli}}{2010}]{MC2010}
{Masset} F.~S.,  {Casoli} J.,  2010, \mn@doi [\apj]
  {10.1088/0004-637X/723/2/1393}, \href
  {http://adsabs.harvard.edu/abs/2010ApJ...723.1393M} {723, 1393}

\bibitem[\protect\citeauthoryear{{Masset} \& {Papaloizou}}{{Masset} \&
  {Papaloizou}}{2003}]{2003ApJ...588..494M}
{Masset} F.~S.,  {Papaloizou} J.~C.~B.,  2003, \mn@doi [\apj] {10.1086/373892},
  \href {http://adsabs.harvard.edu/abs/2003ApJ...588..494M} {588, 494}

\bibitem[\protect\citeauthoryear{{McNally}, {Nelson}  \&
  {Paardekooper}}{{McNally} et~al.}{2018}]{2018MNRAS.477.4596M}
{McNally} C.~P.,  {Nelson} R.~P.,   {Paardekooper} S.-J.,  2018, \mn@doi
  [\mnras] {10.1093/mnras/sty905}, \href
  {http://adsabs.harvard.edu/abs/2018MNRAS.477.4596M} {477, 4596}

\bibitem[\protect\citeauthoryear{{Migaszewski}}{{Migaszewski}}{2015}]{Migaszewski2015}
{Migaszewski} C.,  2015, \mn@doi [\mnras] {10.1093/mnras/stv1739}, \href
  {http://adsabs.harvard.edu/abs/2015MNRAS.453.1632M} {453, 1632}

\bibitem[\protect\citeauthoryear{{Miguel}, {Guilera}  \& {Brunini}}{{Miguel}
  et~al.}{2011}]{Miguel2011}
{Miguel} Y.,  {Guilera} O.~M.,   {Brunini} A.,  2011, \mn@doi [\mnras]
  {10.1111/j.1365-2966.2011.19264.x}, \href
  {http://adsabs.harvard.edu/abs/2011MNRAS.417..314M} {417, 314}

\bibitem[\protect\citeauthoryear{{Mizuno}}{{Mizuno}}{1980}]{Mizuno1980}
{Mizuno} H.,  1980, \mn@doi [Progress of Theoretical Physics]
  {10.1143/PTP.64.544}, \href
  {http://adsabs.harvard.edu/abs/1980PThPh..64..544M} {64, 544}

\bibitem[\protect\citeauthoryear{{Ndugu}, {Bitsch}  \& {Jurua}}{{Ndugu}
  et~al.}{2018}]{Ndugu2018}
{Ndugu} N.,  {Bitsch} B.,   {Jurua} E.,  2018, \mn@doi [\mnras]
  {10.1093/mnras/stx2815}, \href
  {http://adsabs.harvard.edu/abs/2018MNRAS.474..886N} {474, 886}

\bibitem[\protect\citeauthoryear{{Ohtsuki}, {Stewart}  \& {Ida}}{{Ohtsuki}
  et~al.}{2002}]{Ohtsuki.et.al.2002}
{Ohtsuki} K.,  {Stewart} G.~R.,   {Ida} S.,  2002, \mn@doi [\icarus]
  {10.1006/icar.2001.6741}, \href
  {http://adsabs.harvard.edu/abs/2002Icar..155..436O} {155, 436}

\bibitem[\protect\citeauthoryear{{Ormel} \& {Klahr}}{{Ormel} \&
  {Klahr}}{2010}]{Ormel&Klahr2010}
{Ormel} C.~W.,  {Klahr} H.~H.,  2010, \mn@doi [\aap]
  {10.1051/0004-6361/201014903}, \href
  {http://adsabs.harvard.edu/abs/2010A%26A...520A..43O} {520, A43}

\bibitem[\protect\citeauthoryear{{Ormel}, {Dullemond}  \& {Spaans}}{{Ormel}
  et~al.}{2010}]{Ormel.et.al2010}
{Ormel} C.~W.,  {Dullemond} C.~P.,   {Spaans} M.,  2010, \mn@doi [\apjl]
  {10.1088/2041-8205/714/1/L103}, \href
  {http://adsabs.harvard.edu/abs/2010ApJ...714L.103O} {714, L103}

\bibitem[\protect\citeauthoryear{{Paardekooper}}{{Paardekooper}}{2014}]{2014MNRAS.444.2031P}
{Paardekooper} S.-J.,  2014, \mn@doi [\mnras] {10.1093/mnras/stu1542}, \href
  {http://adsabs.harvard.edu/abs/2014MNRAS.444.2031P} {444, 2031}

\bibitem[\protect\citeauthoryear{{Paardekooper} \& {Mellema}}{{Paardekooper} \&
  {Mellema}}{2006}]{PM2006}
{Paardekooper} S.-J.,  {Mellema} G.,  2006, \mn@doi [\aap]
  {10.1051/0004-6361:20066304}, \href
  {http://adsabs.harvard.edu/abs/2006A%26A...459L..17P} {459, L17}

\bibitem[\protect\citeauthoryear{{Paardekooper} \& {Papaloizou}}{{Paardekooper}
  \& {Papaloizou}}{2008}]{PP2008}
{Paardekooper} S.-J.,  {Papaloizou} J.~C.~B.,  2008, \mn@doi [\aap]
  {10.1051/0004-6361:20078702}, \href
  {http://adsabs.harvard.edu/abs/2008A%26A...485..877P} {485, 877}

\bibitem[\protect\citeauthoryear{{Paardekooper}, {Baruteau}, {Crida}  \&
  {Kley}}{{Paardekooper} et~al.}{2010}]{paardekooper.etal2010}
{Paardekooper} S.-J.,  {Baruteau} C.,  {Crida} A.,   {Kley} W.,  2010, \mn@doi
  [\mnras] {10.1111/j.1365-2966.2009.15782.x}, \href
  {http://adsabs.harvard.edu/abs/2010MNRAS.401.1950P} {401, 1950}

\bibitem[\protect\citeauthoryear{{Paardekooper}, {Baruteau}  \&
  {Kley}}{{Paardekooper} et~al.}{2011}]{paardekooper.etal2011}
{Paardekooper} S.-J.,  {Baruteau} C.,   {Kley} W.,  2011, \mn@doi [\mnras]
  {10.1111/j.1365-2966.2010.17442.x}, \href
  {http://adsabs.harvard.edu/abs/2011MNRAS.410..293P} {410, 293}

\bibitem[\protect\citeauthoryear{{Papaloizou} \& {Terquem}}{{Papaloizou} \&
  {Terquem}}{1999}]{Papaloizou.Terquem.1999}
{Papaloizou} J.~C.~B.,  {Terquem} C.,  1999, \mn@doi [\apj] {10.1086/307581},
  \href {http://adsabs.harvard.edu/abs/1999ApJ...521..823P} {521, 823}

\bibitem[\protect\citeauthoryear{{Pierens}}{{Pierens}}{2015}]{2015MNRAS.454.2003P}
{Pierens} A.,  2015, \mn@doi [\mnras] {10.1093/mnras/stv2024}, \href
  {http://adsabs.harvard.edu/abs/2015MNRAS.454.2003P} {454, 2003}

\bibitem[\protect\citeauthoryear{{Pringle}}{{Pringle}}{1981}]{Pringle1981}
{Pringle} J.~E.,  1981, \mn@doi [\araa] {10.1146/annurev.aa.19.090181.001033},
  \href {http://adsabs.harvard.edu/abs/1981ARA%26A..19..137P} {19, 137}

\bibitem[\protect\citeauthoryear{{Rafikov}}{{Rafikov}}{2004}]{Rafikov2004}
{Rafikov} R.~R.,  2004, \mn@doi [\aj] {10.1086/423216}, \href
  {http://adsabs.harvard.edu/abs/2004AJ....128.1348R} {128, 1348}

\bibitem[\protect\citeauthoryear{{Ronco}, {Guilera}  \& {de
  El{\'{\i}}a}}{{Ronco} et~al.}{2017}]{Ronco.et.al.2017}
{Ronco} M.~P.,  {Guilera} O.~M.,   {de El{\'{\i}}a} G.~C.,  2017, \mn@doi
  [\mnras] {10.1093/mnras/stx1746}, \href
  {http://adsabs.harvard.edu/abs/2017MNRAS.471.2753R} {471, 2753}

\bibitem[\protect\citeauthoryear{{Safronov}}{{Safronov}}{1969}]{Safronov1969}
{Safronov} V.~S.,  1969, {Evoliutsiia doplanetnogo oblaka.}

\bibitem[\protect\citeauthoryear{{San Sebasti\'an}, {Guilera}  \&
  {Parisi}}{{San Sebasti\'an} et~al.}{2018}]{SanSebastian.et.al.2018}
{San Sebasti\'an} I.~L.,  {Guilera} O.~M.,   {Parisi} M.~G.,  2018, Submitted
  to \aap

\bibitem[\protect\citeauthoryear{{Shakura} \& {Sunyaev}}{{Shakura} \&
  {Sunyaev}}{1973}]{Shakura1973}
{Shakura} N.~I.,  {Sunyaev} R.~A.,  1973, \aap, \href
  {http://adsabs.harvard.edu/abs/1973A%26A....24..337S} {24, 337}

\bibitem[\protect\citeauthoryear{{Shiraishi} \& {Ida}}{{Shiraishi} \&
  {Ida}}{2008}]{Shiraishi-Ida2008}
{Shiraishi} M.,  {Ida} S.,  2008, \mn@doi [\apj] {10.1086/590226}, \href
  {http://adsabs.harvard.edu/abs/2008ApJ...684.1416S} {684, 1416}

\bibitem[\protect\citeauthoryear{{Stevenson}}{{Stevenson}}{1982}]{Stevenson1982}
{Stevenson} D.~J.,  1982, \mn@doi [\planss] {10.1016/0032-0633(82)90108-8},
  \href {http://adsabs.harvard.edu/abs/1982P%26SS...30..755S} {30, 755}

\bibitem[\protect\citeauthoryear{{Stokes}}{{Stokes}}{1851}]{1851TCaPS...9....8S}
{Stokes} G.~G.,  1851, Transactions of the Cambridge Philosophical Society,
  \href {http://adsabs.harvard.edu/abs/1851TCaPS...9....8S} {9, 8}

\bibitem[\protect\citeauthoryear{{Tanaka} \& {Ida}}{{Tanaka} \&
  {Ida}}{1999}]{Tanaka-Ida1999}
{Tanaka} H.,  {Ida} S.,  1999, \mn@doi [\icarus] {10.1006/icar.1999.6107},
  \href {http://adsabs.harvard.edu/abs/1999Icar..139..350T} {139, 350}

\bibitem[\protect\citeauthoryear{{Tanaka}, {Takeuchi}  \& {Ward}}{{Tanaka}
  et~al.}{2002}]{Tanaka.et.al.2002}
{Tanaka} H.,  {Takeuchi} T.,   {Ward} W.~R.,  2002, \mn@doi [\apj]
  {10.1086/324713}, \href {http://adsabs.harvard.edu/abs/2002ApJ...565.1257T}
  {565, 1257}

\bibitem[\protect\citeauthoryear{{Tanigawa} \& {Ikoma}}{{Tanigawa} \&
  {Ikoma}}{2007}]{TanigawaIkoma2007}
{Tanigawa} T.,  {Ikoma} M.,  2007, \mn@doi [\apj] {10.1086/520499}, \href
  {http://adsabs.harvard.edu/abs/2007ApJ...667..557T} {667, 557}

\bibitem[\protect\citeauthoryear{{Ward}}{{Ward}}{1997}]{Ward1997}
{Ward} W.~R.,  1997, \mn@doi [\icarus] {10.1006/icar.1996.5647}, \href
  {http://adsabs.harvard.edu/abs/1997Icar..126..261W} {126, 261}

\bibitem[\protect\citeauthoryear{{Weidenschilling}}{{Weidenschilling}}{1977}]{Weidenschilling1977}
{Weidenschilling} S.~J.,  1977, \mn@doi [\mnras] {10.1093/mnras/180.1.57},
  \href {http://adsabs.harvard.edu/abs/1977MNRAS.180...57W} {180, 57}

\bibitem[\protect\citeauthoryear{{Whipple}}{{Whipple}}{1972}]{1972fpp..conf..211W}
{Whipple} F.~L.,  1972, in {Elvius} A.,  ed., From Plasma to Planet. p.~211

\bibitem[\protect\citeauthoryear{{Youdin} \& {Goodman}}{{Youdin} \&
  {Goodman}}{2005}]{YoudinGoodman2005}
{Youdin} A.~N.,  {Goodman} J.,  2005, \mn@doi [\apj] {10.1086/426895}, \href
  {http://adsabs.harvard.edu/abs/2005ApJ...620..459Y} {620, 459}

\bibitem[\protect\citeauthoryear{{Youdin} \& {Lithwick}}{{Youdin} \&
  {Lithwick}}{2007}]{Youdin2007}
{Youdin} A.~N.,  {Lithwick} Y.,  2007, \mn@doi [\icarus]
  {10.1016/j.icarus.2007.07.012}, \href
  {http://adsabs.harvard.edu/abs/2007Icar..192..588Y} {192, 588}

\makeatother
\end{thebibliography}


\appendix

\section{Corotation torque prescriptions from 2D and 3D hydrodynamical simulations}
\label{apex-1}

As we mentioned in Sec.~\ref{sec2-4-1}, the corotation torque derived by \citet{paardekooper.etal2011} from 2D hydrodynamical simulations is composed by two components, a barotropic and an entropy one, and thus $\Gamma_{\text{C}}= \Gamma_{\text{C,bar}} + \Gamma_{\text{C,ent}}$. The barotropic contribution is given by 
\begin{equation}
 \Gamma_{\text{C,bar}}= F(p_{\nu})G(p_{\nu})\Gamma_{\text{hs,bar}} + [1-K(p_{\nu})]\Gamma_{\text{C,lin,bar}},
 \label{eq7-sec2-4}
\end{equation}
where $\Gamma_{\text{hs,bar}}$ is the fully unsaturated horseshoe drag part given by 
\begin{equation}
\Gamma_{\text{hs,bar}}= 1.1(1.5 - \alpha' ) \dfrac{\Gamma_0}{\gamma_{\text{eff}}},
\label{eq8-sec2-4}
\end{equation}
and $\Gamma_{\text{C,lin,bar}}$ is the linear barotropic corotation part given by 
\begin{equation}
\Gamma_{\text{C,lin,bar}}= 0.7(1.5 - \alpha' ) \dfrac{\Gamma_0}{\gamma_{\text{eff}}},
\label{eq9-sec2-4}
\end{equation}
and where the function $F$ governs the saturation, while the functions $G$ and $K$ govern the cut-off at high viscosity \citep[see][for the explicit form of such functions]{paardekooper.etal2011}. $p_{\nu}$ is the viscosity saturation parameter given by
\begin{equation}
p_{\nu}= \dfrac{2}{3}\sqrt{\dfrac{R_{\text{P}}^2\Omega_{\text{P}}x_s^3}{2\pi \nu_{\text{P}}}},
\label{eqA4}
\end{equation}
where $\nu_{\text{P}}$ is the viscosity at the planet location and $x_s$ is the non-dimensional half-width of the horseshoe region, given by
\begin{equation}
x_s= \dfrac{1.1}{\gamma_{\text{eff}}} \left( \dfrac{0.4}{b/h} \right)^{-1/4} \sqrt{\dfrac{q}{h}}, 
\label{eqA5}
\end{equation}
and where \citet{paardekooper.etal2011} considered the softening parameter $b/h= 0.4$ in order to mimic 3D effects. 
Finally, the entropy contribution to the corotation torque is given by
\begin{align}
\Gamma_{\text{C,ent}}= & F(p_{\nu})F(p_{\chi})\sqrt{G(p_{\nu})G(p_{\chi})}\Gamma_{\text{hs,ent}} + \nonumber \\
& \sqrt{[1-K(p_{\nu})][1-K(p_{\chi})]} \Gamma_{\text{C,lin,ent}},
\label{eq11-sec2-4}
\end{align}
where $\Gamma_{\text{hs,ent}}$ is the fully unsaturated entropy part, given by
\begin{equation}
\Gamma_{\text{hs,ent}}= 7.9 \left( \dfrac{\xi}{\gamma_{\text{eff}}} \right) \dfrac{\Gamma_0}{\gamma_{\text{eff}}},
\label{eq12-sec2-4}
\end{equation}
with $\xi= \beta' - (\gamma - 1) \alpha'$, and $\Gamma_{\text{C,lin,ent}}$ is the linear entropy part given by 
 \begin{equation}
\Gamma_{\text{C,lin,ent}}=  \left( 2.2 - \dfrac{1.4}{\gamma_{\text{eff}}} \right) \dfrac{\Gamma_0}{\gamma_{\text{eff}}}.
\label{eq13-sec2-4}
\end{equation}
The parameter $p_{\chi}$ is the saturation parameter associated with thermal diffusion, and it is given by
\begin{equation}
p_{\chi}= \sqrt{\dfrac{R_{\text{P}}^2\Omega_{\text{P}}x_s^3}{2\pi \chi_{\text{P}}}}. 
\label{eq10-sec2-4}
\end{equation}

On the other hand, \citet{jm2017} improved the expression of the corotation torque using 3D hydrodynamical simulations. In their expression, the latter has four components $\Gamma_{\text{C}}= \Gamma_{\text{C,vor}}+\Gamma_{\text{C,ent}}+\Gamma_{\text{C,temp}}+\Gamma_{\text{C,vv}}$, associated with the radial gradient of vortensity, entropy, and temperature, and with a viscously created vortensity. The different contributions for the corotation torque are given by  
\begin{equation}
\Gamma_{\text{C,vor}}= \epsilon_b \Gamma_{\text{vor,hs}} + (1-\epsilon_b) \Gamma_{\text{vor,lin}},  
 \label{eq4-sec2-4-1}
\end{equation}
where $\epsilon_b= 1/(1+30hz_{\nu})$, with $z_{\nu}= R_{\text{P}} \nu_{\text{P}}/\Omega_{\text{P}}x_s^3$. A new result from \citet{jm2017} is the improvement in the calculation of the half width of the horseshoe region $x_s$ given now by
\begin{equation}
x_s= \dfrac{1.05\sqrt{q/h'} + 3.4q^{7/3}/h'^6}{1 + 2q^2/h'^6}R_{\text{P}},
\label{eq5-sec2-4-1}
\end{equation}
where $h'= h\sqrt{\gamma}$. $\Gamma_{\text{vor,lin}}$ is the linear part of the vortencity component and is given by  
\begin{equation}
\Gamma_{\text{vor,lin}}= (0.976 - 0.64 \alpha') \dfrac{\Gamma_0}{\gamma}, 
\label{eq6-sec2-4-1}
\end{equation}
while $\Gamma_{\text{vor,hs}}$ is the horseshoe drag part given by 
\begin{equation}
\Gamma_{\text{vor,hs}}= \mathcal{F}_v \Gamma_{\text{vor,uhs}}, 
\label{eq7-sec2-4-1}
\end{equation}
where $\mathcal{F}_v$ is the saturation function for the vortensity component of the horseshoe drag \citep[see][for the explicit form of $\mathcal{F}_v$]{jm2017}, and 
\begin{equation}
\Gamma_{\text{vor,uhs}}= \dfrac{3}{4}\left(\dfrac{3}{2}-\alpha'\right)\Sigma_{\text{g,P}}\Omega_{\text{P}}^2 x_s^4,
\label{eq8-sec2-4-1}
\end{equation}
is the vortensity unsaturated horseshoe drag component. The entropy component of the corotation torque is given by
 \begin{equation}
\Gamma_{\text{C,ent}}= \epsilon_{\nu} \epsilon_{\chi} \Gamma_{\text{ent,hs}} + (1-\epsilon_{\nu} \epsilon_{\chi}) \Gamma_{\text{ent,lin}},  
 \label{eq9-sec2-4-1}
\end{equation}
where $\epsilon_{\nu}= 1/(1+6hz_{\nu})^2$ and $\epsilon_{\chi}= 1/(1+15hz_{\chi})^2$, with $z_{\chi}= R_{\text{P}} \chi_{\text{P}}/\Omega_{\text{P}}x_s^3$. The linear part is given by 
\begin{equation}
\Gamma_{\text{ent,lin}}= 0.8 \xi \dfrac{\Gamma_0}{\gamma}, 
\label{eq10-sec2-4-1}
\end{equation}
where $\xi= \beta' - 0.4\alpha' -0.64$. The horseshoe drag term of the entropy component is given by 
\begin{equation}
\Gamma_{\text{ent,hs}}= \mathcal{F}_S \Gamma_{\text{ent,uhs}}, 
\label{eq11-sec2-4-1}
\end{equation}
where $\mathcal{F}_S$ is saturation function of the entropy corotation torque \citep[see][for its explicit form]{jm2017} and the unsaturated horseshoe drag contribution is given by 
\begin{equation}
\Gamma_{\text{ent,uhs}}= 3.3 \xi \Sigma_{\text{g,P}}\Omega_{\text{P}}^2 x_s^4. 
\label{eq12-sec2-4-1}
\end{equation}
The temperature component is given by 
\begin{equation}
\Gamma_{\text{C,temp}}= \epsilon_{\nu} \Gamma_{\text{temp,hs}} + (1-\epsilon_{\nu}) \Gamma_{\text{temp,lin}},  
 \label{eq13-sec2-4-1}
\end{equation}
being the linear contribution 
\begin{equation}
\Gamma_{\text{temp,lin}}= 1.0 \beta' \dfrac{\Gamma_0}{\gamma}, 
\label{eq14-sec2-4-1}
\end{equation}
and the temperature horseshoe drag  
\begin{equation}
\Gamma_{\text{temp,hs}}= \mathcal{F}_T \Gamma_{\text{temp,uhs}}, 
\label{eq15-sec2-4-1}
\end{equation}
where $\mathcal{F}_T$ is the saturation function \citep{jm2017}, and the unsaturated horseshoe drag component is given by 
\begin{equation}
\Gamma_{\text{ent,uhs}}= 0.73 \beta' \Sigma_{\text{g,P}}\Omega_{\text{P}}^2 x_s^4. 
\label{eq16-sec2-4-1}
\end{equation}
Finally, the last contribution to the corotation torque is given by 
\begin{equation}
\Gamma_{\text{C,vv}}= \dfrac{4\pi\xi}{\gamma} \Sigma_{\text{g,P}}\Omega_{\text{P}}^2 x_s^4 \epsilon_b z_{\nu} g(z_{\nu},z_{\chi}),
\label{eq17-sec2-4-1}
\end{equation}
where the explicit form of $g$ can be found in \citet[][Eq.~64]{jm2017}.

\begin{figure}
    \centering
    \includegraphics[width= 0.345\textwidth,angle= -90]{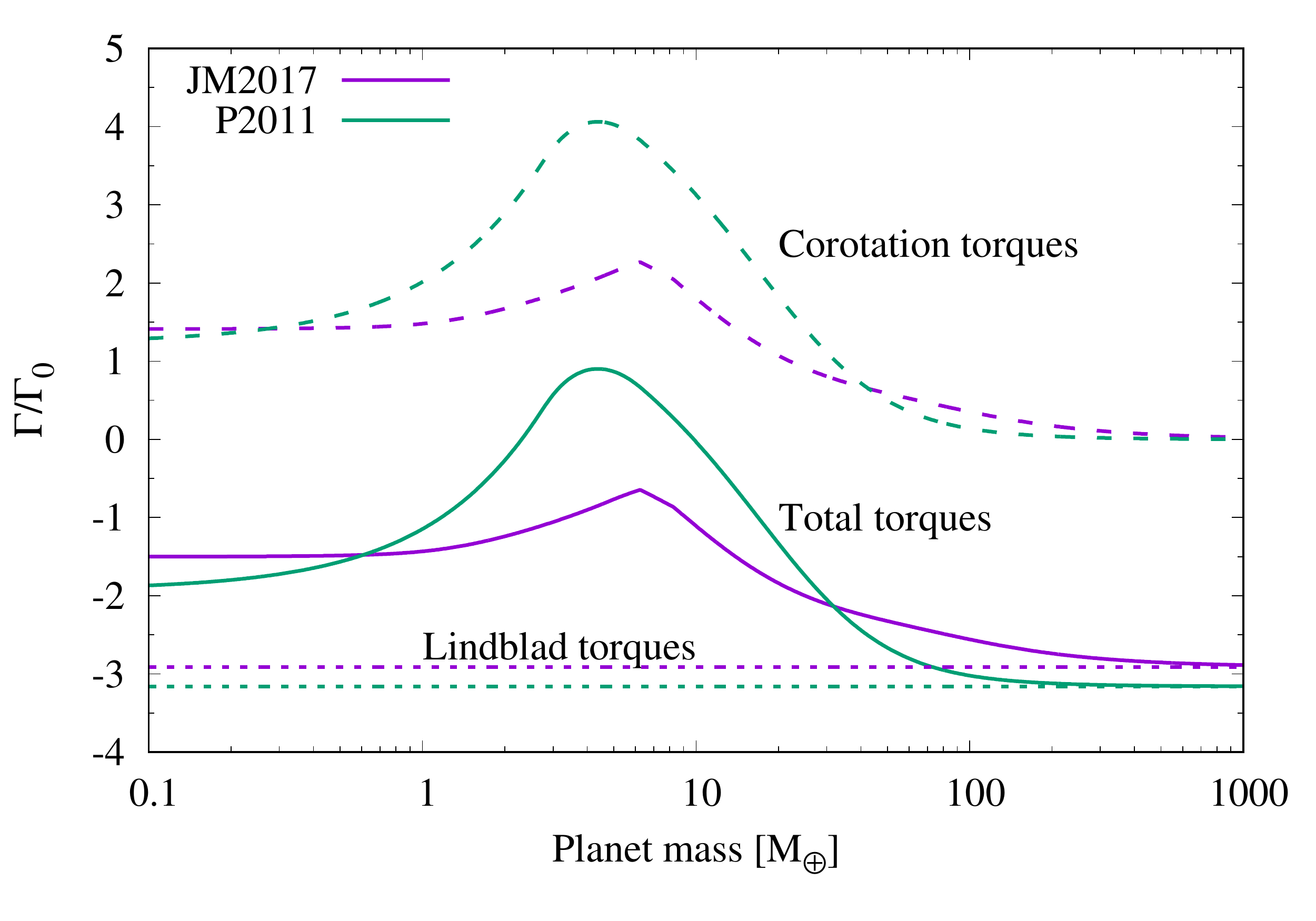} 
    \caption{Comparison between the normalised Lindblad, corotation and total torques as a function of the planet mass and for a fixed planet location of 3.5~au for the case of the fiducial disc at 50~Kyr.}
 \label{fig:fig12}
\end{figure}

\begin{figure}
    \centering
    \includegraphics[width= 0.345\textwidth,angle= -90]{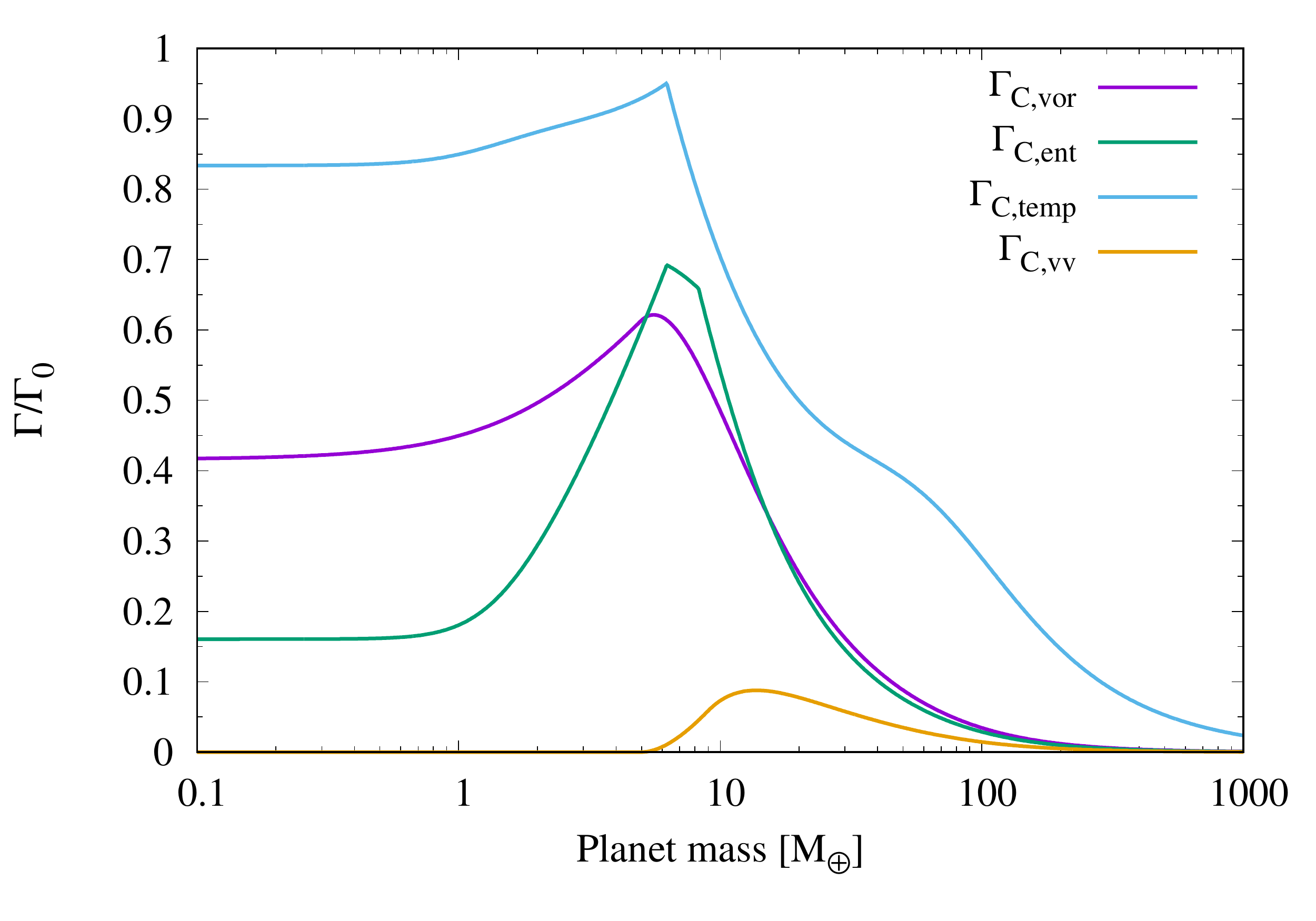} 
    \caption{The components of the corotation torque given by \citet{jm2017} as a function of the planet mass and for a fixed planet location of 3.5~au for the case of the fiducial disc at 50~Kyr.}
 \label{fig:fig13}
\end{figure}

In Fig.~\ref{fig:fig12}, we plot the normalised Lindblad, corotation and total torques as a function of the planet mass for a fixed planet location of 3.5~au, using both torque recipes for the case of the fiducial disc at 50~Kyr. We see that while the Lindblad torques remain very similar between both models, the corotation torque given by \citet{paardekooper.etal2011} overestimates the corresponding one from \citet{jm2017}. Finally, in Fig.~\ref{fig:fig13} we plot the components of the corotation torque given by \citet{jm2017}. The functional forms of such components are very similar to the one presented by \citet{jm2017} (right panel of their  figure~8). The differences in the values that reach the different components with respect to \citet{jm2017} arise from the difference in the disc model adopted.    

\bsp	
\label{lastpage}
\end{document}